\documentclass[prx,twocolumn,nofootinbib,citeautoscript,longbibliography,notitlepage,superscriptaddress]{revtex4-1}

\usepackage{graphicx}
\usepackage{dcolumn}
\usepackage{bm}
\usepackage{color}
\usepackage{amsmath}
\usepackage{tabularx,graphicx}
\usepackage{epstopdf}
\usepackage{latexsym}
\usepackage{amssymb}
\usepackage{amsmath}
\usepackage{color, colortbl}
\usepackage{psfrag}
\usepackage{bbm}
\usepackage{bm}
\usepackage{titlesec}
\usepackage{dsfont}
\usepackage{feynmp}
\usepackage{slashed}
\usepackage{multirow}
\usepackage[tight]{subfigure}

\usepackage[papersize={8.5in,11in}]{geometry}

\usepackage{color}

\usepackage{physics}

\renewcommand{\epsilon}{\varepsilon}

\allowdisplaybreaks[3]

\newcommand{\red}[1]{\textcolor{red}{#1}}

\begin{document}



\title{Effects of the interplay between fermionic interactions and disorders in the nodal-line superconductors}

\date{\today}

\author{Wen-Hao Bian}
\affiliation{Department of Physics, Tianjin University, Tianjin 300072, P.R. China}
\author{Xiao-Zhuo Chu}
\affiliation{Department of Physics, Tianjin University, Tianjin 300072, P.R. China}
\author{Jing Wang}
\altaffiliation{Corresponding author: jing$\textunderscore$wang@tju.edu.cn}
\affiliation{Department of Physics, Tianjin University, Tianjin 300072, P.R. China}
\affiliation{Tianjin Key Laboratory of Low Dimensional Materials Physics and
Preparing Technology, Tianjin University, Tianjin 300072, P.R. China}

\begin{abstract}

We carefully study the interplay between short-range fermion-fermion interactions
and disorder scatterings beneath the superconducting dome of the noncentrosymmetric
nodal-line superconductors. With the application of renormalization group, the energy-dependent
coupled flows of all these associated interaction parameters are established after taking into account
the potential low-energy physical ingredients including both kinds of fermionic interactions and disorder
couplings as well as their competitions on the same footing. Encoding the low-energy information from these entangled
evolutions gives rise to several interesting behaviors in the low-energy regime.
At the clean limit, fermion-fermion interactions decrease with lowering the energy
scales but conversely fermion velocities climb up and approach certain saturated values.
This yields a slight decrease or increase in the anisotropy of fermion velocities depending upon
their initial ratio. After bringing out four kinds of disorders designated by the random
charge ($\Delta_{1}$), random mass ($\Delta_{2}$), random axial chemical potential ($\Delta_{3}$),
and spin-orbit scatterers ($\Delta_{4}$) based on their own unique features, we begin
with presenting the distinct low-energy fates of these disorders.
For the presence of sole disorder, its strength becomes either relevant ($\Delta_{1,4}$)
or irrelevant($\Delta_{2,3}$) in the low-energy regime. However,
the competition for multiple sorts of disorders is capable of qualitatively reshaping
the low-energy properties of disorders $\Delta_{2,3,4}$. Besides, it can generate an initially absent
disorder as long as two of $\Delta_{1,2,3}$ are present. In addition, the fermion-fermion couplings
are insensitive to the presence of disorder $\Delta_4$ but rather substantially modified by $\Delta_1$,
$\Delta_2$, or $\Delta_3$, and they evolve toward zero or certain finite nonzero values under
the coexistence of distinct disorders. Furthermore, the fermion velocities flow toward certain
finite saturated value for the only presence of $\Delta_{2,3}$ and vanish for all other situations.
As to their ratio, it acquires a little increase once the disorder is subordinate to fermion-fermion
interactions, otherwise keeps some fixed constant.
\end{abstract}

\pacs{\red{74.20.Mn, 74.20.Rp}}

\maketitle


\section{Introduction}

\begin{figure}
\centering
\includegraphics[width=2.5in]{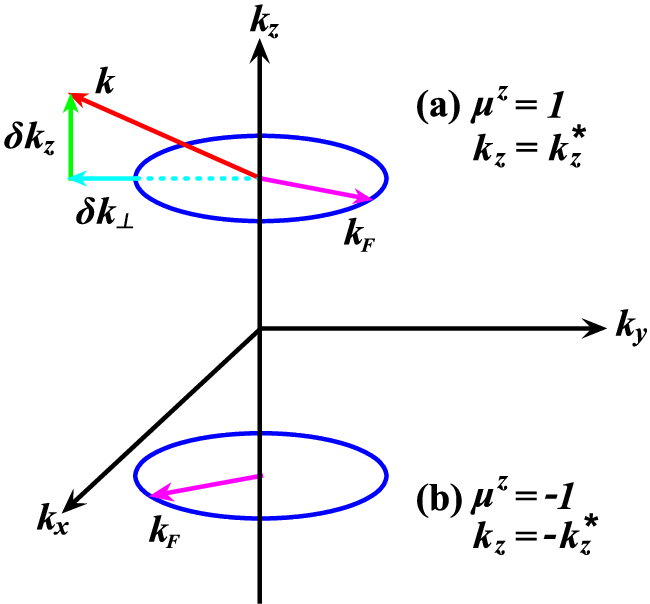}
\vspace{0.15cm}
\caption{(Color online) Schematic structure for low-energy
nodal-line excitations.}\label{Fig_fermion_surface}
\end{figure}

Since the superconductivity was discovered in the beginning of the last century,
it has been attracting a vast amount of both experimental and theoretical efforts and
becoming one of the most important and hottest topics
in contemporary condensed matter physics.
Principally, conventional superconductors that are well understood by the celebrated BCS theory~\cite{BCS1957PR}
possess an isotropic $s$-wave gap so that the gapless fermionic excitations are not allowed in
the low-energy regime due to the absence of any nodal points around the Fermi surface~\cite{Tinkham1996Book,Anderson1997Book,Larkin2005Book}.
In marked comparison, most of the cuprate high-temperature superconductors (HTSCs) can be regarded
as effectively (quasi) two-dimensional compounds~\cite{Lee2006RMP,Berezinskii1971JETP,
Kosterlitz_Thouless1973JPC}. Particularly, these HTSCs are commonly
equipped with a $d$-wave superconducting gap which allows four gapless nodal points on the Fermi  surface.~\cite{Lee2006RMP,Ding1996Nature,Loeser1996Science,
Valla1999Science,Yoshida2003PRL,Ronning2003PRL}.
As a result, the gapless fermionic quasiparticles can be always excited even at the lowest-energy limit
to participate in potential physical processes and induce a plethora of critical
behaviors~\cite{Lee2006RMP}. Besides these very nodal-point materials, a number of
groups advocated that there are several three-dimensional (3D) compounds
including the heavy-fermion superconductors $\mathrm{CePtSi_3}$~\cite{Bonalde2005PRL,Izawa2005PRL,Tateiwa2005JPSJ},
UCoGe~\cite{Slooten2009PRL,Gasparini2010JLTP} and pnictides superconductors  $\mathrm{Ba(Fe_{1-x}Co_x)_2As_2}$~\cite{Reid2010PRB,Reid2010PRL}, FeSe~\cite{Song2011Science,Watashige2015PRX} and $\mathrm{(Ba_{1-x}K_x)Fe_2As_2}$~\cite{Reid2016PRL},
which share an analogous but revised version of nodal structure with cuprate HTSCs, namely,
owning the nodal-line points as illustrated in Fig.~\ref{Fig_fermion_surface},
and are consequently dubbed the nodal-line superconductors.

Without loss of generality, there at least exist three major ingredients including the dispersion of
low-energy excitations as well as fermion-fermion interactions and disorder
scatters, which are expected to play an crucial role in
determining the low-energy physical properties involving the
ground states, transport quantities, etc.~\cite{Castro2009RMP,Hasan2010RMP,Qi2011RMP,Das2011RMP,
Kotov2012RMP,Altland2002PR,Lee2006RMP,Fradkin2010ARCMP,Sachdev2011book}.
Recently, the significant roles of electronic correlations and their couplings with other physical degrees
in pinning down the low-energy fates and underlying phase transitions of fermionic materials
have gradually attracted a great attention~\cite{Fradkin2009PRL,Vafek2012PRB,
Vafek2014PRB,Herbut2014PRB,Herbut2014PRL,Herbut2015PRB,Herbut2016PRB,Roy-Slager2018PRX,Roy2018PRX,Wang2018-2019}.
Additionally, disorder scatterings that are always present in the real compounds
have been verified to induce a multitude of singular behaviors in the low-energy
regime~\cite{Novoselov2005Nature, Castro2009RMP, Hasan2010RMP, Altland2002PR,
Fradkin2010ARCMP, Das2011RMP, Kotov2012RMP,Wang2011PRB, Wang2013PRB,
Aleiner2006PRL,Foster2008PRB,Lee2017arXiv,Nandkishore2017PRB,Wang-Nandkishore2017PRB,Nandkishore2014PRB,
Roy2017PRB-96,Roy1812.05615,Roy2016SR,Nersesyan1995NPB,Qi2011RMP,Stauber2005PRB,Wang2017QBCP,Wang2018-2019,
Wang2020PRB,Wang2021NPB,Mandal2018AP}.
Considering unique topologies of their Fermi surfaces~\cite{Sigrist1991RMP,Matsuda2006JPCM,Sur2016NJP},
the gapless fermionic excitations can always be excited from the nodal lines
of nodal-line superconductors. On the one hand, such gapless excitations
are naturally interacted with each other~\cite{Vojta2000PRL,Vojta2000IJMPB,
Vojta2000PRB,Sachdev2011book}. On the other hand, they can intimately entangle
with disorder scatterings. These accordingly can result in a plethora of unusual but interesting
behaviors with lowering the energy scales~\cite{Vojta2000PRL,Vojta2000IJMPB,Vojta2000PRB,
Sachdev2011book,Vojta2003RPP,Lohneysen2007RMP,Wang2018-2019}. It is, therefore,
imperative to unbaisedly take into account both fermion-fermion interactions and
impurities to capture more physical information in the low-energy regime. A question
naturally raises how the fermion-fermion interactions and disorders influence
the physical behaviors.

Stimulated by these, we within this work endeavor to investigate
the effects of interplay between short-range fermion-fermion interactions
and disorder scatterings on the low-energy fates in the superconducting dome of
nodal-line superconductors. After carrying both attentively analytical and numerical studies,
several interesting results are obtained with the help of renormalization
group (RG) approach~\cite{Wilson1975RMP,Polchinski9210046,Shankar1994RMP} that treats all the physical
ingredients on the equal footing.

For completeness, four distinct types of disorders are taken into account, which are dubbed the random
charge ($\Delta_{1}$), random mass ($\Delta_{2}$), random axial chemical potential ($\Delta_{3}$),
and spin-orbit scatterers ($\Delta_{4}$) as defined in Eq.~(\ref{Eq_S_dis}) based on their own unique features~\cite{Roy-Saram2016PRB,Roy2018PRX}.
As for the clean limit, we find that the fermionic couplings tend to decrease with lowering the energy
scale but instead the fermion velocities designated in Eq.~(\ref{Eq_S_eff}) climb up
and eventually become saturated attesting to the contributions from
fermion-fermion interactions. Besides, the anisotropy of fermion velocities characterized by $v_z/v_p$
alters and obtains a slight decrease or increase for the initial condition
$v_{z0}/v_{p0}<1$ or $v_{z0}/v_{p0}>1$, respectively.

In comparison, the phenomena are more interesting
under the competition between fermion-fermion and disorders interactions.
At first, for the presence of sole disorder, we notice that the disorder strength
gradually decreases and eventually vanishes with lowering the energy
scale ($\Delta_{2,3}$), and becomes relevant and goes toward
divergence ($\Delta_{1,4}$), respectively. Such divergence of disorder
may turn the system into a disorder-dominated diffusive metallic state~\cite{Fradkin1986PRB,Foster2008PRB,Kobayashi2014PRL,Lai-1409.8675,Goswami2011PRL,Moon-1409.0573,Wang2015PLA}.
As to the presence of multiple sorts of disorders, the disorder $\Delta_1$ strength shares the similar
evolution with its sole presence's but rather the fates of other three types of disorders
are qualitatively reformulated, with $\Delta_4$ being driven irrelevant and $\Delta_2$ (or $\Delta_3$)
being changed from vanishment to divergence. It is also of particular importance to highlight
that an additional sort of disorder which is absent initially can be generated due to the interplay of multi-type
disorders as long as any two of three sorts of disorders $\Delta_1$, $\Delta_2$, and
$\Delta_3$ are present at the starting point. In addition, we figure out that, compared to
their clean-limit counterparts, the evolutions of fermion-fermion interactions are insusceptible
to the sole presence of disorder $\Delta_4$. In contrast, the single presence of $\Delta_1$,
$\Delta_2$, or $\Delta_3$ is able to dramatically modify the fates of fermion-fermion interactions
and hence make them more significant in the low-energy regime and even divergent at certain
critical energy after involving the disorder contributions. However, the furious competition
among distinct types of disorders can be harmful to the divergence of fermionic couplings
and render them flowing toward zero or certain finite nonzero values.
Furthermore, under the competition between fermionic interactions and disorder scatterings,
the fermion velocities are prone to evolving toward zero for the sole
presence of $\Delta_{1}$ (or $\Delta_{4}$) or more than two kinds of disorders but certain
saturated value for the single presence of $\Delta_{2}$ (or $\Delta_{3}$).
As to the ratio $v_z/v_p$, it nearly remains some fixed value once the fermion-fermion
interactions are subordinate to the disorder contribution but instead obtains a little increase while
the disorder becomes less and less important, and the fermion-fermion interaction can provide more
significant contributions in the low-energy regime.

We organize the rest of this paper as follows.
The microscopic model is introduced to establish the effective field
theory in Sec.~\ref{Sec_model}. Then, Sec.~\ref{Sec_RG_analysis} is
followed to carry out the RG analysis and derive the coupled evolution equations of all fermion-fermion
interaction parameters and disorder strengths. Then, we within Sec.~\ref{Sec_clean}
perform a warm-up for the clean limit situation. Thereafter,
Sec.~\ref{Sec_disorder} presents the fates of disorder couplings as well as
fermion-fermion interactions and fermion velocities.
At last, Sec.~\ref{Sec_summary} provides a short summary of our main results.

\section{Microscopic model and effective action}\label{Sec_model}

\subsection{Free microscopic model}

We put our focus on the topological nodal-line superconductors that
satisfy a common symmetry group $\mathcal{G}=C_{4v}\times\mathcal{T}
\times\mathcal{P}$ with $\mathcal{T}$ and $\mathcal{P}$, respectively,
corresponding to the time-reversal and particle-hole symmetries~\cite{Matsuura2013NJP},
and begin with the following noninteracting Hamiltonian to characterize
the low-energy excitations from their nodal-line
points~\cite{Matsuura2013NJP,Moon2017PRB},
\begin{eqnarray}
H_0=\sum_{\mathbf{k}}\Psi^\dagger_{\mathbf{k}}\left[h(\mathbf{k})\tau^z
+\Delta(\mathbf{k})\tau^x\right]\Psi_{\mathbf{k}},\label{Eq_H_0}
\end{eqnarray}
where the four component spinor is designated as $\Psi^\dagger_{\mathbf{k}}
=(\chi^\dagger_{\mathbf{k}},i\sigma^y\chi^T_{-\mathbf{k}})$ with $\chi^\dagger_{\mathbf{k}}
=(c^*_{\mathbf{k}\uparrow},c^*_{\mathbf{k}\downarrow})$~\cite{Moon2017PRB}.
In addition, the Pauli matrices $\tau^{x,y,z}$ and $\sigma^{x,y,z}$
apply to the particle-hole space and spin space, respectively.
Furthermore, the normal-state term $h(\mathbf{k})$ and pairing gap function
$\Delta(\mathbf{k})$ are expressed as~\cite{Frigeri2004PRL,Brydon2011PRB,Moon2017PRB}
\begin{eqnarray}
h(\mathbf{k})&=&\epsilon(\mathbf{k})-\mu+\alpha \mathbf{l}(\mathbf{k})\cdot{\bm \sigma},\label{Eq_h}\\
\Delta(\mathbf{k})&=&\Delta_s+\Delta_t\mathbf{d}(\mathbf{k})
\cdot{\bm \sigma},
\end{eqnarray}
where the spin-orbital coupling is specified by the parameter $\alpha$ in Eq.~(\ref{Eq_h})
and the energy dispersion reads
\begin{eqnarray}
\epsilon(\mathbf{k})&=&-2t(\cos k_x+\cos k_y+\cos k_z)\nonumber\\
&\approx&t(k^{2}_{x}+k^{2}_{y}+k^{2}_{z})-6t,
\end{eqnarray}
with the parameter $t$ being the hopping constant.
Additionally, the pairing amplitudes $\Delta_s$ and $\Delta_t$
are assumed to be real and positive due to the $\mathcal{T}$ symmetry
of topological nodal-line superconductors~\cite{Frigeri2004PRL,Brydon2011PRB,Moon2017PRB}.

It is of particular necessity to bear in mind that the nodal-line points can only be realized while the
negative eigenvalue of $\mathbf{l}(\mathbf{k})\cdot{\bm \sigma}$ with $-|\mathbf{d}(\mathbf{k})|$
is taken into account~\cite{Moon2017PRB}. In addition, one for the sake of simplicity can consider that
the spin-orbital and pairing term share the same direction~\cite{Moon2017PRB,Frigeri2004PRL},
\begin{eqnarray}
\mathbf{d}(\mathbf{k})=\mathbf{l}(\mathbf{k})=(\sin(k_{x}), \sin(k_{y}), 0)=\mathbf{k}_{\bot}.
\label{Eq_specific d}
\end{eqnarray}
After performing the Taylor expansion neighboring the nodal-line points
as displayed in Fig.~\ref{Fig_fermion_surface} which consists of two analogous nodal rings
dubbed by $\mu^z=1$ and $\mu^z=-1$, the effective
Hamiltonian can be reformulated as~\cite{Moon2017PRB}
\begin{eqnarray}
H_{0}&=&\int\frac{d^3\mathbf{k}}{(2\pi)^3}\psi_{\mathbf{k}}^{\dag}(v_{z}\delta k_{z}
\Sigma_{03}+v_{p}\delta k_{\bot}\Sigma_{01})\psi_{\mathbf{k}}.\label{Eq_H_0-3}
\end{eqnarray}
Herein, we adopt the transformation $\zeta\delta k_{\bot}+v_{z}
\delta k_{z}\rightarrow v_{z}\delta k_{z}$ (i.e., $\delta k_{z}\rightarrow \delta k_{z}-
\zeta\delta k_{\bot}/v_{z}$) and introduce two fermion velocities $v_{z}$ and $v_{p}$
that are tied to the microscopic parameters 
$k^{\ast}_{z}$, $m$, as well as $\Delta_{t}$~\cite{Moon2017PRB}.
In addition, the $4\times4$ matrix $\Sigma_{\mu\nu}$ is designated as
$\Sigma_{\mu\nu}\equiv\sigma_\mu\otimes\tau_\nu$ with $\mu,\nu=0,1,2,3$.
Hereby, the four-component spinor $\psi_{\mathbf{k}}^{T}=
(c_{\mathbf{k},\uparrow},c_{-\mathbf{k},\uparrow},
c_{\mathbf{k},\downarrow},c_{-\mathbf{k},\downarrow})$ is nominated to specify the excited nodal fermions
surrounding the upper ($\mu^{z}=1$) and lower ($\mu^{z}=-1$) nodal rings in
Fig.~\ref{Fig_fermion_surface} with $k_F$ serving as the radius of the nodal
line. Further, $\delta k_{z}$ and $\delta k^2_\perp=\delta k^2_x+\delta k^2_y$ designate the transfer
momenta of the low-energy fermionic excitations in the $k_z$ direction and \emph{$k_{x}-k_{y}$} plane
around the nodal line, respectively.

\subsection{Fermion-fermion interactions and disorder scatterings}

To proceed, we bring out the short-ranged fermion-fermion interactions
to characterize the potential role played by the low-energy excitations
from nodal lines~\cite{Vafek2012PRB,Vafek2014PRB,Wang2017QBCP,Moon2017PRB,
Roy-Sau2016PRB,Roy-Saram2016PRB,Roy-Sau2017PRL,Roy2021PRB},
\begin{eqnarray}
S_{\mathrm{ff}}\!&=&\!\!\!\!\sum^3_{\mu\nu=0}\!\!\!\lambda_{\mu\nu}\!\!\prod^4_{j=1}\!\!
\int\!\!\frac{d^3\mathbf{k}_jd\omega_j}{(2\pi)^4}
\psi^\dagger_{\mathbf{k}_1,\omega_1}\!
\Sigma_{\mu\nu}\psi_{\mathbf{k}_2,\omega_2}
\psi^\dagger_{\mathbf{k}_3,\omega_3}\!\Sigma_{\mu\nu}
\psi_{\mathbf{k}_4,\omega_4}\nonumber\\
\!\!\!\!\!\!&&\times\delta(\mathbf{k}_1+\mathbf{k}_2
+\mathbf{k}_3-\mathbf{k}_4)\delta(\omega_1+\omega_2
+\omega_3-\omega_4),\label{Eq_S_int}
\end{eqnarray}
where the vertex matrices $\Sigma_{\mu\nu}\equiv\sigma_\mu\otimes\tau_\nu$ as aforementioned and $\lambda_{\mu\nu}$
are employed to measure the strengths of fermion-fermion couplings
with $\mu,\nu=0,1,2,3$.

In principle, these 16 kinds of fermion-fermion interactions are not all independent among each other.
In order to examine and pick out the independent ones, we resort to the Fierz identity~\cite{Herbut2009PRB,
Herbut2016PRB,Roy-Saram2016PRB}. With the spirit of such an identity, a general interacting term can
be expressed as,
\begin{eqnarray}
&&(\psi^\dag \mathcal{M}\psi)(\psi^\dag \mathcal{N}\psi)\nonumber\\
&&=-\frac{1}{16}\sum_{a=1}^{16}\mathrm{Tr}(\mathcal{M}\Gamma^{a}\mathcal{N}\Gamma^{b})
(\psi^\dag\Gamma^{b}\psi)(\psi^\dag\Gamma^{a}\psi),\label{Eq_Fiertz_ID}
\end{eqnarray}
where $\mathcal{M}$, $\mathcal{N}$, and $\Gamma$ denote certain $4\times4$ matrices with requiring $(\Gamma^{a})^\dag=\Gamma^{a}=(\Gamma^{a})^{-1}$. This indicates that any sort of interactions
appearing in Eq.~(\ref{Eq_S_int}) is allowed to be written by the combinations of parts of these couplings.

To be specific, we hereby following the strategy~\cite{Herbut2009PRB,
Herbut2016PRB,Roy-Saram2016PRB,Dzero2021JPCM} introduce
the interaction vector $\mathcal{V}
=\{(\psi^\dag\Sigma_{00}\psi)^2,(\psi^\dag\Sigma_{01}\psi)^2,(\psi^\dag\Sigma_{02}\psi)^2,
(\psi^\dag\Sigma_{03}\psi)^2, (\psi^\dag\Sigma_{10}\psi)^2,\\
(\psi^\dag\Sigma_{11}\psi)^2,(\psi^\dag\Sigma_{12}\psi)^2,(\psi^\dag\Sigma_{13}\psi)^2,
(\psi^\dag\Sigma_{20}\psi)^2,(\psi^\dag\Sigma_{21}\psi)^2,\\
(\psi^\dag\Sigma_{22}\psi)^2,(\psi^\dag\Sigma_{23}\psi)^2,(\psi^\dag\Sigma_{30}\psi)^2,
(\psi^\dag\Sigma_{31}\psi)^2,(\psi^\dag\Sigma_{32}\psi)^2,\\(\psi^\dag\Sigma_{33}\psi)^2\}$.
After adopting Eq.~(\ref{Eq_Fiertz_ID}), we are left with $\sum_{j=1}^{16}\frac{1}{4}\mathcal{F}_{ij}\mathcal{V}_{j}$=0
where $\mathcal{F}$ takes the form of
\begin{tiny}
\begin{eqnarray}
\mathbf{\mathcal{F}\!=\!}
\left(
              \begin{array}{cccccccccccccccc}
                5 & 1 & 1 & 1 & 1 & 1 & 1 & 1 & 1 & 1 & 1 & 1 & 1 & 1 & 1 & 1 \\
                1 & 5 & -1 & -1 & 1 & 1 & -1 & -1 & 1 & 1 & -1 & -1 & 1 & 1 & -1 & -1 \\
                1 & -1 & 5 & -1 & 1 & -1 & 1 & -1 & 1 & -1 & 1 & -1 & 1 & -1 & 1 & -1 \\
                1 & -1 & -1 & 5 & 1 & -1 & -1 & 1 & 1 & -1 & -1 & 1 & 1 & -1 & -1 & 1 \\
                1 & 1 & 1 & 1 & 5 & 1 & 1 & 1 & -1 & -1 & -1 & -1 & -1 & -1 & -1 & -1 \\
                1 & 1 & -1 & -1 & 1 & 5 & -1 & -1 & -1 & -1 & 1 & 1 & -1 & -1 & 1 & 1 \\
                1 & -1 & 1 & -1 & 1 & -1 & 5 & -1 & -1 & 1 & -1 & 1 & -1 & 1 & -1 & 1 \\
                1 & -1 & -1 & 1 & 1 & -1 & -1 & 5 & -1 & 1 & 1 & -1 & -1 & 1 & 1 & -1 \\
                1 & 1 & 1 & 1 & -1 & -1 & -1 & -1 & 5 & 1 & 1 & 1 & -1 & -1 & -1 & -1 \\
                1 & 1 & -1 & -1 & -1 & -1 & 1 & 1 & 1 & 5 & -1 & -1 & -1 & -1 & 1 & 1 \\
                1 & -1 & 1 & -1 & -1 & 1 & -1 & 1 & 1 & -1 & 5 & -1 & -1 & 1 & -1 & 1 \\
                1 & -1 & -1 & 1 & -1 & 1 & 1 & -1 & 1 & -1 & -1 & 5 & -1 & 1 & 1 & -1 \\
                1 & 1 & 1 & 1 & -1 & -1 & -1 & -1 & -1 & -1 & -1 & -1 & 5 & 1 & 1 & 1 \\
                1 & 1 & -1 & -1 & -1 & -1 & 1 & 1 & -1 & -1 & 1 & 1 & 1 & 5 & -1 & -1 \\
                1 & -1 & 1 & -1 & -1 & 1 & -1 & 1 & -1 & 1 & -1 & 1 & 1 & -1 & 5 & -1 \\
                1 & -1 & -1 & 1 & -1 & 1 & 1 & -1 & -1 & 1 & 1 & -1 & 1 & -1 & -1 & 5
              \end{array}
            \right)\!\!\!.\nonumber
\end{eqnarray}
\end{tiny}

With the help of linear algebra, this gives rise to $\mathrm{rank Null}\,(\mathcal{F})=6$
signaling only six independent couplings. Without loss of generality, we hereafter select six
representative couplings in Eq.~(\ref{Eq_S_int}) as the effective fermion-fermion
interactions and henceforth reformulate the interaction terms into
\begin{eqnarray}
S_{\mathrm{int}}\!\!&=&\!\!\!\sum^6_{i=1}\!\lambda_i\!\prod^4_{j=1}\!\!
\int\!\!\frac{d^3\mathbf{k}_jd\omega_j}{(2\pi)^4}
\psi^\dagger_{\mathbf{k}_1,\omega_1}
\mathcal{M}_i\psi_{\mathbf{k}_2,\omega_2}
\psi^\dagger_{\mathbf{k}_3,\omega_3}\mathcal{M}_i
\psi_{\mathbf{k}_4,\omega_4}\nonumber\\
\!\!\!\!\!\!&&\times\delta(\mathbf{k}_1+\mathbf{k}_2
+\mathbf{k}_3-\mathbf{k}_4)\delta(\omega_1+\omega_2
+\omega_3-\omega_4),\label{Eq_S_int-2}
\end{eqnarray}
where  $\mathcal{M}_i$ corresponds to $\Sigma_{01}$, $\Sigma_{03}$,
$\Sigma_{23}$, $\Sigma_{30}$, $\Sigma_{31}$, and $\Sigma_{32}$
as well as $\lambda_i$ represents the associated strengths with $i$
running from $1$ to $6$.

Besides the fermion-fermion interactions, we also take into account the effects
of a quenched, Gaussian white-noise disorder, which obeys the following
restrictions~\cite{Nersesyan1995NPB,Stauber2005PRB,Wang2011PRB, Mirlin2008RMP,Coleman2015Book,Roy2018PRX},
\begin{eqnarray}
\langle \mathcal{D}(\mathbf{x})\rangle=0,\hspace{0.5cm}\langle \mathcal{D}(\mathbf{x})
\mathcal{D}(\mathbf{x}')\rangle
=\Delta\delta^{(3)}(\mathbf{x}-\mathbf{x}').\label{Eq_S_d-d}
\end{eqnarray}
Hereby,  $\mathcal{D}$ denotes the impurity field, and the parameter
$\Delta$ serves as the concentration of the
impurity. Averaging over the random
impurity potential by virtue of so-called replica method~\cite{Anderson1975JPE,Lee1985RMP,Lerner0307471,
Wang2015PLA,Roy2018PRX}, the fermion-disorder coupling due to disorder scatterings
can be written as follows~\cite{Roy-Saram2016PRB,Roy2018PRX}
\begin{eqnarray}
S_{\mathrm{dis}}
&=&\sum_i\Delta_i\int d\mathbf{x}d\tau d\tau'
\psi^{\dagger}_\alpha(\mathbf{x},\tau)\Gamma_i\psi_\alpha(\mathbf{x},\tau)\nonumber\\
&&\times\psi^{\dagger}_\beta(\mathbf{x},\tau')
\Gamma_i\psi_\beta(\mathbf{x},\tau'),\label{Eq_S_dis}
\end{eqnarray}
where $\Gamma_1=\sigma_0\otimes\tau_0$,
$\Gamma_2=\sigma_0\otimes\tau_2$, $\Gamma_3=\sigma_0\otimes\tau_1$, and
$\Gamma_{4j}=\sigma_j\otimes\tau_3$ with $j=1,2,3$ characterize the random
charge, random mass, random axial chemical potential,
and spin-orbit scatters, respectively~\cite{Roy-Saram2016PRB,Roy2018PRX}.
In addition, $\alpha,\beta$ label the replica indices, and $\Delta_i$ is adopted to
specify the corresponding strength of fermion-impurity
coupling.

\subsection{Effective theory}

After combining the free model~(\ref{Eq_H_0-3}) and the
short-ranged fermion-fermion interactions~(\ref{Eq_S_int-2}) in tandem
with the disorder scatterings~(\ref{Eq_S_dis}), we can obtain our low-energy effective theory
\begin{eqnarray}
S_{\mathrm{eff}}\!&=&\!\!\!\int\frac{d^3\mathbf{k}d\omega}
{(2\pi)^4}\psi^\dagger_{\mathbf{k},\omega}
(-i\omega+v_{z}\delta k_{z}\Sigma_{03}+v_{p}\delta k_{\bot}\Sigma_{01})\psi_{\mathbf{k},\omega}\nonumber\\
&&+S_{\mathrm{int}}+S_{\mathrm{dis}}.\label{Eq_S_eff}
\end{eqnarray}
As a consequence, we can extract the free fermion propagator from the free terms
\begin{eqnarray}
G_0(k)=\frac{1}{-i\omega+v_{z}\delta k_{z}\Sigma_{03}+v_{p}\delta k_{\bot}\Sigma_{01}}.
\end{eqnarray}
Given the physical degrees of freedom with small momenta
can be excited in the low-energy regime, we
only put our focus on the limit
\emph{$|\delta\mathbf{k}|\ll k_F$} and adopt the following
approximation~\cite{Frigeri2004PRL,Brydon2011PRB,Moon2017PRB}
\begin{eqnarray}
\int\!\frac{d^3\mathbf{k}d\omega}{(2\pi)^4}\approx\int\frac{d\delta k_z}{2\pi}
\!\int k_F\frac{d\delta k_\perp}{2\pi}\!\int \frac{d\theta_\mathbf{k}}{2\pi}\!\int\frac{d\omega}{2\pi},
\end{eqnarray}
where the radius of nodal ring $k_F$ is schematically presented in Fig.~\ref{Fig_fermion_surface}.

Afterwards, we consider the effective action~(\ref{Eq_S_eff}) as our starting point and
are going to derive the coupled RG evolutions of all related
parameters appearing in Eq.~(\ref{Eq_S_eff}) after involving the intimate relations between
fermion-fermion interactions and disorder scatterings, and then examine their consequences on
the low-energy physical behaviors in the looming sections.

\section{RG analysis of fermion-fermion interactions and disorder strengths}\label{Sec_RG_analysis}

On the basis of Wilsonian momentum-shell RG formalism~\cite{Wilson1975RMP,Polchinski9210046,Shankar1994RMP},
the energy-dependent interaction parameters in the effective action~(\ref{Eq_S_eff}) that
carry the low-energy information can be established. To this end, we are required to integrate
out the fast modes of fields within the momentum shell $b\Lambda<k<\Lambda$,
where $\Lambda$ denotes the energy scale, and the variable parameter $b$ is
written as $b=e^{-l}$ with the running energy scale $l>0$ describing
the changes of energy scales~\cite{Shankar1994RMP,Kim2008PRB,Huh2008PRB,
She2010PRB,She2015PRB,Wang2011PRB,Wang2013PRB,Wang2014PRD,Wang2015PRB,Wang2017PRB,Wang2017QBCP,
Vafek2012PRB,Vafek2014PRB,Wang2007.14981}. For convenience, it is helpful to rescale the momenta and energy by $\Lambda_0$ that
is associated with the lattice constant, i.e., $k\rightarrow k/\Lambda_0$ and
$\omega\rightarrow\omega=\omega/\Lambda_0$.

Following the spirit of RG approach~\cite{Wilson1975RMP,Polchinski9210046,Shankar1994RMP}, we subsequently
consider the free term of effective theory as an initial fixed point that is invariant under the
RG transformation. As a consequence, the RG rescaling transformations of momenta, energy, and fermionic
fields, which are employed to connect continuous steps of RG processes, can be derived as follows~\cite{Shankar1994RMP,Huh2008PRB,She2010PRB,She2015PRB,Wang2011PRB,Wang2022NPB}
\begin{eqnarray}
\omega&\rightarrow&\omega e^{-l},\label{Eq_RG-scaling-1}\\
\delta k_z&\rightarrow&\delta k_ze^{-l},\\
\delta k_\perp&\rightarrow&\delta k_\perp e^{-l},\\
\psi_{\mathbf{k},\omega}&\rightarrow&\psi_{\mathbf{k},\omega}e^{\frac{1}{2}\int^l_0dl(4-\eta_f)}.\label{Eq_RG-scaling-2}
\end{eqnarray}
Hereby, the parameter $\eta_f$ serves as the anomalous fermion dimension and
it collects the intimate contributions from one-loop corrections due to the
interplay between fermion-fermion interactions
and impurities. As presented in Appendix~\ref{Sec_appendix-one-loop-corrections},
the fermionic self-energy receives nontrivial corrections in the presence of
fermion-disorder interplay, which are explicitly provided in Eq.~(\ref{Eq_self-energy}).
Combining the RG rescalings~(\ref{Eq_RG-scaling-1})-(\ref{Eq_RG-scaling-2})
and the free fixed point gives rise to
\begin{eqnarray}
\eta_f=\mathcal{C}_{2}
(\Delta_{1}+\Delta_{2}+\Delta_{3}
+\Delta_{41}+\Delta_{42}+\Delta_{43}).
\end{eqnarray}

With these in hand, we are able to dwell on the RG equations of all interaction parameters
appearing in Eq.~(\ref{Eq_S_eff}). After long but straightforwardly algebraic calculations,
all the one-loop corrections are obtained and presented in
Eqs.~(\ref{Eq_one-loop-corrections-lambda-1})-(\ref{Eq_one-loop-corrections-dis})
of Appendix~\ref{Sec_appendix-one-loop-corrections}, which contain the interplay between
fermion-fermion interactions and disorder scatterings. Afterwards, adopting these one-loop
corrections and parallelling the analogous procedures of
RG analysis~\cite{Shankar1994RMP,Kim2008PRB,Huh2008PRB,
She2010PRB,She2015PRB,Wang2011PRB} with the help of rescaling
transformations~(\ref{Eq_RG-scaling-1})-(\ref{Eq_RG-scaling-2}) yield
the coupled RG flow equations of all coupling parameters,
\begin{widetext}
\begin{small}
\begin{eqnarray}
\frac{dv_z}{dl}
&=&[4\mathcal{C}_{1}\lambda_{2}-\mathcal{C}_{2}
(\Delta_{1}+\Delta_{2}+\Delta_{3}
+\Delta_{41}+\Delta_{42}+\Delta_{43})]v_z,\label{Eq_RG_v_z}\\
\frac{dv_p}{dl}
&=&[4\mathcal{C}_{1}\lambda_{1}-\mathcal{C}_{2}
(\Delta_{1}+\Delta_{2}+\Delta_{3}
+\Delta_{41}+\Delta_{42}+\Delta_{43})]v_p,\label{Eq_RG_v_perp}\\
\frac{d\lambda_1}{dl}
&=&2\lambda_1\Bigl[-\frac{1}{2}+\mathcal{C}_{3}
\left(
-3\lambda_{1}-2\lambda_{2}-\lambda_{3}+\lambda_{4}
+\lambda_{5}-\lambda_{6}
\right)+
\mathcal{C}_{7}(\Delta_{1}
-\Delta_{2}
-7\Delta_{3}
-\Delta_{41}
-\Delta_{42}
-\Delta_{43})\nonumber\\
&&-\mathcal{C}_{2}
(\Delta_{1}+\Delta_{2}+\Delta_{3}
+\Delta_{41}+\Delta_{42}+\Delta_{43})\Bigr],\\
\frac{d\lambda_2}{dl}
&=&2\lambda_2\Bigl[-\frac{1}{2}+\mathcal{C}_{4}
\left(
-2\lambda_{1}-3\lambda_{2}+\lambda_{3}+\lambda_{4}
-\lambda_{5}-\lambda_{6}
\right)+
\mathcal{C}_{7}(-\Delta_{1}
+\Delta_{2}
+\Delta_{3}
-\Delta_{41}
-\Delta_{42}
-\Delta_{43}
)\nonumber\\
&&-\mathcal{C}_{2}
(\Delta_{1}+\Delta_{2}+\Delta_{3}
+\Delta_{41}+\Delta_{42}+\Delta_{43})\Bigr],\\
\frac{d\lambda_3}{dl}
&=&2\lambda_3\Bigl[-\frac{1}{2}+\mathcal{C}_{4}
\left(
-2\lambda_{1}+\lambda_{2}-3\lambda_{3}-\lambda_{4}
+\lambda_{5}+\lambda_{6}
\right)
+
\mathcal{C}_7(-\Delta_1
+\Delta_2
+\Delta_3
+\Delta_{41}
+7\Delta_{42}
+\Delta_{43})\nonumber\\
&&
-\mathcal{C}_{2}
(\Delta_{1}+\Delta_{2}+\Delta_{3}
+\Delta_{41}+\Delta_{42}+\Delta_{43})\Bigr]
-4\mathcal{C}_6\lambda_5\Delta_{41},\\
\frac{d\lambda_4}{dl}
&=&
\lambda_4[-1-4\mathcal{C}_{2}
(\Delta_{41}+\Delta_{42})]
-4\mathcal{C}_5(\lambda_5\Delta_2
+\lambda_6\Delta_3)
-2\lambda_{6}(\mathcal{C}_{4}\lambda_{2}
+\mathcal{C}_{3}\lambda_{1}),\\
\frac{d\lambda_5}{dl}
&=&2\lambda_5
\Bigl[-\frac{1}{2}+
\mathcal{C}_3
(\lambda_1-2\lambda_2+\lambda_3
+\lambda_4-3\lambda_5-\lambda_6
)
+
\mathcal{C}_7(\Delta_1
-\Delta_2
+\Delta_3
+\Delta_{41}
+\Delta_{42}
-\Delta_{43}
)\nonumber\\
&&
-\mathcal{C}_{2}
(\Delta_{1}+\Delta_{2}+\Delta_{3}
+\Delta_{41}+\Delta_{42}+\Delta_{43})\Bigr]
+2\mathcal{C}_1\lambda_2\lambda_6-
4(\mathcal{C}_6\lambda_3\Delta_{41}
+\mathcal{C}_5\lambda_4\Delta_2
+\mathcal{C}_5\lambda_6\Delta_1),\\
\frac{d\lambda_6}{dl}
&=&2\lambda_6\Bigl[-\frac{1}{2}+
\mathcal{C}_1
(-\lambda_1-\lambda_2+\lambda_3
+\lambda_4-\lambda_5-3\lambda_6)
-(\mathcal{C}_3\lambda_2+\mathcal{C}_4\lambda_1)
-2\mathcal{C}_2(\Delta_1
+\Delta_2
+\Delta_{41}
+\Delta_{42}
)\Bigr]\nonumber\\
&&+2\mathcal{C}_1\lambda_2\lambda_5
-4\mathcal{C}_5(\lambda_4\Delta_3
+\lambda_5\Delta_1),\label{Eq_RG_lambda_6}\\
\frac{d\Delta_1}{dl}
&=&
2\mathcal{C}_2\Delta_1
(\Delta_1+\Delta_2+\Delta_3
+\Delta_{41}+\Delta_{42}+\Delta_{43})
+8\mathcal{C}_5\Delta_2\Delta_3,\\
\frac{d\Delta_2}{dl}
&=&\Delta_2\Big[2\mathcal{C}_{2}
(-3\Delta_{1}-3\Delta_{2}+\Delta_{3}
+\Delta_{41}+\Delta_{42}+\Delta_{43})+
\mathcal{C}_1
(\lambda_1+\lambda_2+\lambda_3
-\lambda_4+\lambda_5-\lambda_6)
\Bigr]+8\mathcal{C}_5\Delta_1\Delta_3,\label{Eq_RG_Delta-2}\\
\frac{d\Delta_{3}}{dl}
&=&\Delta_{3}\Big[
4\mathcal{C}_7
(\Delta_1-\Delta_2+\Delta_3
-\Delta_{41}-\Delta_{42}-\Delta_{43})
-2\mathcal{C}_{2}
(\Delta_{1}+\Delta_{2}+\Delta_{3}
+\Delta_{41}+\Delta_{42}+\Delta_{43})\nonumber\\
&&+
\mathcal{C}_3
(-\lambda_1+\lambda_2+\lambda_3
-\lambda_4-\lambda_5+\lambda_6)
\Bigr]+8\mathcal{C}_5\Delta_1\Delta_2,\\
\frac{d\Delta_{41}}{dl}
&=&\Delta_{41}\Big[
4\mathcal{C}_7
(-\Delta_1+\Delta_2+\Delta_3
-\Delta_{41}+\Delta_{42}+\Delta_{43})
-2\mathcal{C}_{2}
(\Delta_{1}+\Delta_{2}+\Delta_{3}
+\Delta_{41}+\Delta_{42}+\Delta_{43})\nonumber\\
&&+
\mathcal{C}_4
(\lambda_1-\lambda_2+\lambda_3
+\lambda_4-\lambda_5-\lambda_6)
\Bigr]+8\mathcal{C}_5\Delta_{42}\Delta_{43},\\
\frac{d\Delta_{42}}{dl}
&=&\Delta_{42}\Big[
4\mathcal{C}_7
(-\Delta_1+\Delta_2+\Delta_3
+\Delta_{41}-\Delta_{42}+\Delta_{43})
-2\mathcal{C}_{2}
(\Delta_{1}+\Delta_{2}+\Delta_{3}
+\Delta_{41}+\Delta_{42}+\Delta_{43})\nonumber\\
&&+
\mathcal{C}_4
(\lambda_1-\lambda_2-\lambda_3
+\lambda_4-\lambda_5-\lambda_6)
\Bigr]+
8\mathcal{C}_5\Delta_{41}\Delta_{43},\\
\frac{d\Delta_{43}}{dl}
&=&\Delta_{43}\Big[
4\mathcal{C}_7
(-\Delta_1+\Delta_2+\Delta_3
+\Delta_{41}+\Delta_{42}-\Delta_{43})
-2\mathcal{C}_{2}
(\Delta_{1}+\Delta_{2}+\Delta_{3}
+\Delta_{41}+\Delta_{42}+\Delta_{43})\nonumber\\
&&+
\mathcal{C}_4
(\lambda_1-\lambda_2+\lambda_3
-\lambda_4+\lambda_5+\lambda_6)
\Bigr]+
8\mathcal{C}_5\Delta_{41}\Delta_{42},\label{Eq_RG_Delta}
\end{eqnarray}
\end{small}
\end{widetext}
where the coefficients $\mathcal{C}_i$ with $i=1-7$ are nominated as follows
\begin{eqnarray}
\mathcal{C}_{1}
&\equiv&\frac{1}{(2\pi)^3}
\int_{0}^{\pi}d\theta
\frac{\pi}
{(\upsilon_{z}^2\sin^2\theta+\upsilon_{p}^2\cos^2\theta)^{1/2}},\label{Eq_coeff-C1}\\
\mathcal{C}_{2}
&\equiv&\frac{1}{(2\pi)^3}
\int_{0}^{\pi}d\theta
\frac{2\pi}
{\upsilon_{z}^2\sin^2\theta+\upsilon_{p}^2\cos^2\theta},\\
\mathcal{C}_{3}
&\equiv&\frac{1}{(2\pi)^3}
\int_{0}^{\pi}d\theta
\frac{\pi\upsilon_{z}^2\sin^2\theta}
{(\upsilon_{z}^2\sin^2\theta+\upsilon_{p}^2\cos^2\theta)^{3/2}},\\
\mathcal{C}_{4}
&\equiv&\frac{1}{(2\pi)^3}
\int_{0}^{\pi}d\theta
\frac{\pi\upsilon_{p}^2\cos^2\theta}
{(\upsilon_{z}^2\sin^2\theta+\upsilon_{p}^2\cos^2\theta)^{3/2}},\\
\mathcal{C}_{5}
&\equiv&\frac{1}{(2\pi)^3}
\int_{0}^{\pi}d\theta
\frac{4\pi\upsilon_{z}^2\sin^2\theta}
{(\upsilon_{z}^2\sin^2\theta+\upsilon_{p}^2\cos^2\theta)^2},\\
\mathcal{C}_{6}
&\equiv&\frac{1}{(2\pi)^3}
\int_{0}^{\pi}d\theta
\frac{4\pi\upsilon_{p}^2\cos^2\theta}
{(\upsilon_{z}^2\sin^2\theta+\upsilon_{p}^2\cos^2\theta)^2},\\
\mathcal{C}_{7}
&\equiv&\frac{1}{(2\pi)^3}
\int_{0}^{\pi}d\theta
\frac{2\pi(\upsilon_{p}^2\cos^2\theta-\upsilon_{z}^2\sin^2\theta)}
{(\upsilon_{z}^2\sin^2\theta+\upsilon_{p}^2\cos^2\theta)^2}.\label{Eq_coeff-C7}
\end{eqnarray}

In this sense, these above entangled evolutions of interaction
parameters~(\ref{Eq_RG_v_z})-(\ref{Eq_RG_Delta}) bear out that
the fermion-fermion interactions as well as fermion-disorder couplings
are pertinently coupled and mutually intertwined with each other upon
varying of the energy scales. In particular, the fermion velocities
including $v_z$ and $v_p$ are also manifestly involved into the coupled equations.
Under this circumstance, such energy-dependent couplings can play a direct or indirect
role in essentially pinning down the low-energy fates of fermion velocities and interaction parameters.
Specifically, we will endeavor to address the clean limit situation
in next section~\ref{Sec_clean} and defer the disorder effects to Sec.~\ref{Sec_disorder}.

\begin{figure}
\centering
\includegraphics[width=3.35in]{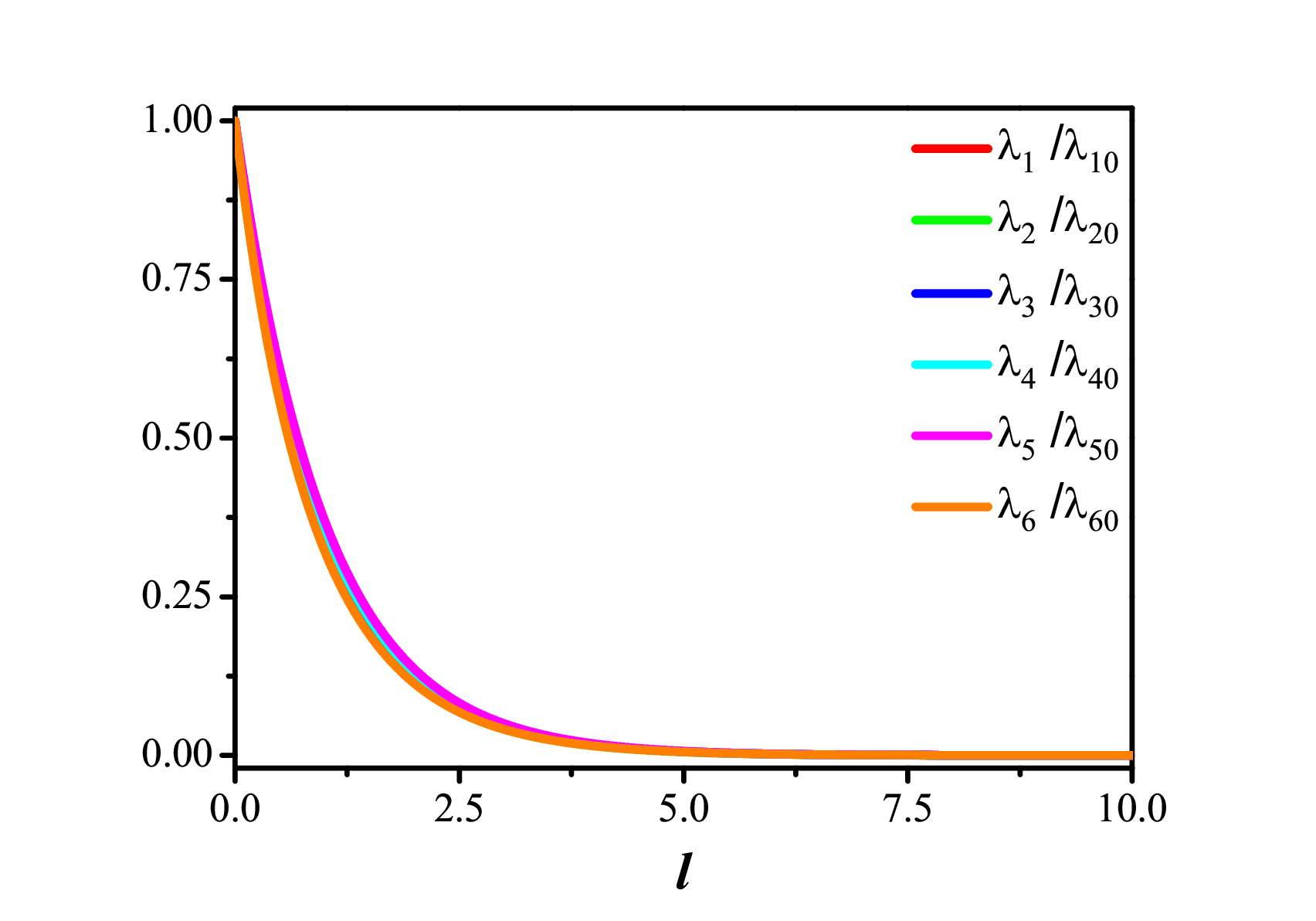}
\vspace{-0.39cm}
\caption{(Color online) Energy-dependent evolutions of fermion-fermion interaction
with $v_{z0}/v_{p0}=0.1$ and $\lambda_{0}=10^{-3}$ (the qualitative results are
insensitive to the initial values of interaction parameters).}\label{lambdai-0.1-M3-clean}
\end{figure}

\begin{figure}
\centering
\includegraphics[width=3in]{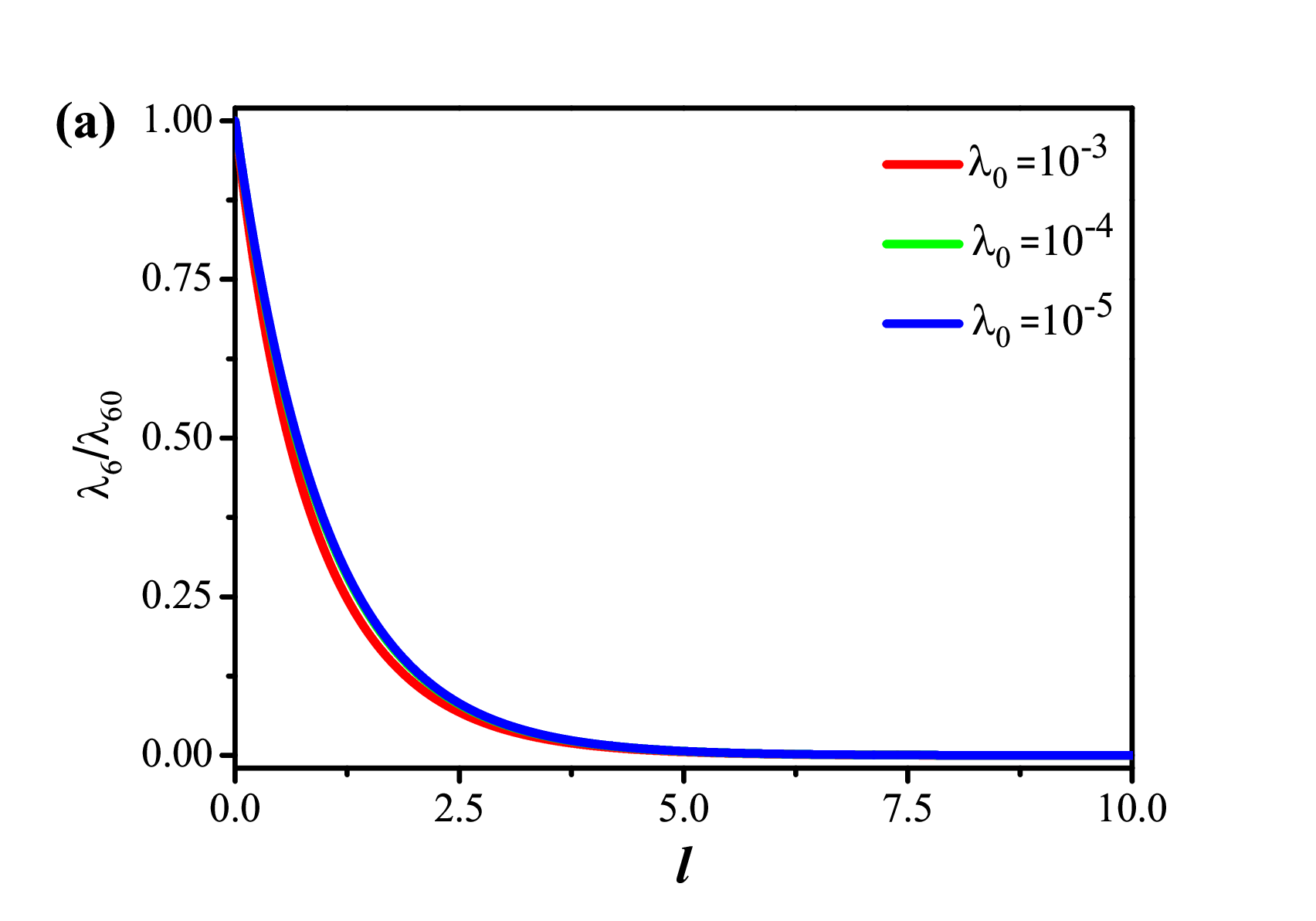}\\ \vspace{-0.3cm}
\includegraphics[width=3in]{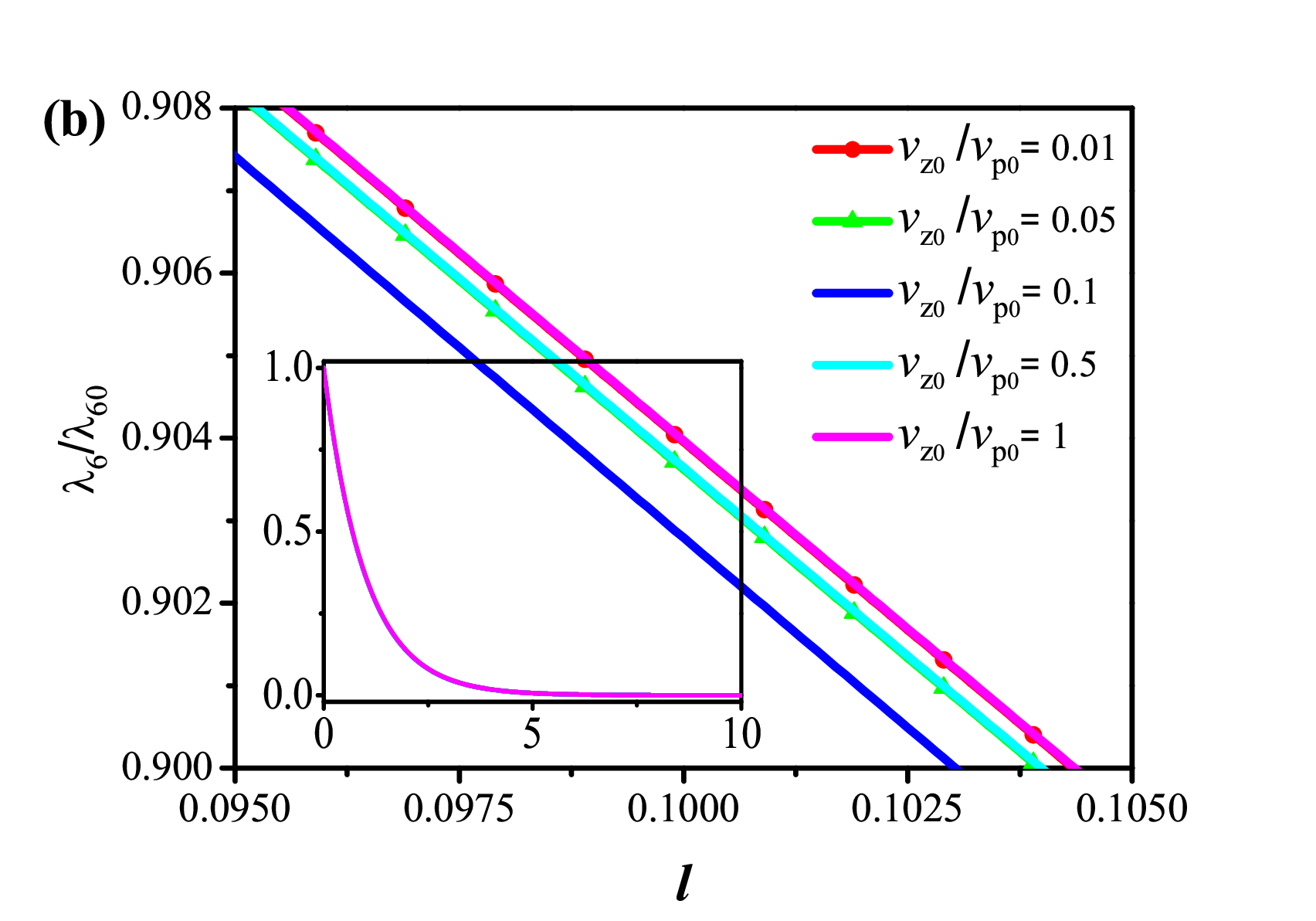}
\vspace{-0.1cm}
\caption{(Color online) Energy-dependent evolutions of fermion-fermion
interaction $\lambda_6/\lambda_{60}$ for (a) $v_{z0}/v_{p0}=0.1$ and distinct
values of $\lambda_{0}$, and (b) $\lambda_0=10^{-4}$
and distinct values of anisotropy $v_{z0}/v_{p0}$ (the qualitative results are
insensitive to the initial values of interaction parameters).}\label{lambda6-many-M4-clean}
\end{figure}


\section{Clean limit}\label{Sec_clean}

Before moving to the interplay between fermion-fermion interactions
and disorder scatterings, we within this section perform a warm-up for the
clean limit situation and compare with its disorder counterpart in next
section~\ref{Sec_disorder}. Concretely, we primarily account for the consequences
of fermion-fermion interactions on the fermion velocities in the absence of disorders.

\subsection{Fates of fermion-fermion interactions}\label{sub-Sec_clean-A}

After carrying out the numerical analysis of the energy-dependent coupled RG equations
in the absence of disorders, we realize from Fig.~\ref{lambdai-0.1-M3-clean}
for certain representative initial conditions that all of fermion-fermion
interactions quickly decrease as the energy scale is lowered and go toward zero
at the lowest-energy limit.

Although the starting conditions do not alter the basic fates of fermion-fermion interactions,
they can quantitatively modify the energy-dependent behaviors as shown in Fig.~\ref{lambda6-many-M4-clean}
for $\lambda_6$ as an instance, which shares the similar results with the other five types of interaction
couplings. On the one hand, we find that Fig.~\ref{lambda6-many-M4-clean}(a) with $v_{z0}/v_{p0}=0.1$
displays that the fermionc strength decreases a little quickly with the increase in initial strength of
fermion-fermion interactions.  On the other hand, in order to examine the role of initial ratio
of fermion velocities in fermionic couplings, we select several representative groups of initial
conditions that are distinguished with varying the anisotropy of
fermion velocities characterized by $v_{z0}/v_{p0}$. The related results are displayed
in Fig.~\ref{lambda6-many-M4-clean}(b). This apparently suggests that the fermion-fermion couplings
are insensitive to the anisotropy of fermion velocities, which only slightly impact their values during the
intermediate stage with an optimal anisotropy at $v_{z0}/v_{p0}\approx0.1$ for the decrease in parameters.

On the basis of above analysis, we figure out that these fermion-fermion interactions are
irrelevant at the clean limit in the language of RG framework~\cite{Shankar1994RMP,Kim2008PRB,Huh2008PRB,
She2010PRB,She2015PRB,Wang2011PRB,Wang2013PRB,Wang2014PRD,Wang2015PRB,Wang2017PRB,Wang2017QBCP,
Vafek2012PRB,Vafek2014PRB}. Despite their decrease with lowering the energy scale, they still bring interesting
corrections to fermion velocities, which are two of particular importance quantities in our
effective theory. In addition, it is noteworthy that the irrelevant fates of fermionic interactions
at the clean limit would be qualitatively changed under the
influence of disorder scatterings, which will be carefully presented
in Sec.~\ref{Sec_disorder}.

\begin{figure}
\centering
\includegraphics[width=3.35in]{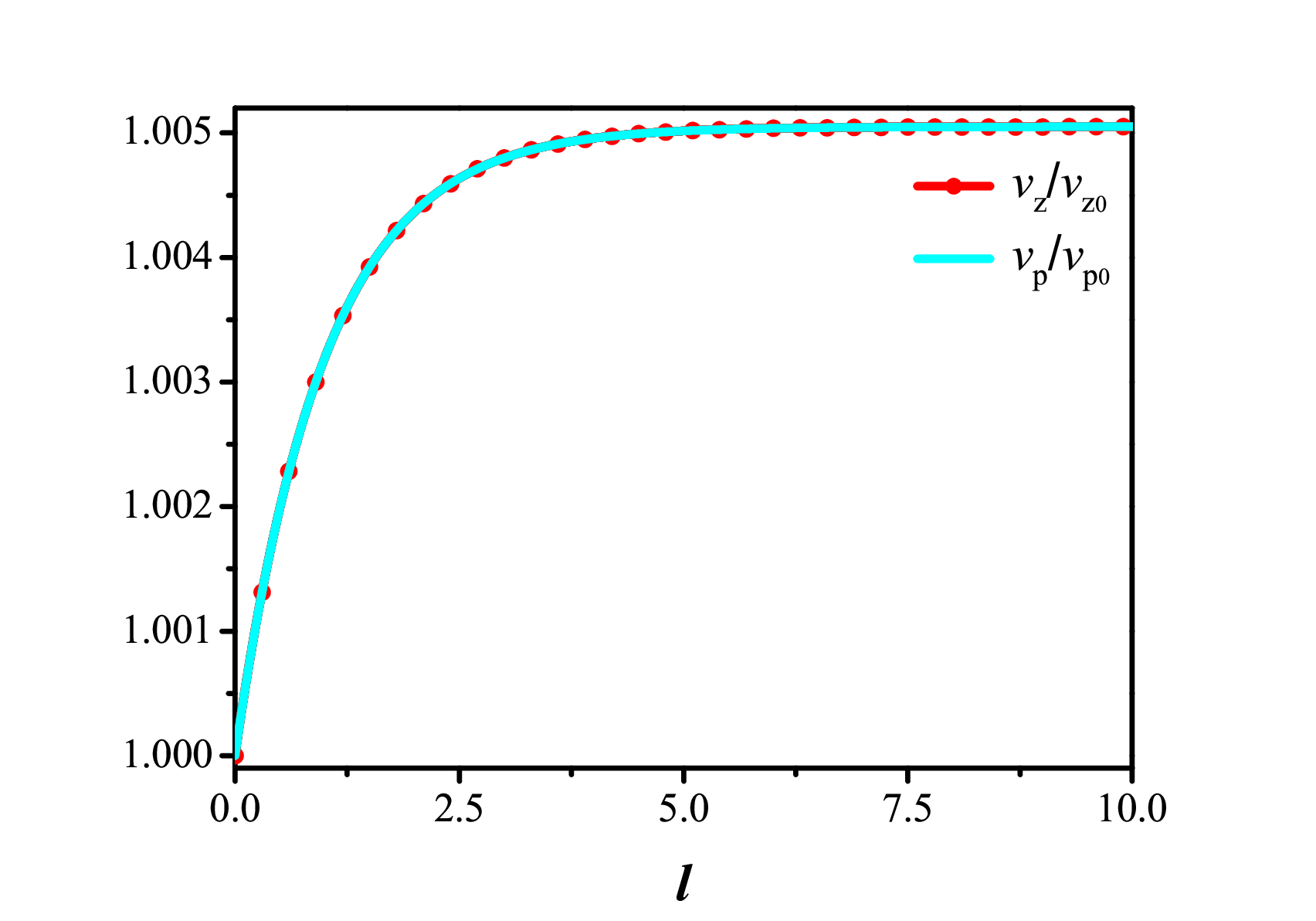}\\
\vspace{-0.1cm}
\caption{(Color online) Energy-dependent evolutions of fermion velocities for the isotropic case
($v_{z0}/v_{p0}$) with $\lambda_0=10^{-4}$ (the qualitative results are
insensitive to the initial values of interaction parameters).}\label{vzandvp-1-M4-clean}
\end{figure}

\begin{figure}
\centering
\includegraphics[width=3in]{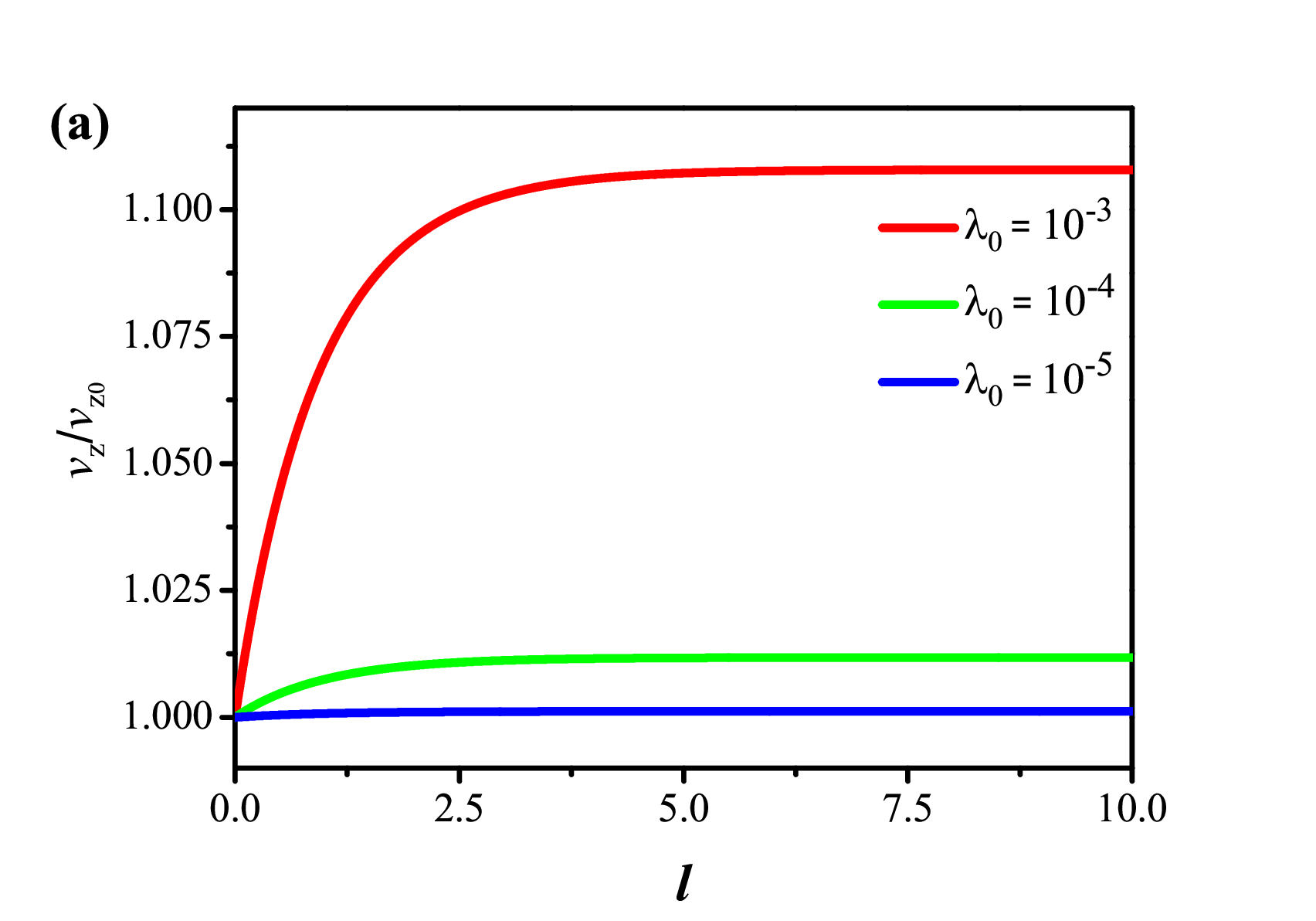}\\ \vspace{-0.3cm}
\includegraphics[width=3in]{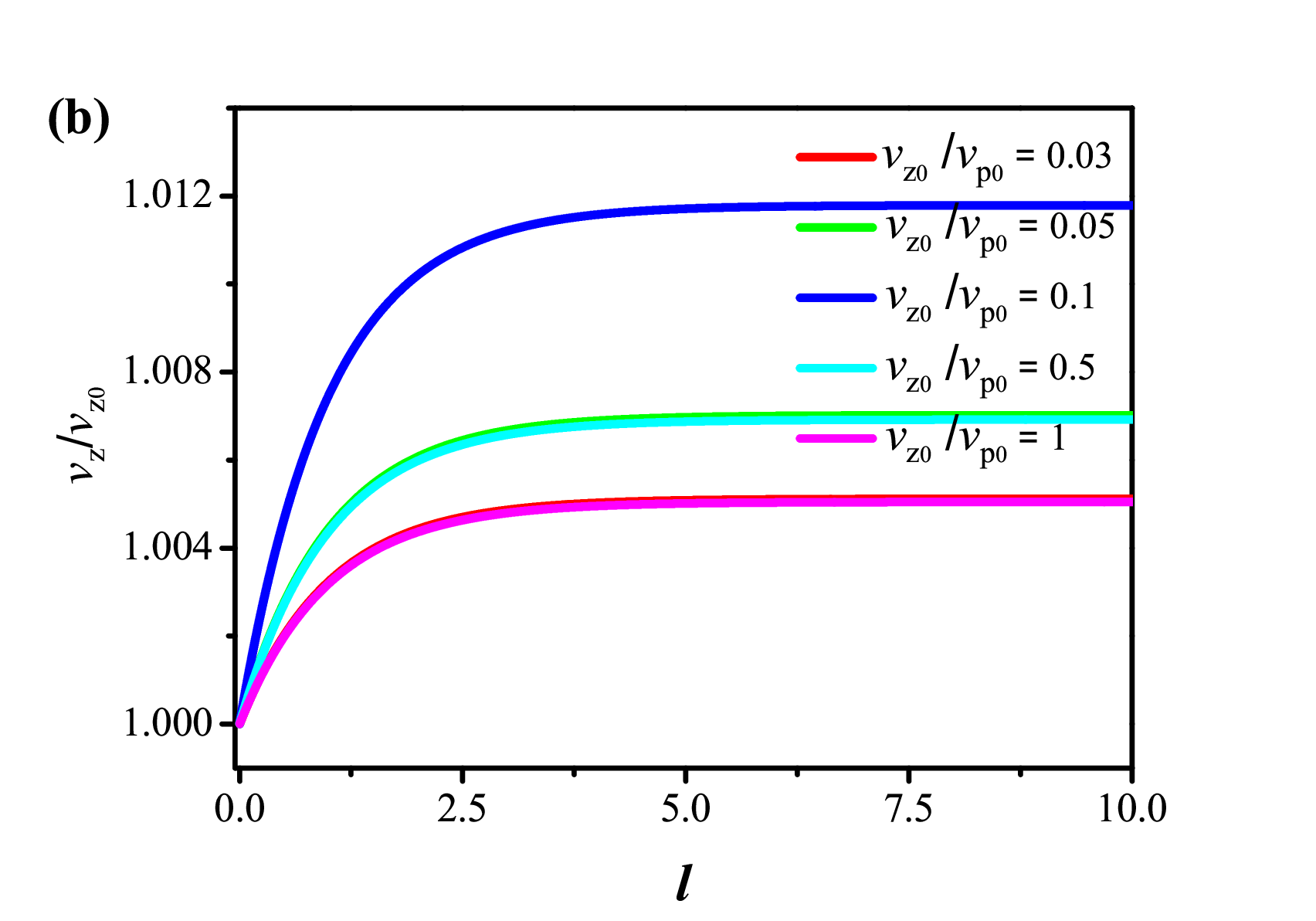}
\vspace{-0.1cm}
\caption{(Color online) Energy-dependent evolutions of $v_z/v_{z0}$ under:
(a) different values of initial fermion-fermion couplings with $v_{z0}/v_{p0}=0.1$,
and (b) different values initial of anisotropy $v_{z0}/v_{p0}$ with $\lambda_0=10^{-4}$ (the qualitative
results are insensitive to the initial values of interaction parameters).}\label{vz-0.1-M345-clean}
\end{figure}

\begin{figure}
\centering
\includegraphics[width=3in]{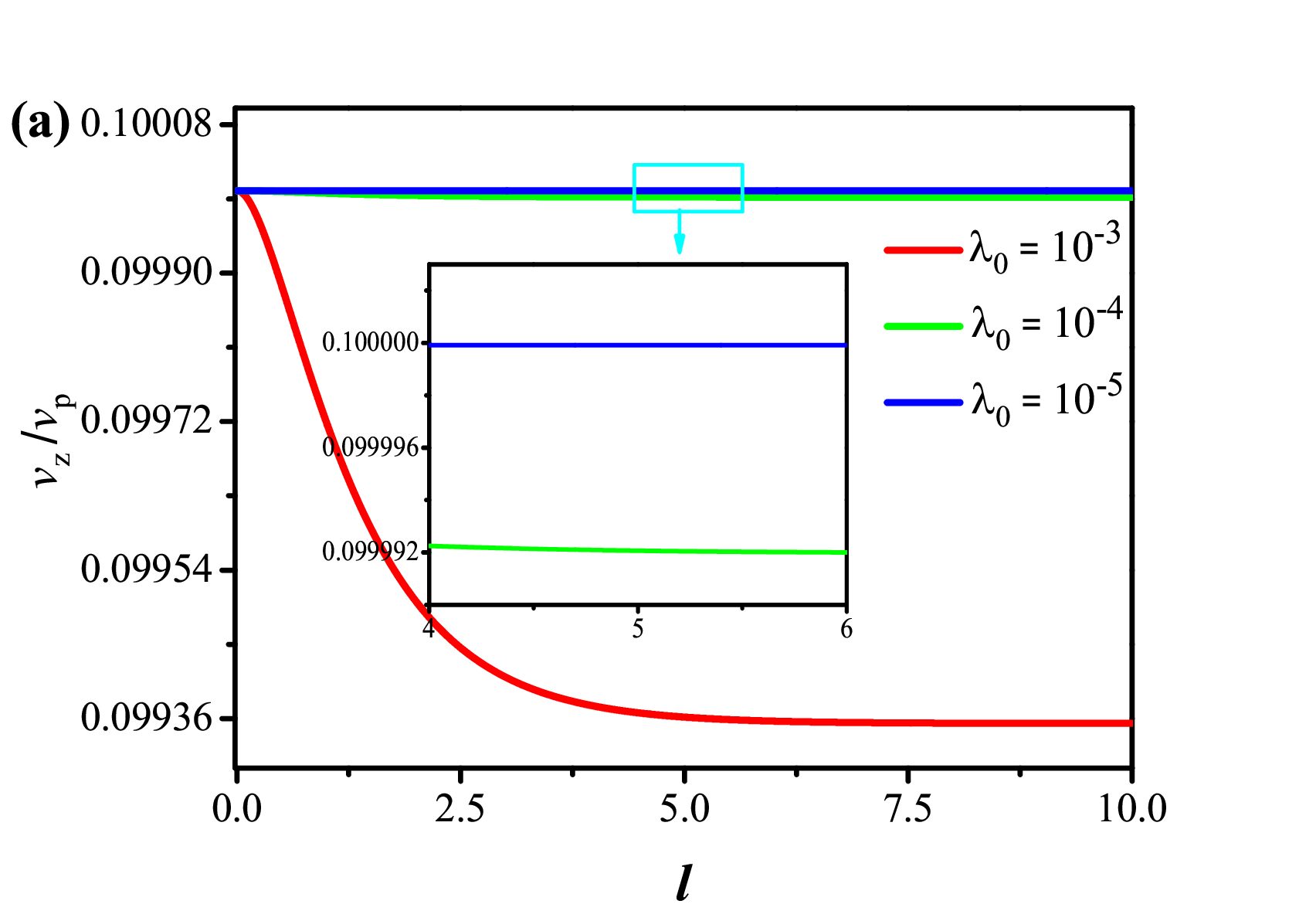}\\ \vspace{-0.3cm}
\includegraphics[width=3in]{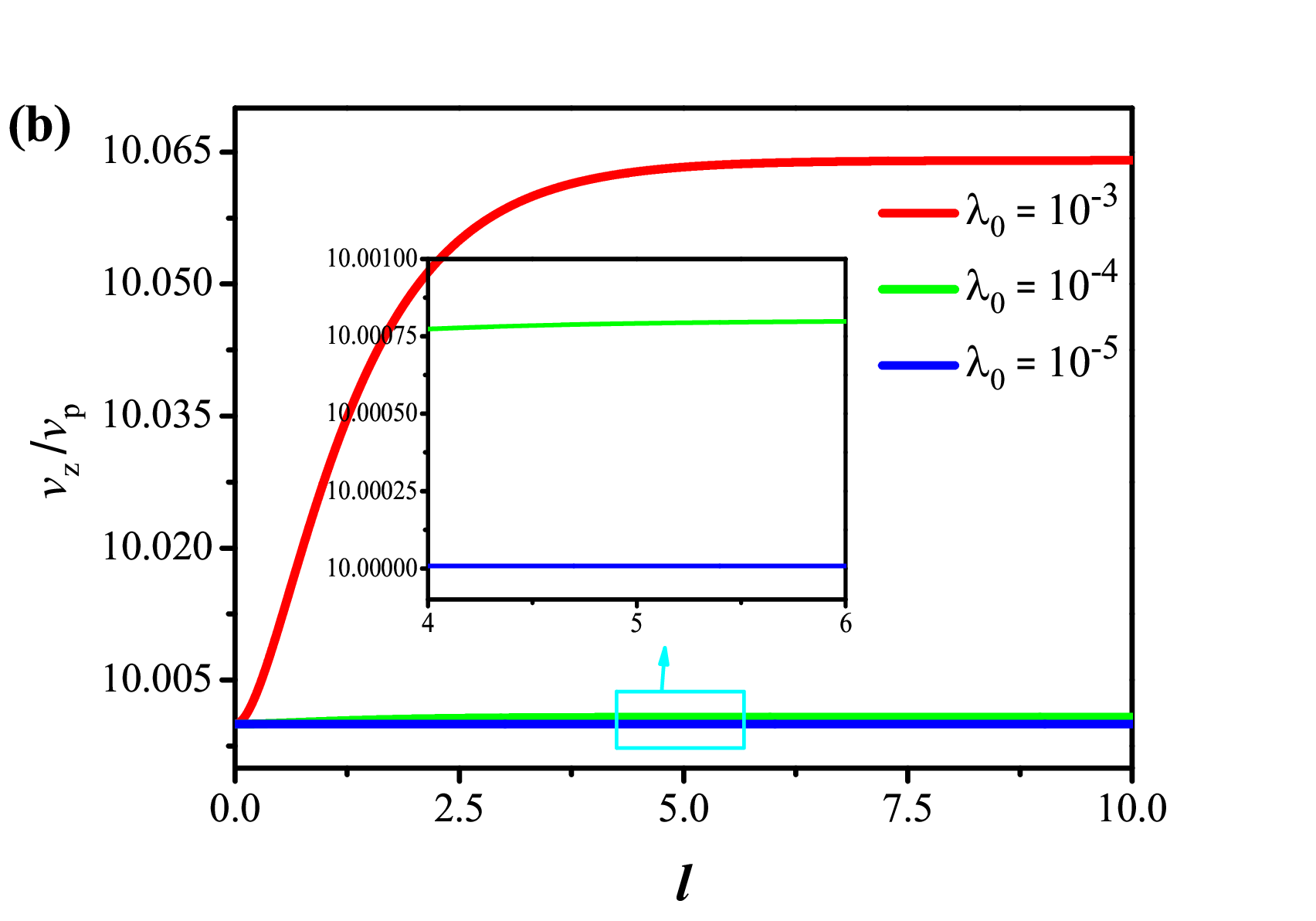}
\vspace{-0.1cm}
\caption{(Color online) Energy-dependent evolutions of $v_z/v_{p}$ under
different values of initial fermion-fermion couplings with (a) $v_{z0}/v_{p0}=0.1$
and (b) $v_{z0}/v_{p0}=10$ (the qualitative results are insensitive to the initial
values of interaction parameters).}\label{vzvp-0.1-M345-clean}
\end{figure}

\subsection{Evolutions of fermion velocities}\label{Subsec-velocities-clean}

The two fermion velocities $v_{z}$ and $v_{p}$ in our model~(\ref{Eq_S_eff}) are of close relevance to
the low-energy properties. Accordingly, one needs to study their energy-dependent evolutions that are
determined by the coupled RG equations in Sec.~\ref{Sec_RG_analysis}. We hereby endeavor
to unveil their low-energy properties at the clean limit. With the help of numerical calculations
for the energy-dependent coupled RG equations in the absence of disorders,
we provide the key energy-dependent tendencies of $v_{z}$ and $v_{p}$ as well as
$v_z/v_p$ in Figs.~\ref{vzandvp-1-M4-clean}-\ref{vzvp-0.1-M345-clean}.

Learning from these results, we realize that the fermion-fermion interactions and their effects
heavily hinge upon the initial value of anisotropy of fermion velocities.
With respect to the initial isotropic fermion velocities ($v_{z0}/v_{p0}$) shown in Fig.~\ref{vzandvp-1-M4-clean},
we find that both $v_z$ and $v_p$ gain a slight increase and then be saturated in the sufficiently
low-energy scale. In such a circumstance, the ratio of fermion velocities with $v_z/v_p=1$ is robust against
the fermion-fermion interactions with $\lambda_0=10^{-4}$ for instance and the basic results are independent of
initial values of fermion-fermion couplings. As to the situation with an initial anisotropy of fermion velocities,
Fig.~\ref{vz-0.1-M345-clean} and Fig.~\ref{vzvp-0.1-M345-clean} show that fermion velocities exhibit more
interesting behaviors with variations of the fermion-fermion interactions and the ratio of fermion velocities, respectively.
Along with Fig.~\ref{vz-0.1-M345-clean}(a), we find that $v_z$
climbs up quickly and arrives at a certain constant
with lowering the energy scale, which obtains much more increase compared to its isotropic counterpart.
Particularly, its saturated value is lifted and ends with some bigger values due to
tuning up the starting strengths of fermion-fermion interactions.
Additionally, the fermion velocity also depends upon the anisotropy of velocities and
aforementioned in Sec.~\ref{sub-Sec_clean-A} $v_z$ gains the biggest enhancement with an optimal anisotropy at
$v_{z0}/v_{p0}\approx0.1$ as delineated in Fig.~\ref{vz-0.1-M345-clean}(b).
In comparison, we realize that $v_p$ does not receive the same contribution from fermionic interactions, indicating
the change of the initial ratio of $v_z$ and $v_p$, i.e., initial anisotropy of fermion velocities.
To be concrete, it increases a little more than $v_z$ at $v_{z0}/v_{p0}<1$.
As a consequence, $v_z/v_p$ decreases with lowering the energy scales and
the increase in the fermionic coupling makes the ratio smaller as displayed
in Fig.~\ref{vzvp-0.1-M345-clean}(a). On the contrary, as the velocity $v_p$ receives a little
less corrections by lowering the energy scales, we figure out that the ratio of fermion velocities $v_z/v_p$
shown in Fig.~\ref{vzvp-0.1-M345-clean}(b) is driven up at $v_{z0}/v_{p0}>1$ and goes toward bigger values
with increasing the initial fermion-fermion couplings.

To recapitulate, both fermion velocities $v_z$ and $v_p$ are increased as
the energy scale is lowered in the absence of disorder scatterings.
However, the fate of their ratio $v_z/v_p$ in the low-energy regime is of close association with
its initial condition, which can either fall down at $v_{z0}/v_{p0}<1$ or climb up at $v_{z0}/v_{p0}>1$.
Although the fermion-fermion interactions are irrelevant and cannot
change the basic results, they are able to provide quantitative corrections
to the fermion velocities and their ratio in the low-energy regime.

\section{Consequences of disorder effects}\label{Sec_disorder}

After addressing the clean-limit situation in the previous
section, we hereby endeavor to investigate the effects of
disorder scatterings, which are always present in the realistic
systems and expected to play an essential role in the
low-energy properties of fermionic systems~\cite{Wang2011PRB, Wang2013PRB,Novoselov2005Nature,
Castro2009RMP, Hasan2010RMP, Qi2011RMP,Altland2002PR,Fradkin2010ARCMP, Das2011RMP, Sachdev2011book, Kotov2012RMP,
 Aleiner2006PRL, Foster2008PRB, Lee2017arXiv,Nersesyan1995NPB,Stauber2005PRB,Nandkishore2017PRB,
 Wang-Nandkishore2017PRB,Nandkishore2014PRB,Roy2017PRB-96,Roy1812.05615,Roy2016SR,Wang2017QBCP,Mandal2018AP}.
 With switching on the disorders, on the one hand, the
disorder strengths $\Delta_i$ with $i=1-4$ introduced in Eq.~(\ref{Eq_S_dis}) are forced
to interact with each other and fight for their low-energy
fates under the ferocious competitions among themselves.
On the other hand, they are able to provide considerable
influences to the energy-dependent behaviors of both fermion-fermion interactions
and fermion velocities by virtue of participating in the
entangled RG flows~(\ref{Eq_RG_v_z})-(\ref{Eq_RG_Delta}).
Afterwards, we are going to address these interesting items
one by one in the rest of this section.

\subsection{Low-energy properties of disorder strengths}\label{Sub-Sec_dis-A}

At the outset, we would like to put our focus on the
low-energy behaviors of disorder strengths in that one
will figure out later in Sec.~\ref{Subsec_fates_fermion-fermion}
and Sec.~\ref{Subsec_behaviors_fermion-velocities} that the disorder
scatterings can be considered as the impulse or catalyst
to trigger a plethora of unusual but interesting behaviors
with lowering the energy scales. For the sake of completeness,
both the presence of the single type of disorder and
multi-type disorders would be carefully delivered as follows.

\begin{figure}
\centering
\includegraphics[width=3in]{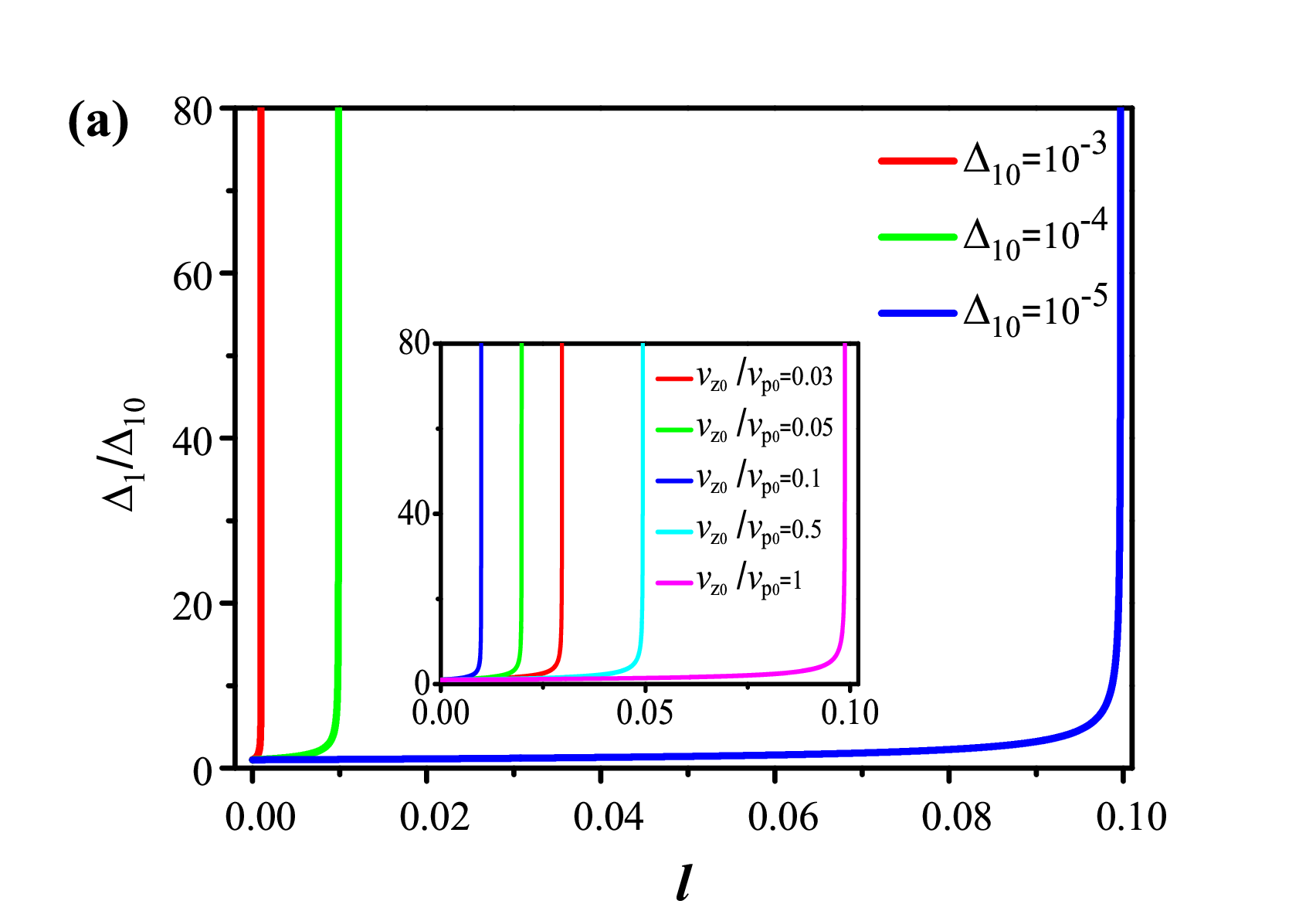}\\ \vspace{-0.3cm}
\includegraphics[width=3in]{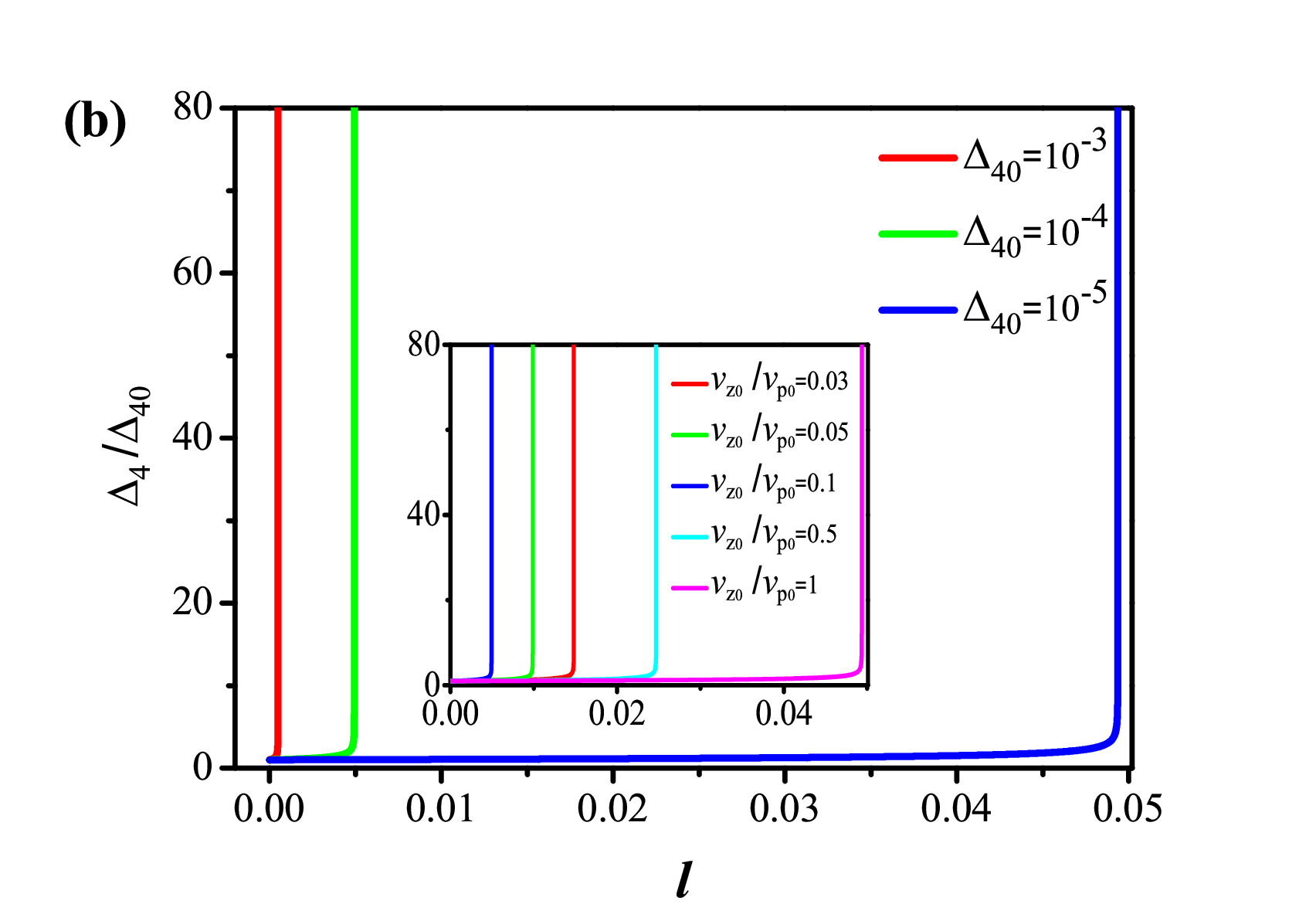}
\vspace{-0.1cm}
\caption{(Color online) Energy-dependent evolutions of disorder strengths
under several representative values of initial strengths at
$v_{z0}/v_{p0}=0.1$ for (a) $\Delta_1/\Delta_{10}$ with $\lambda_0=10^{-3}$
and (b) $\Delta_4/\Delta_{40}$ with $\lambda_0=10^{-4}$. Insets: the evolutions for
different values of initial anisotropy $v_{z0}/v_{p0}$ (the qualitative results are
insensitive to the initial values of interaction parameters).}\label{sole-Delta1_or_Delta4}
\end{figure}

\begin{figure}
\centering
\includegraphics[width=3in]{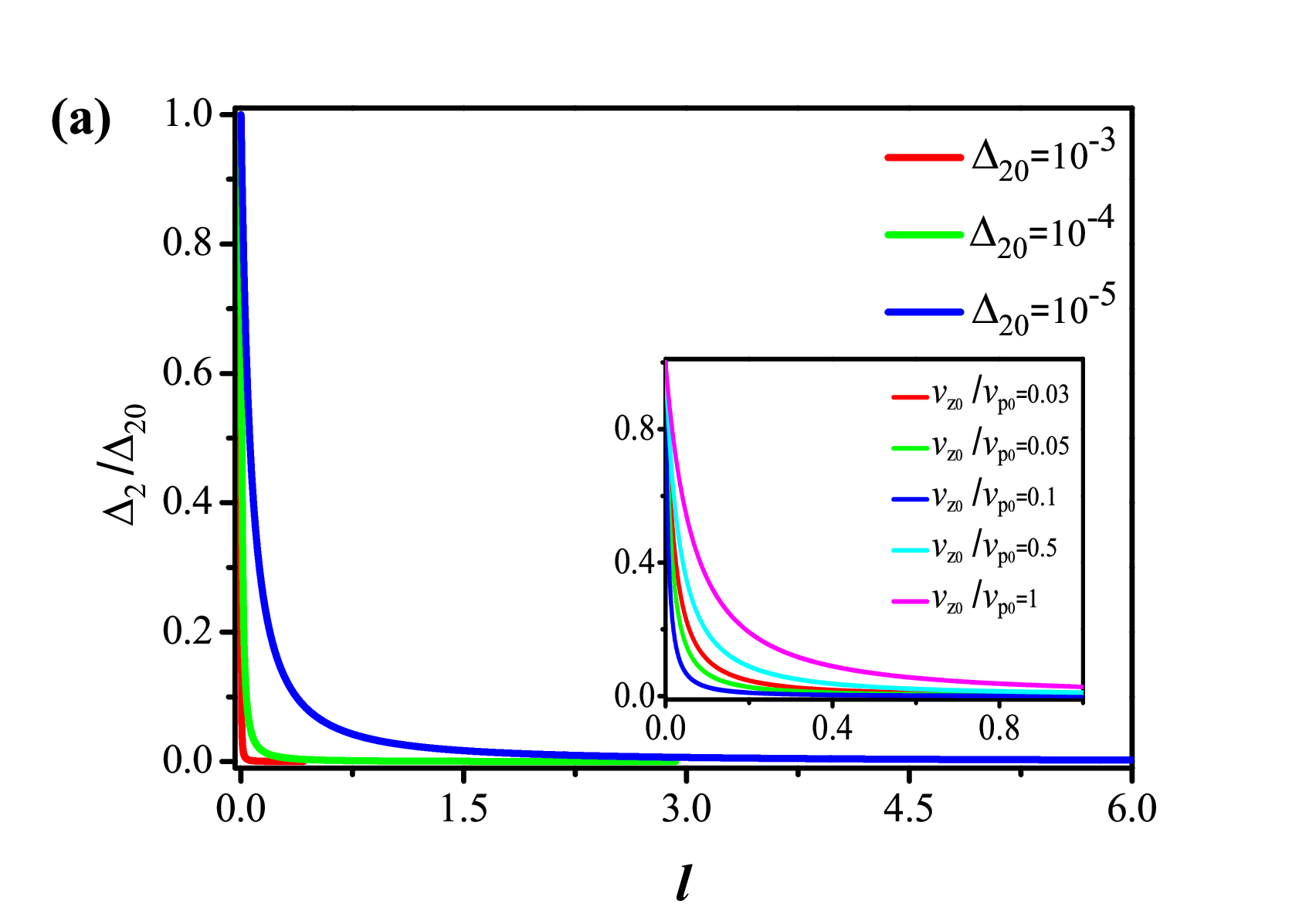}\\ \vspace{-0.3cm}
\includegraphics[width=3in]{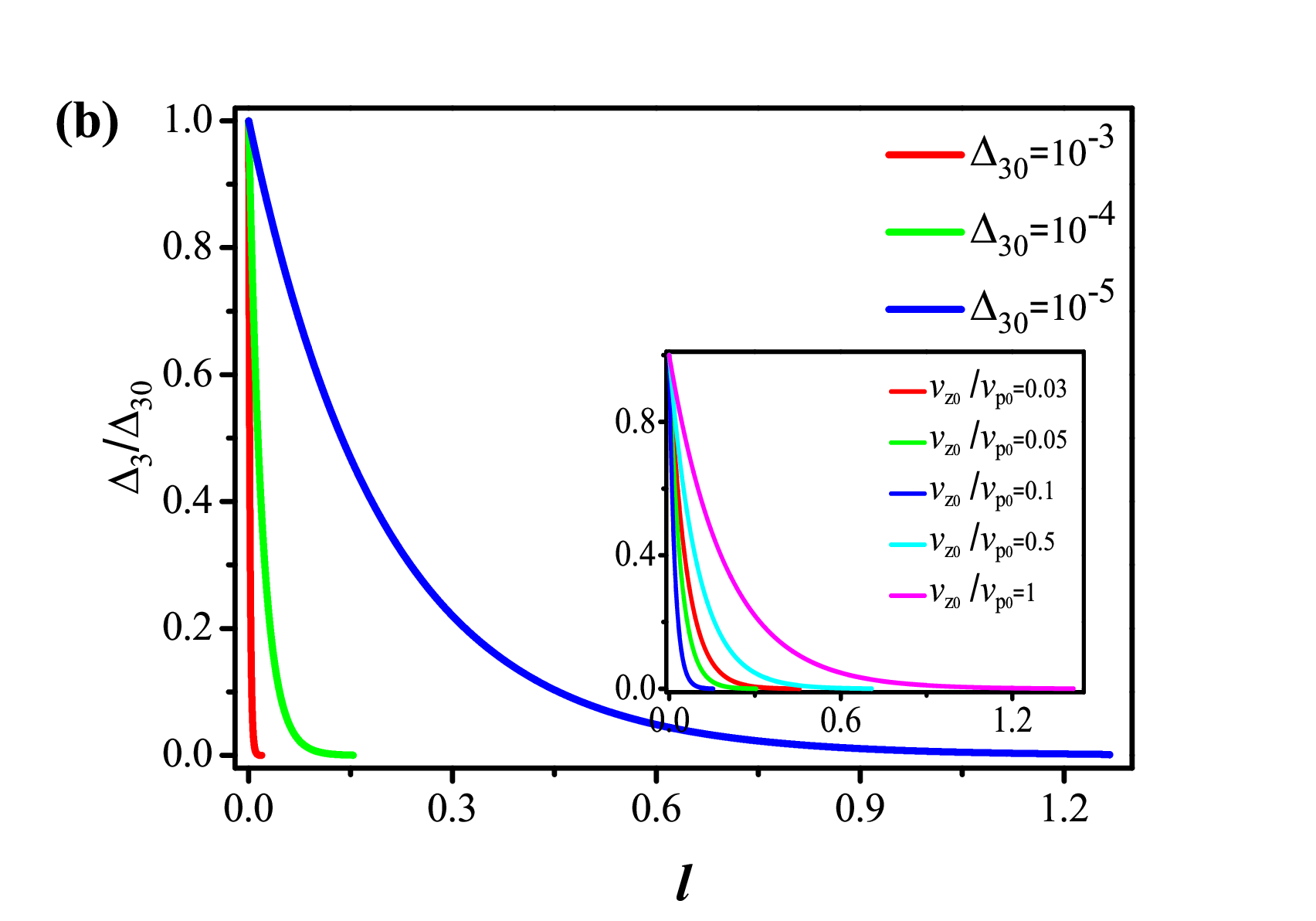}
\vspace{-0.1cm}
\caption{(Color online) Energy-dependent evolutions of disorder strengths
under several representative values of initial strengths at $v_{z0}/v_{p0}=0.1$ for (a)
$\Delta_2/\Delta_{20}$ with $\lambda_0=10^{-3}$ and (b) $\Delta_3/\Delta_{30}$
with $\lambda_0=10^{-4}$. Insets: the evolutions for
different values of initial anisotropy $v_{z0}/v_{p0}$ (the qualitative results are
insensitive to the initial values of interaction parameters).}\label{sole-Delta12_or_Delta3}
\end{figure}

\subsubsection{Certain sole type of disorder}\label{Subsubsection_sole-disorder}

Let us begin with the situation of a single sort of
disorder. Under such a circumstance, there exists only one
equation of disorder strength that survives in the coupled
equations~(\ref{Eq_RG_v_z})-(\ref{Eq_RG_Delta}). Carrying out the numerical calculations
indicates that four kinds of disorders fall into two
totally different fates at the lowest-energy limit once one
of them inheres alone. Learning from Fig.~\ref{sole-Delta1_or_Delta4}, it is worth
pointing out that either the disorder $\Delta_1$ or $\Delta_4$ (as the three components $\Delta_{4i}$
with $i=1,2,3$ share with the similar conclusion,
we from now on utilize $\Delta_4$ to represent $\Delta_{4i}$)
increases upon lowering the energy scale and eventually
goes toward diverging at certain critical energy denoted
by $l=l_c$. On the contrary, Fig.~\ref{sole-Delta12_or_Delta3} for the sole presence
of $\Delta_2$ or $\Delta_3$ displays that the disorder strength prefers
to gradually decrease and finally vanishes at the lowest-energy limit.

\begin{figure}
\centering
\includegraphics[width=3in]{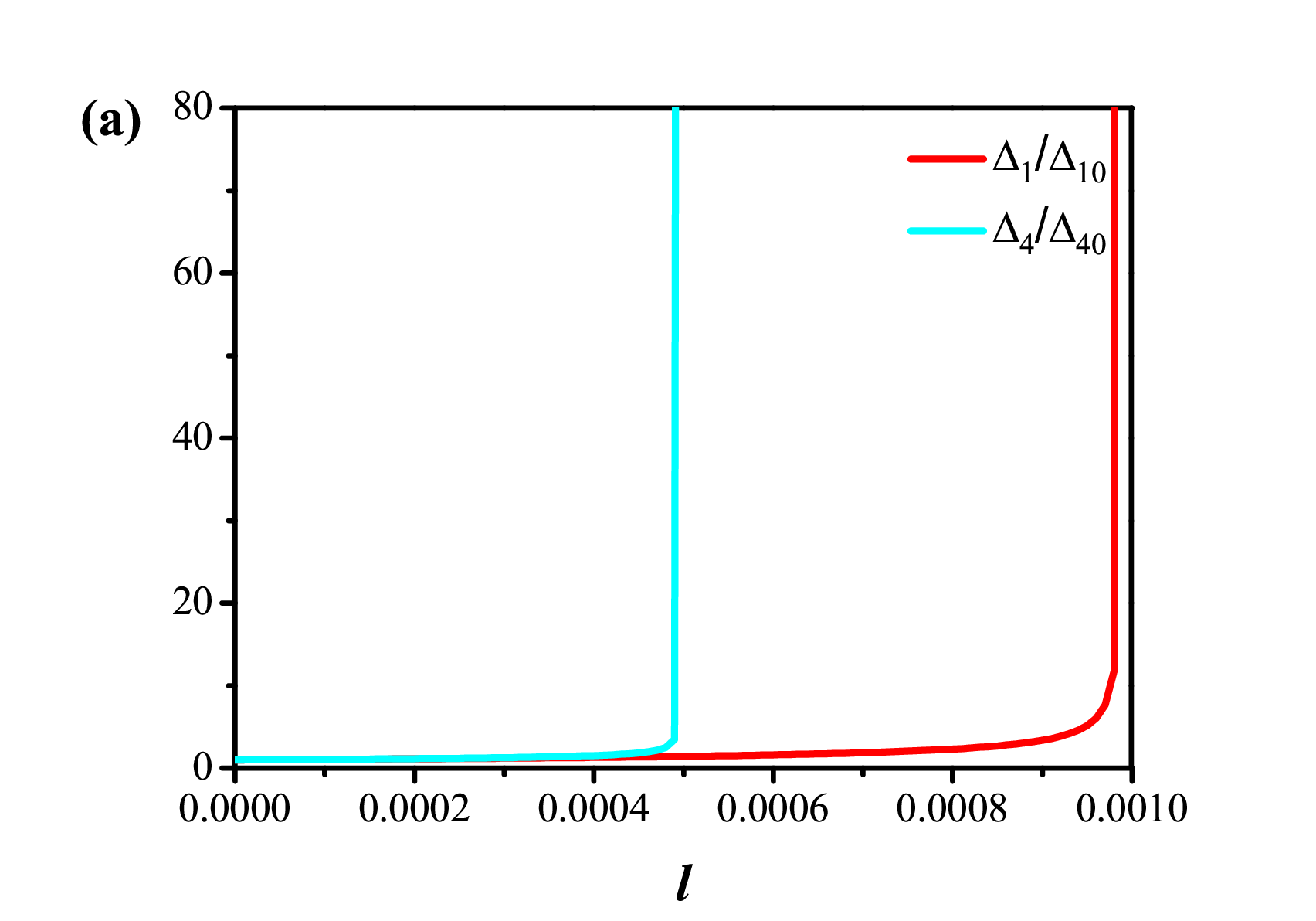}\\ \vspace{-0.3cm}
\includegraphics[width=3in]{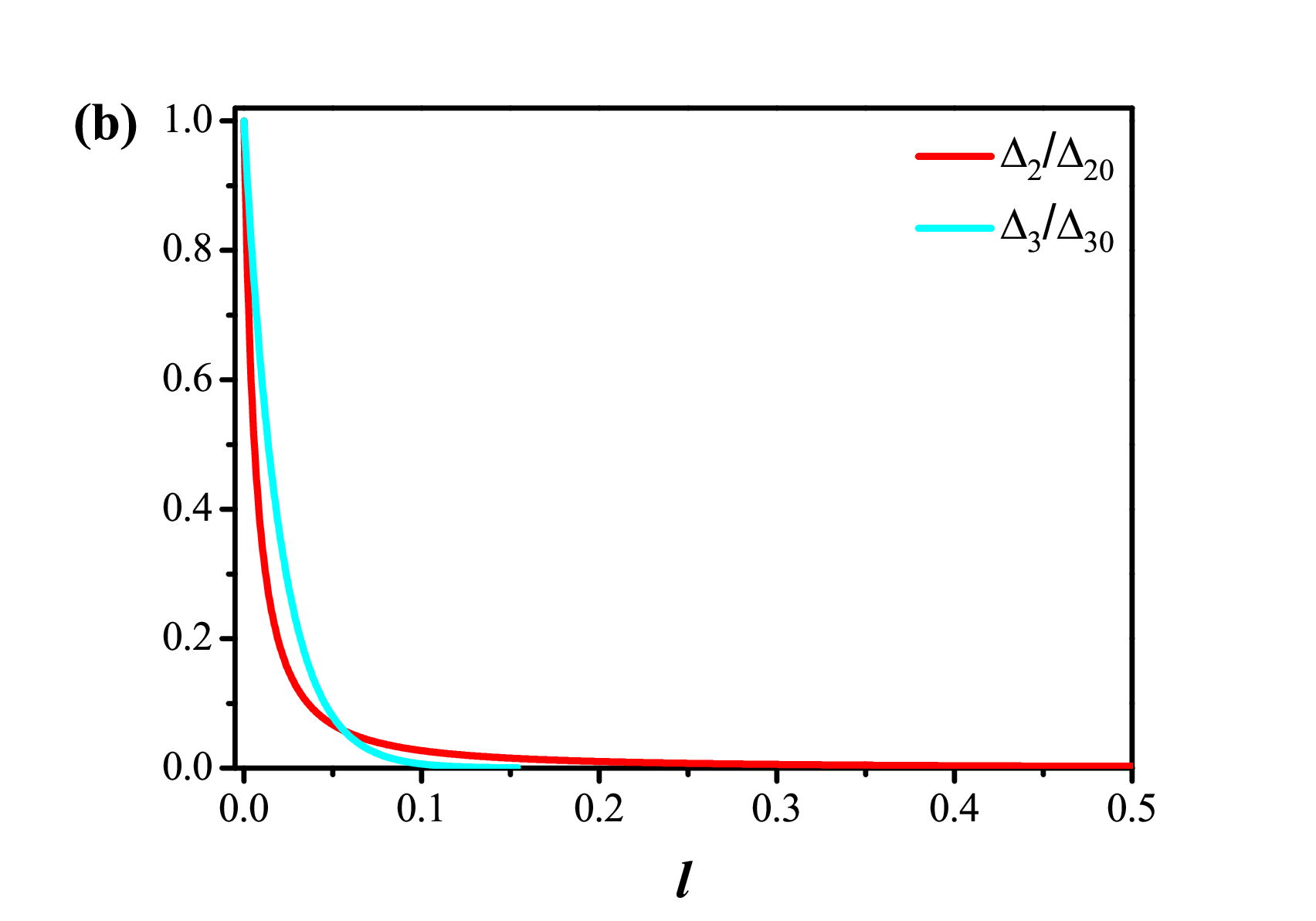}
\vspace{-0.1cm}
\caption{(Color online) Evolution comparisons with lowering energy scales
at $v_{z0}/v_{p0}=0.1$ and $\lambda_0=10^{-4}$
for (a) $\Delta_1/\Delta_{10}$ and $\Delta_4/\Delta_{40}$ with $\Delta_{10}=\Delta_{40}=10^{-3}$,
as well as (b) $\Delta_2/\Delta_{20}$ and $\Delta_3/\Delta_{30}$
with $\Delta_{20}=\Delta_{30}=10^{-4}$ (the qualitative results are insensitive
to the initial values of interaction parameters).}\label{sole_Delta_comparison}
\end{figure}

\begin{figure}
\centering
\includegraphics[width=3in]{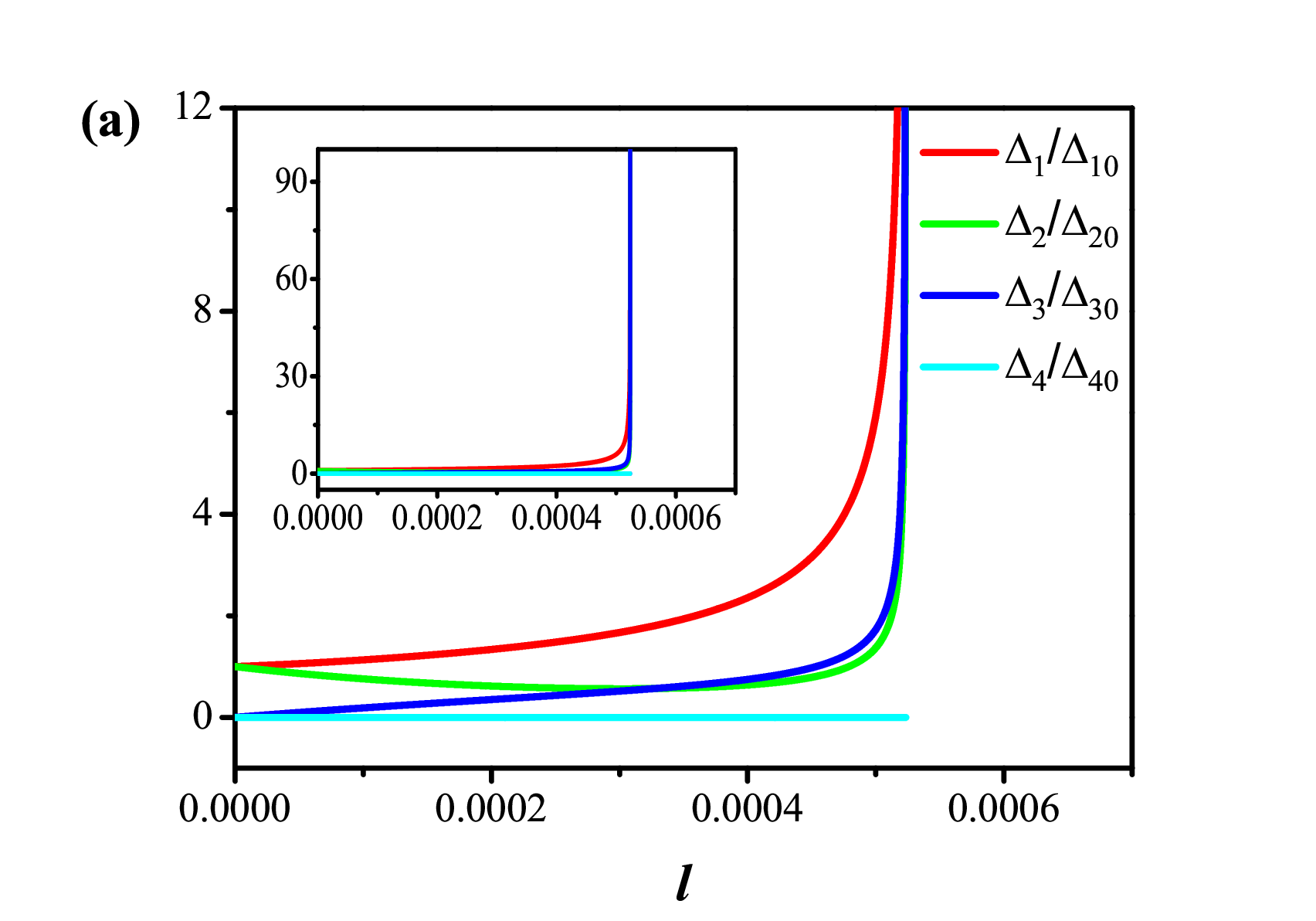}\\ \vspace{-0.3cm}
\includegraphics[width=3in]{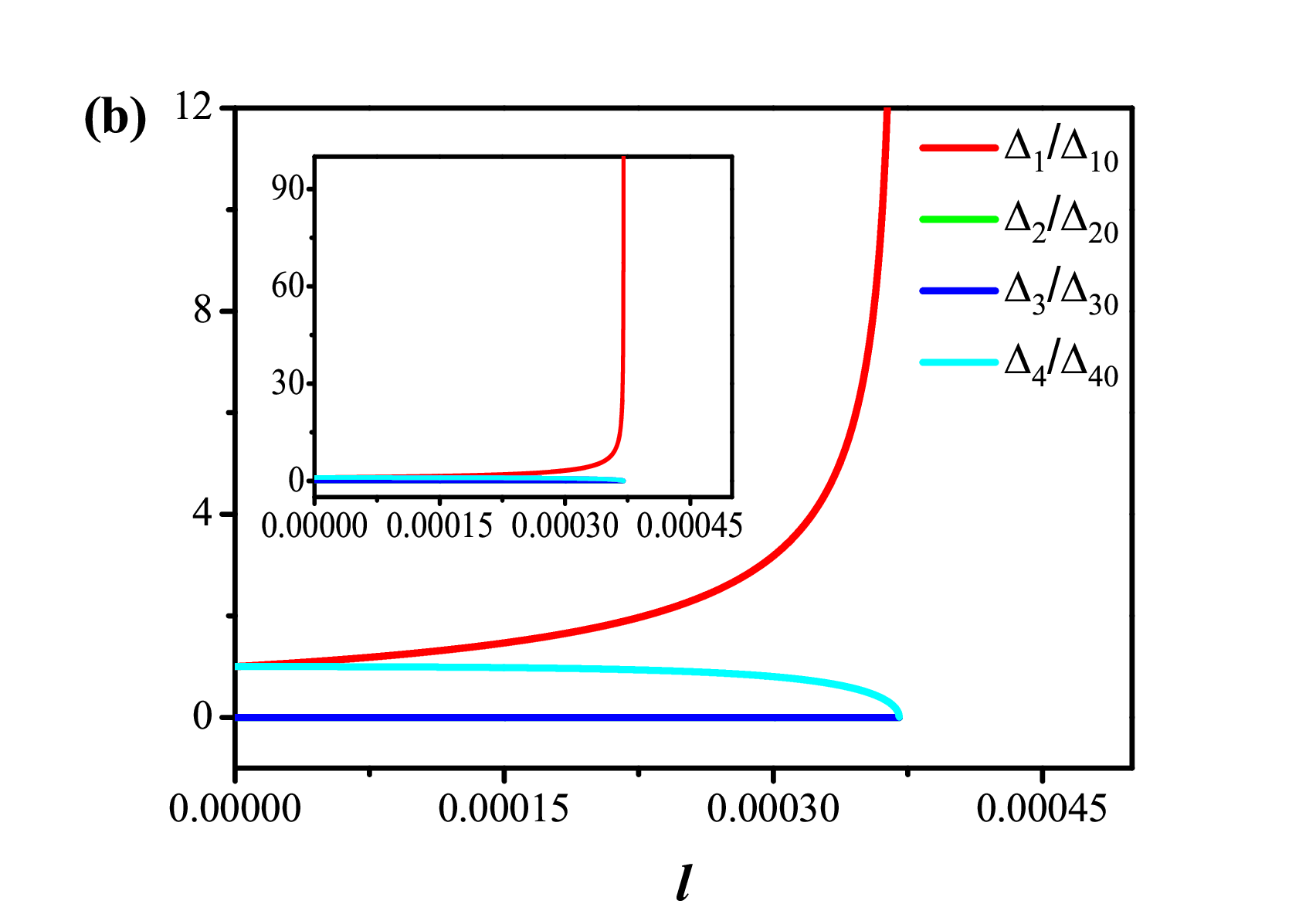}\\ \vspace{-0.3cm}
\includegraphics[width=3in]{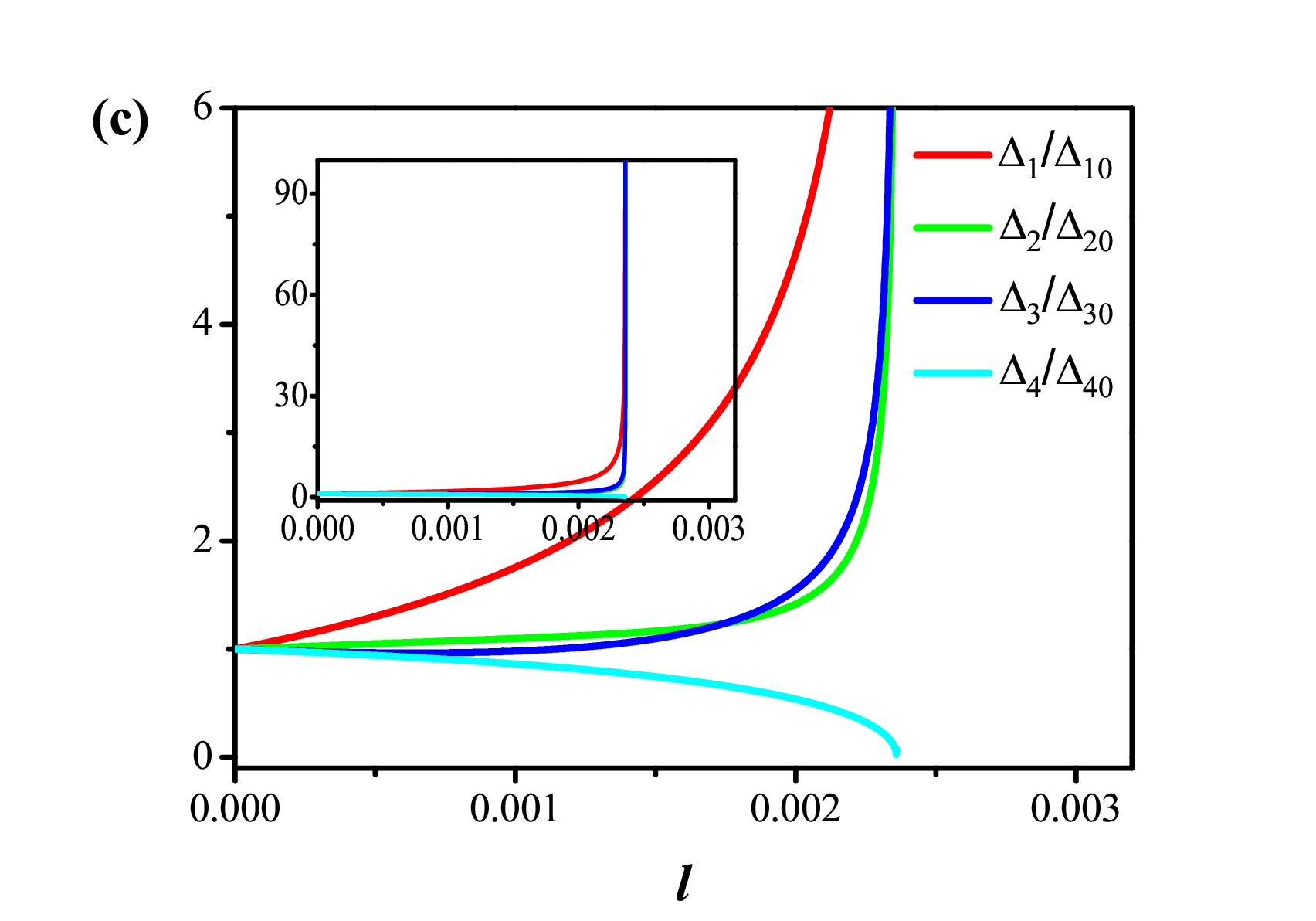}\\
\vspace{-0.1cm}
\caption{(Color online) Energy-dependent evolutions of disorder strengths at
$v_{z0}/v_{p0}=0.1$ and $\lambda_0=10^{-3}$ for the presence of
(a) $\Delta_1$ and $\Delta_2$ with $\Delta_{10}=\Delta_{20}=10^{-3}$ at the beginning
which cause the generation of the $\Delta_3$, (b) $\Delta_1$
and $\Delta_4$ (its three components are nearly overlapped and hence only of them is presented)
cannot induce any type of disorder, and (c) all sorts of disorders
with $\Delta_{10}=\Delta_{20}=\Delta_{30}=\Delta_{40}=10^{-4}$
(the qualitative results are insensitive to the initial values of
interaction parameters).}\label{mutiple-Delta1234-one}
\end{figure}

As to the presence of disorder $\Delta_1$ or $\Delta_4$ which is prone
to divergence as the energy scale is lower enough, we realize that
both the starting disorder strengths and initial anisotropy of fermion
velocities can be capable of providing considerable impacts on their energy-dependent
evolutions and the very critical energy scale at which
the instability is triggered. To be specific, one reading
from Fig.~\ref{sole-Delta1_or_Delta4} would manifestly figure out that either the
initial value of $\Delta_1$ or $\Delta_4$ is favorable to increase
the disorder strength with reducing the energy scale. This accordingly
causes the disorder strength divergences at certain smaller
$l_c$ indicating of a higher critical energy scale.
Compared to the beginning values of disorder strengths,
the fermion velocities play a more subtle role in the behaviors
of disorders. While the anisotropy of fermion velocities is strong
with $v_{z0}/v_{p0}$ taking a comparatively big or small value,
it is somewhat harmful to the enhancement of disorder
strength as clearly shown in Fig.~\ref{sole-Delta1_or_Delta4}.
In contrast, one can learn from Fig.~\ref{sole-Delta1_or_Delta4}
that a moderate anisotropy is helpful to the increase in disorder strength.
The optimal anisotropy is reached at $v_{z0}/v_{p0}\approx 0.1$,
which gives rise to a much smaller $l_c$. With respect to the other two
sorts of disorders $\Delta_2$ and $\Delta_3$, we notice that they exhibit
totally different tendencies. Fig.~\ref{sole-Delta12_or_Delta3} suggests that either
$\Delta_2$ or $\Delta_3$ falls off much more rapidly and quickly flows
toward zero as long as their initial values are tuned up.
Regarding the role played by the anisotropy of fermion
velocities, one can readily draw a conclusion from Fig.~\ref{sole-Delta12_or_Delta3}
that the disorder strengths for the sole presence of $\Delta_2$
or $\Delta_3$ climb down more quickly with an intermediate
anisotropy than its counterpart equipped with a relatively
strong anisotropy. What is more, we examine and realize that the critical energy scale represented
by $l_c$ is insusceptible to the initial values of fermion-fermion
interactions. Notwithstanding sharing with the same energy-dependent tendencies,
$\Delta_{1,4}$ or $\Delta_{2,3}$ still exhibit interesting distinctions between them.
Under the same initial conditions with lowering the energy scales,
one can obviously learn from Fig.~\ref{sole_Delta_comparison} that
the disorder strength $\Delta_4$ compared to $\Delta_1$ flows toward
divergence at some higher energy scale associated with a
smaller $l_c$. Meanwhile, in contrast with $\Delta_3$, the disorder $\Delta_2$
is a little difficult to be decreased and driven to zero at
a much lower energy scale. In other words, the disorder
$\Delta_4$ is more relevant but instead $\Delta_3$ becomes more irrelevant in
the low-energy regime. Although the numerical results
are evaluated on the basis of several representative initial
parameters of our theory, we would like to emphasize
that the basic conclusions are insusceptible to their
concrete values. Hence, we from now on would not highlight
this issue unless it is necessary.

\begin{figure}
\includegraphics[width=2.5in]{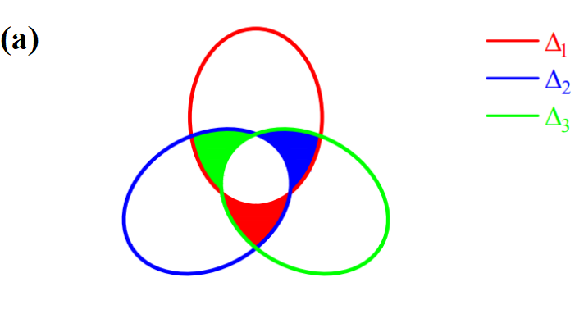} \\
\vspace{0.15cm}
\includegraphics[width=2.5in]{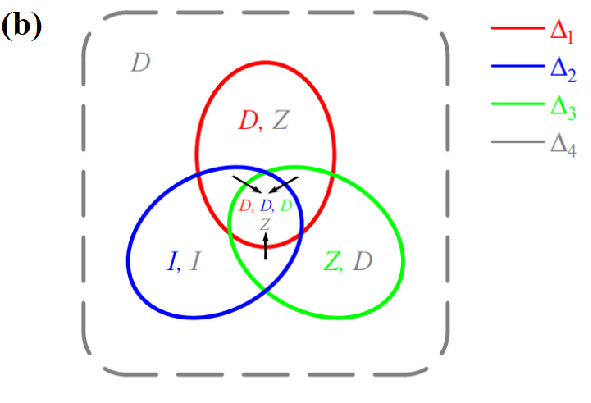}
\\
\vspace{-0.05cm}
\caption{(Color online) Schematic presentations for the behaviors of distinct kinds of disorders
with the regions circled by the red, blue, green, and gray curves denoting the
presence of $\Delta_{1,2,3,4}$ at the beginning,
respectively. (a) Dynamical-generated disorders:
the single presence of disorders $\Delta_{1,2,3}$ cannot develop another
type of disorder but instead any two of them are able to induce the third one
as labeled by red, blue and green regions (the basic results are insensitive
to $\Delta_4$ which hence is not shown). (b) Fates of disorders: the disorder strength for the sole
presence of $\Delta_{1,4}$ flows divergently and conversely
$\Delta_{2,3}$ goes toward zero.  In comparison, they are
capable of going either divergently or evolve toward zero as well as
hardly changing their initial values (Here $D$, $Z$ and $I$ equipped with the
colors to being associated with different kinds of disorders correspond to
being divergent, going toward zero, and approaching the initial values, respectively).}\label{Fig_dis_summary}
\end{figure}

\subsubsection{Multiple types of disorders}\label{Subsec_multi_disorder}

Subsequently, we move to the situations for the simultaneous presence of more
than one sort of disorders in which the related equations of disorder
strengths (\ref{Eq_RG_v_z})-(\ref{Eq_RG_Delta}) are accompanied. After performing the numerical
analysis, we are left with a number of interesting results
that are addressed as follows.

First of all, we highlight that the interplay of multi-type
disorders can generate an additional sort of disorder
that is out of the disorders appearing initially.
In other words, a new type of disorder without initial
strength would receive a finite disorder strength attesting
to the evolutions of the coupled equations of disorders.
It is then convenient to nominate such phenomenon
as the dynamical-generated disorder for further discussions.
We have examined that these basic conclusions
are robust against the starting values of fermion-fermion
interactions, which, therefore, in principle are rooted in the
intimate competition among distinct types of disorders.
To be concrete, certain dynamical-generated disorder is
always realized as long as two of disorders $\Delta_1$, $\Delta_2$, and
$\Delta_3$ are present at the starting point. Taking the presence of disorders
$\Delta_1$ and $\Delta_2$ for an instance, Fig.~\ref{mutiple-Delta1234-one}(a) displays
that the disorder $\Delta_3$ develops as the energy scale is lowered
due to the dynamical-generated mechanism of disorder scatterings.
However, it is of particular importance to point out that such dynamical-generated mechanism
cannot be activated once two of disorders $\Delta_1$, $\Delta_2$, and
$\Delta_3$ are absent at the initial point as delineated in Fig.~\ref{mutiple-Delta1234-one}(b).
In sharp contrast with disorders $\Delta_{1,2,3}$, we notice that the
disorder $\Delta_4$ shown in Fig.~\ref{mutiple-Delta1234-one}(b) cannot contribute to the
dynamical-generated scenario. This means that it does
neither participate in the generation of certain disorder nor
can be formed by the other sorts of disorders. Accordingly,
we argue that the disorders $\Delta_1$, $\Delta_2$, and $\Delta_3$ are
closely entangled with each other but instead the disorder
$\Delta_4$ seems to be independent of other disorders.
This may be rooted in the combination of the unique dispersions of nodal-line fermions, and
the distinguished features of disorder themselves as well as
intimate interplay with fermion-fermion interactions.

In addition, we assume all of four sorts of disorders are
present and shed light on the low-energy fates of disorders
under the close competition among themselves with
the variations of initial conditions. One can clearly infer
from Fig.~\ref{mutiple-Delta1234-one}(c) that the disorders
$\Delta_1$ and $\Delta_2$ together with $\Delta_3$ quickly climb
up via lowering the energy scale and go toward divergence as
approaching the critical scale at $l=l_c$ (the basic results for $\Delta_2$
and $\Delta_3$ are similar). Conversely, the strength of disorder $\Delta_4$
is gradually diminished and eventually vanishes at the lowest-energy
limit. On the one side, this signals that the disorder $\Delta_4$ is
insusceptible to the interplay among disorders and thus
indeed shares the analogous low-energy properties with the circumstance
where it presents alone as shown in Fig.~\ref{sole-Delta1_or_Delta4},
which is in well consistent with the basic conclusion in the
previous paragraph. On the other side, the tendencies of
both the disorders $\Delta_2$ and $\Delta_3$, which are driven to vanish
if only one of them is present, are heavily reshaped by
the competition among disorders.

Last but not the least important, it is necessary to address several comments on
the critical energy scale represented by $l_c$ in the presence of all types of disorders.
After paralleling the similar numerical analysis, we find that initial values of fermion-fermion
interactions as behaved for the presence of sole type of disorder
nearly do not influence the critical energy scale.
In sharp contrast, the bigger initial strengths are in favor of supporting the
increase in disorders and make the critical energy scale
a little bigger causing the instability to occur in advance, which is also
analogous to its sole-disorder counterpart illustrated in Fig.~\ref{sole-Delta1_or_Delta4}.
As for the beginning anisotropy of fermion velocities, it bears strong similarities to
the single type of disorder displayed in Fig.~\ref{sole-Delta12_or_Delta3} as well.
One can notice that the strong anisotropy where $v_{z0}/v_{p0}$ is taken as either
a big or small value is preferable to hamper the divergence of disorder.
However, there exists certain moderate anisotropy with $v_{z0}/v_{p0}\approx 0.1$
tends to facilitate the increase of disorder strength
with the considerably bigger critical energy scales.

Before moving further, we would like to give a short
summary for the low-energy fates of all kinds of disorders.
As schematically summarized in Fig.~\ref{Fig_dis_summary},
we find that certain dynamical-generated disorder can be generated owing to
the disorder competitions which are primarily involved
the interplay of disorders $\Delta_1$, $\Delta_2$, and $\Delta_3$ but not the disorder $\Delta_4$.
Compared to the sole presence of disorder in Sec.~\ref{Subsubsection_sole-disorder},
the fates of $\Delta_2$ plus $\Delta_3$ influenced by the disorder
competitions are qualitatively reformulated from vanishment to divergence
in the low-energy regime and conversely for the $\Delta_4$, but instead the $\Delta_1$'s is not changed.
It is noteworthy that the divergence of disorder may point to a disorder-dominated diffusive metallic state~\cite{Fradkin1986PRB,Foster2008PRB,Kobayashi2014PRL,Lai-1409.8675,Goswami2011PRL,Moon-1409.0573,Wang2015PLA}.
Regarding the initial conditions, both beginning values of disorder strengths and
anisotropy of fermion velocities bring out the substantial impacts on the
low-energy behaviors whereas the effects of starting fermion-fermion interactions
are negligible.

\subsection{Fates of fermion-fermion interactions}\label{Subsec_fates_fermion-fermion}

With the low-energy fates of the disorders in hand, we are, therefore, suitable to consider
and reveal the interesting properties of fermion-fermion interactions.

\begin{figure}
\includegraphics[width=3in]{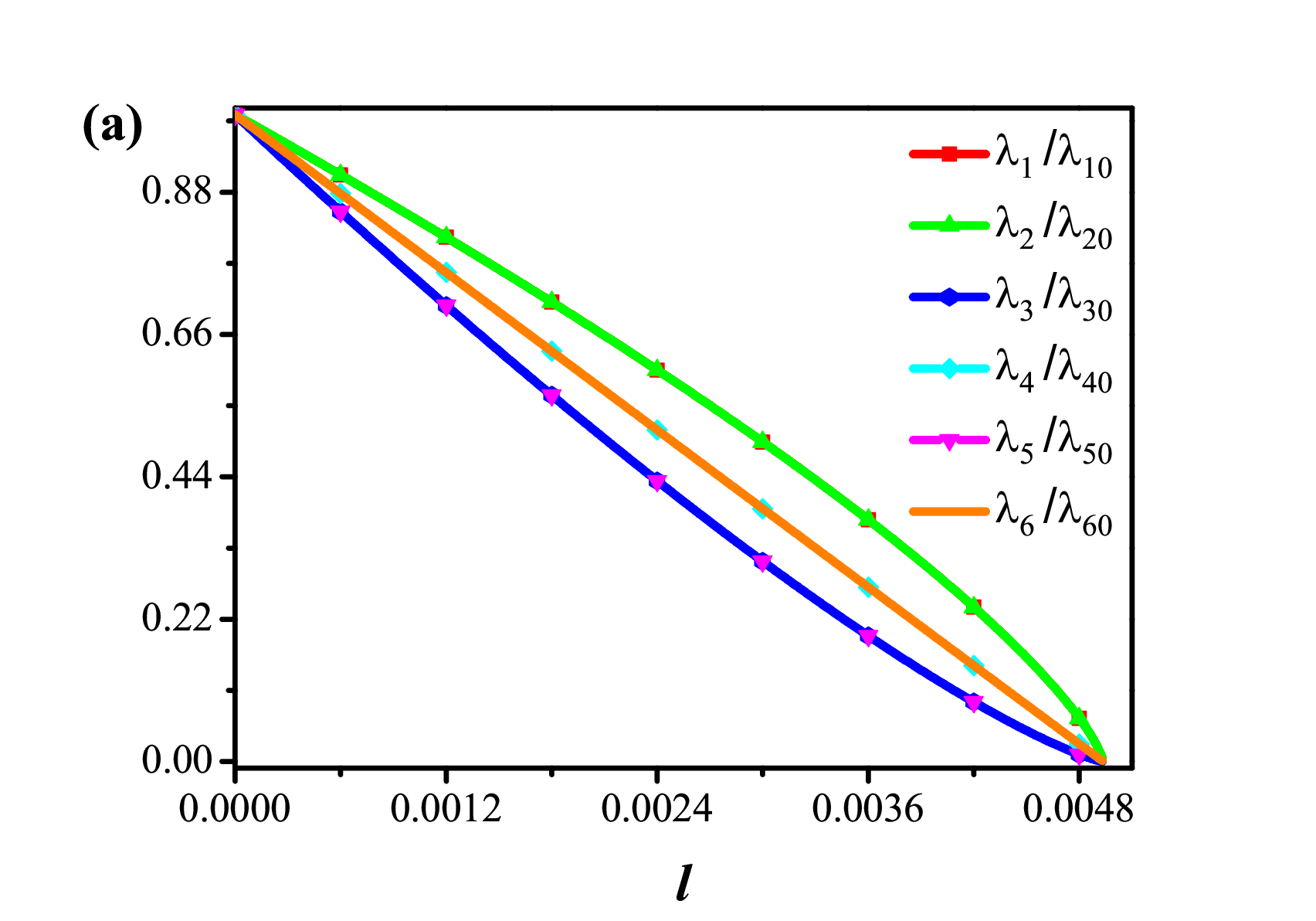}\vspace{-0.3cm}
\includegraphics[width=3in]{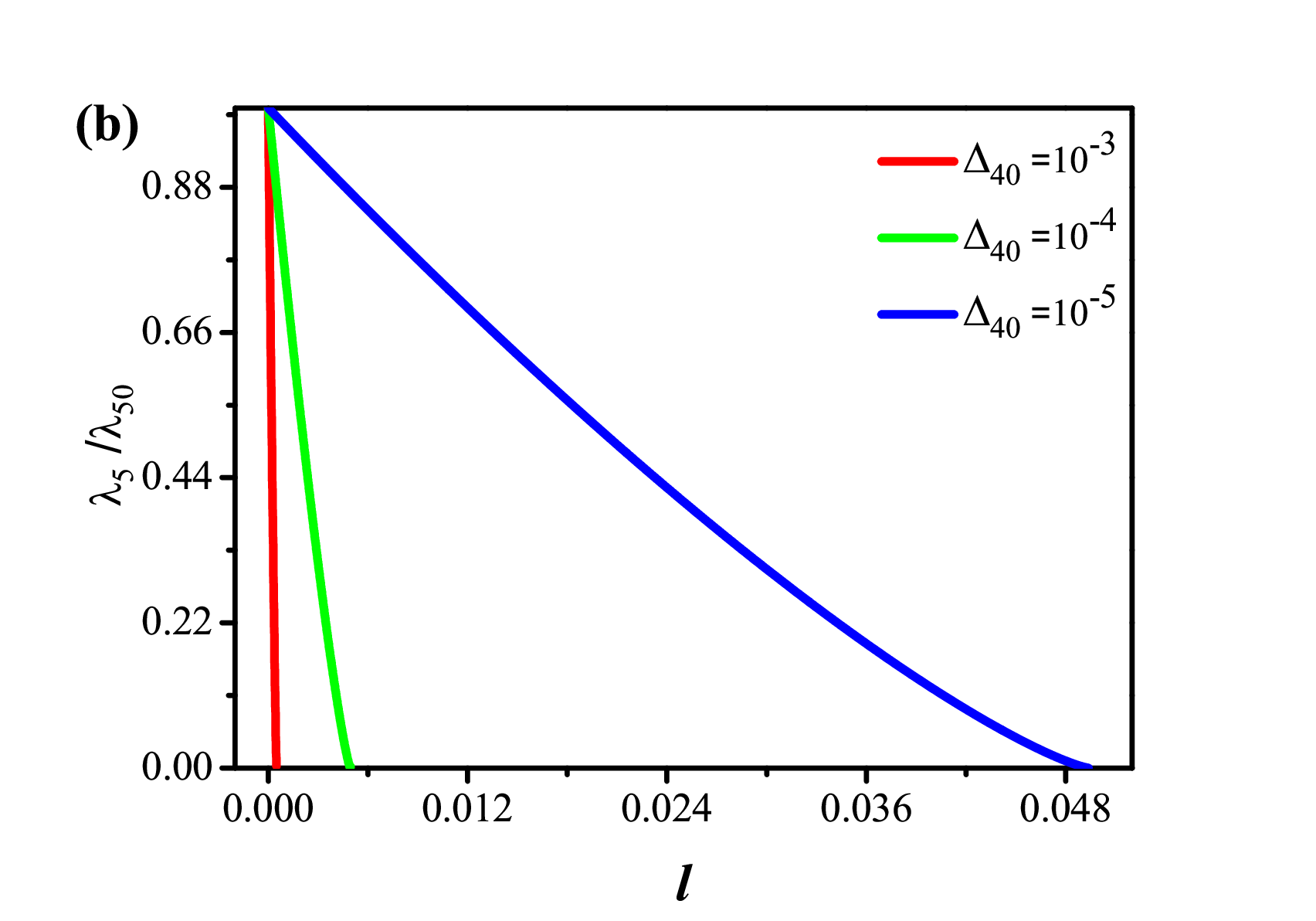}\vspace{-0.3cm}
\includegraphics[width=3in]{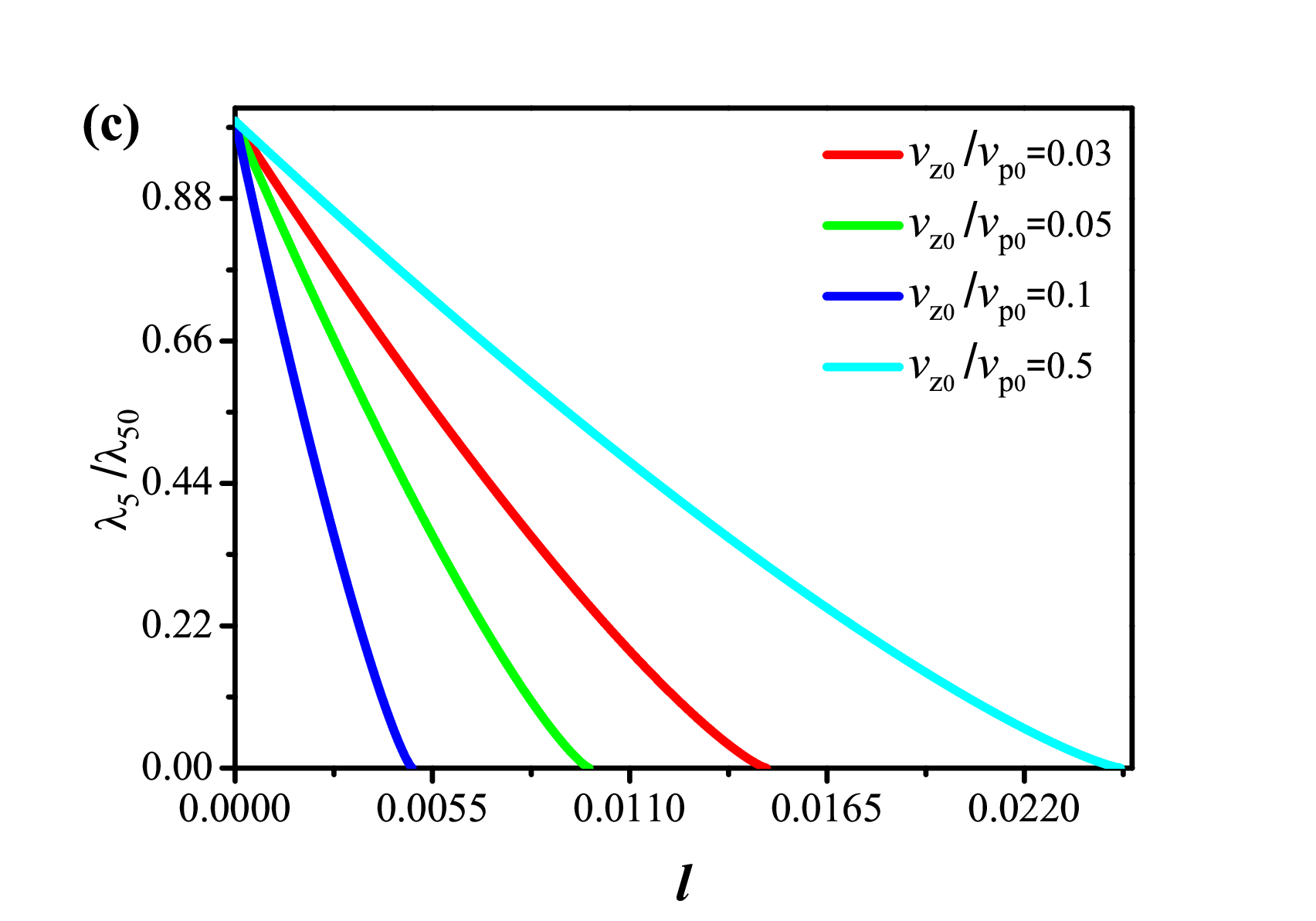}\\
\vspace{-0.1cm}
\caption{(Color online) Energy-dependent evolutions of (a) $\lambda_i/\lambda_{i0}$
($i=1-6$) with $v_{z0}/v_{p0}=0.1$, $\lambda_0=10^{-4}$ and $\Delta_{40}=10^{-4}$;
(b) $\lambda_5/\lambda_{50}$ at $\lambda_0=10^{-4}$ and $v_{z0}/v_{p0}=0.1$ with
several representative values of initial strengths of $\Delta_4$;
and (c) $\lambda_5/\lambda_{50}$ at $\Delta_{40}=10^{-4}$ with
different initial values of anisotropy $v_{z0}/v_{p0}$.}\label{dis4-f-f-interaction}
\end{figure}

\begin{figure}
\includegraphics[width=3in]{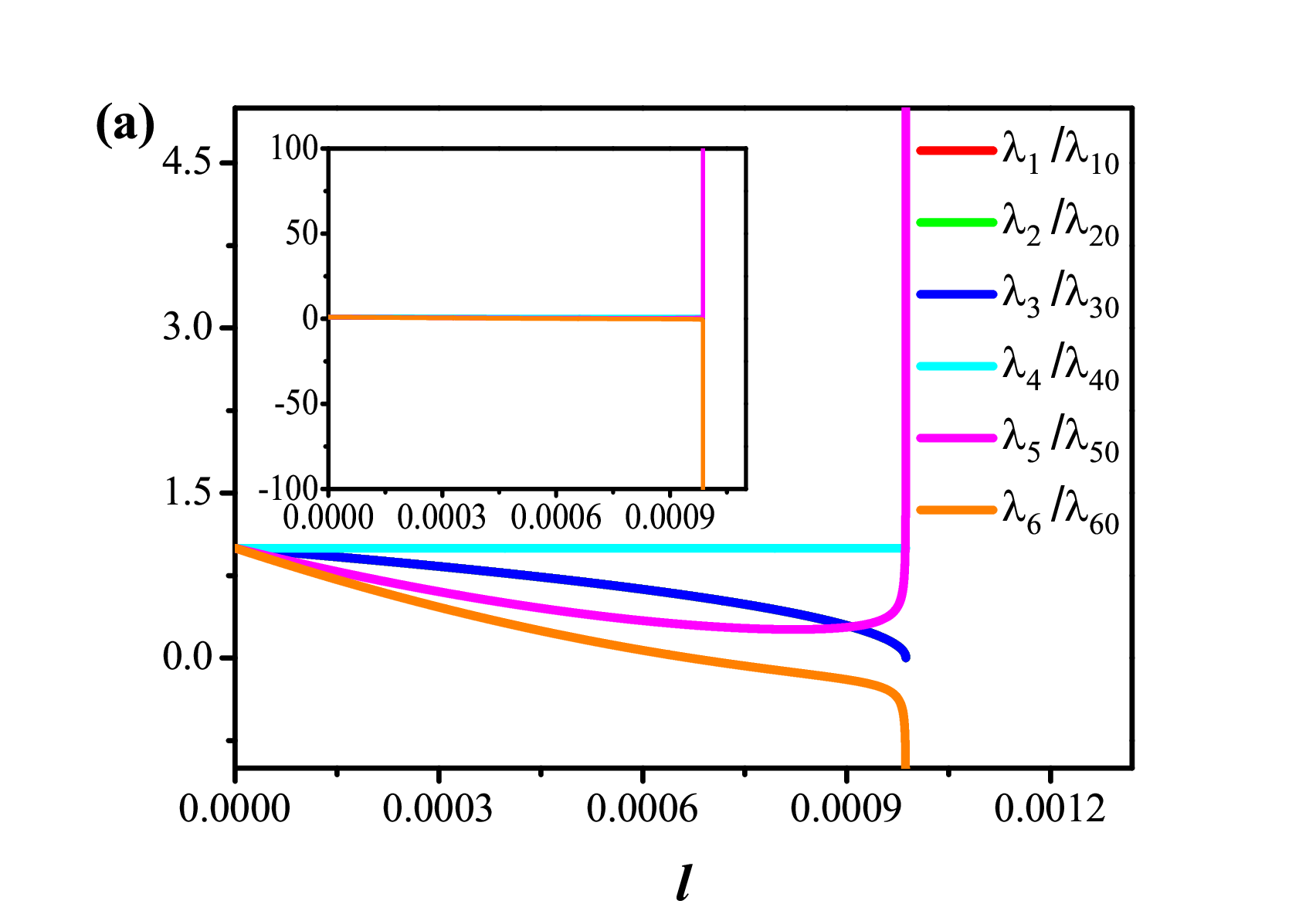}\vspace{-0.3cm}
\includegraphics[width=3in]{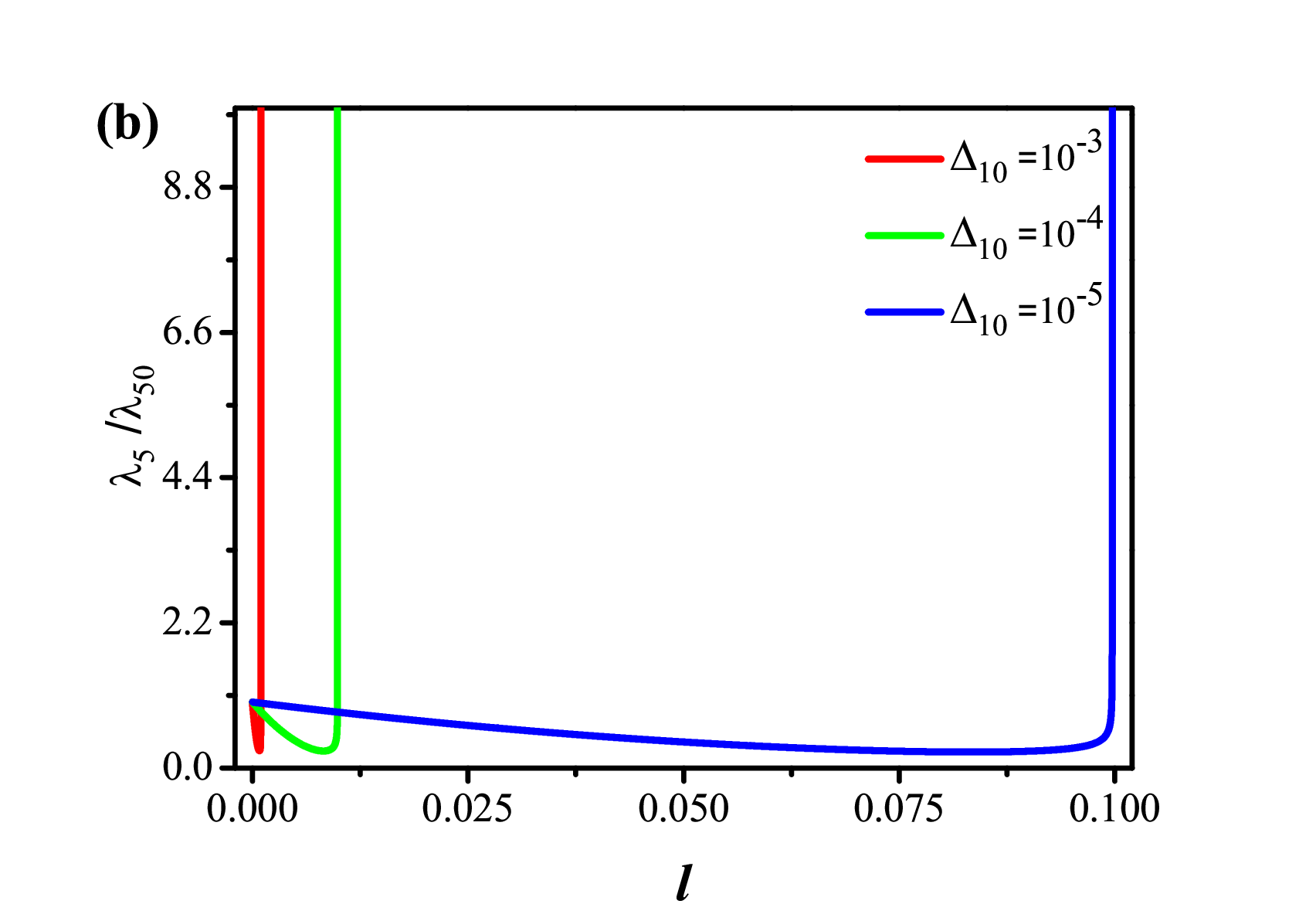}\vspace{-0.3cm}
\includegraphics[width=3in]{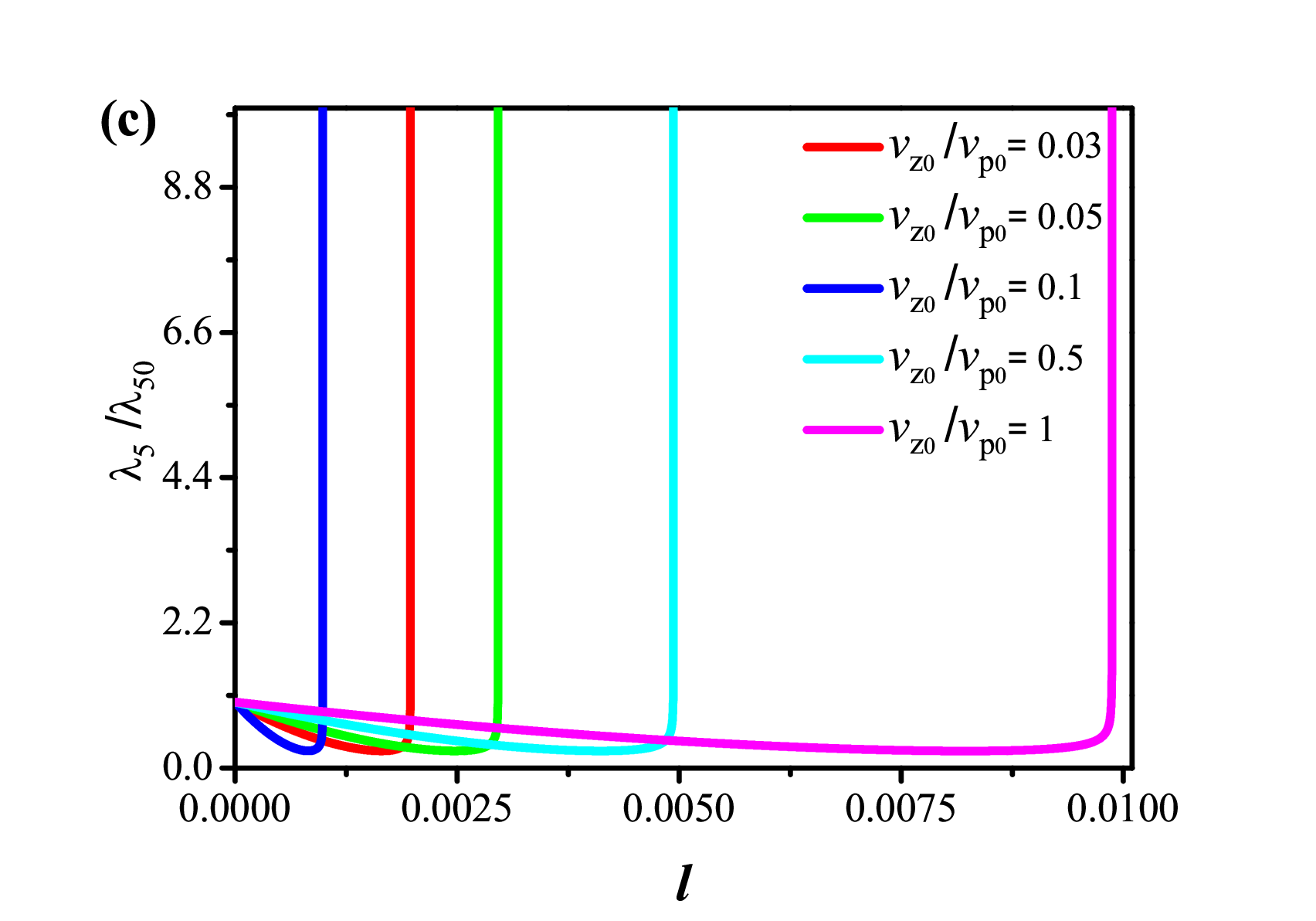}\\
\vspace{-0.1cm}
\caption{(Color online) Energy-dependent evolutions of (a) $\lambda_i/\lambda_{i0}$
($i=1-6$) with $v_{z0}/v_{p0}=0.1$, $\lambda_0=10^{-3}$ and $\Delta_{10}=10^{-3}$;
(b) $\lambda_5/\lambda_{50}$ at $\lambda_0=10^{-3}$ and $v_{z0}/v_{p0}=0.1$ with
several representative values of initial strengths of $\Delta_1$;
and (c) $\lambda_5/\lambda_{50}$ at $\Delta_{10}=10^{-3}$ with
different initial values of anisotropy $v_{z0}/v_{p0}$.}\label{dis1-f-f-interaction}
\end{figure}

\begin{figure}
\includegraphics[width=3in]{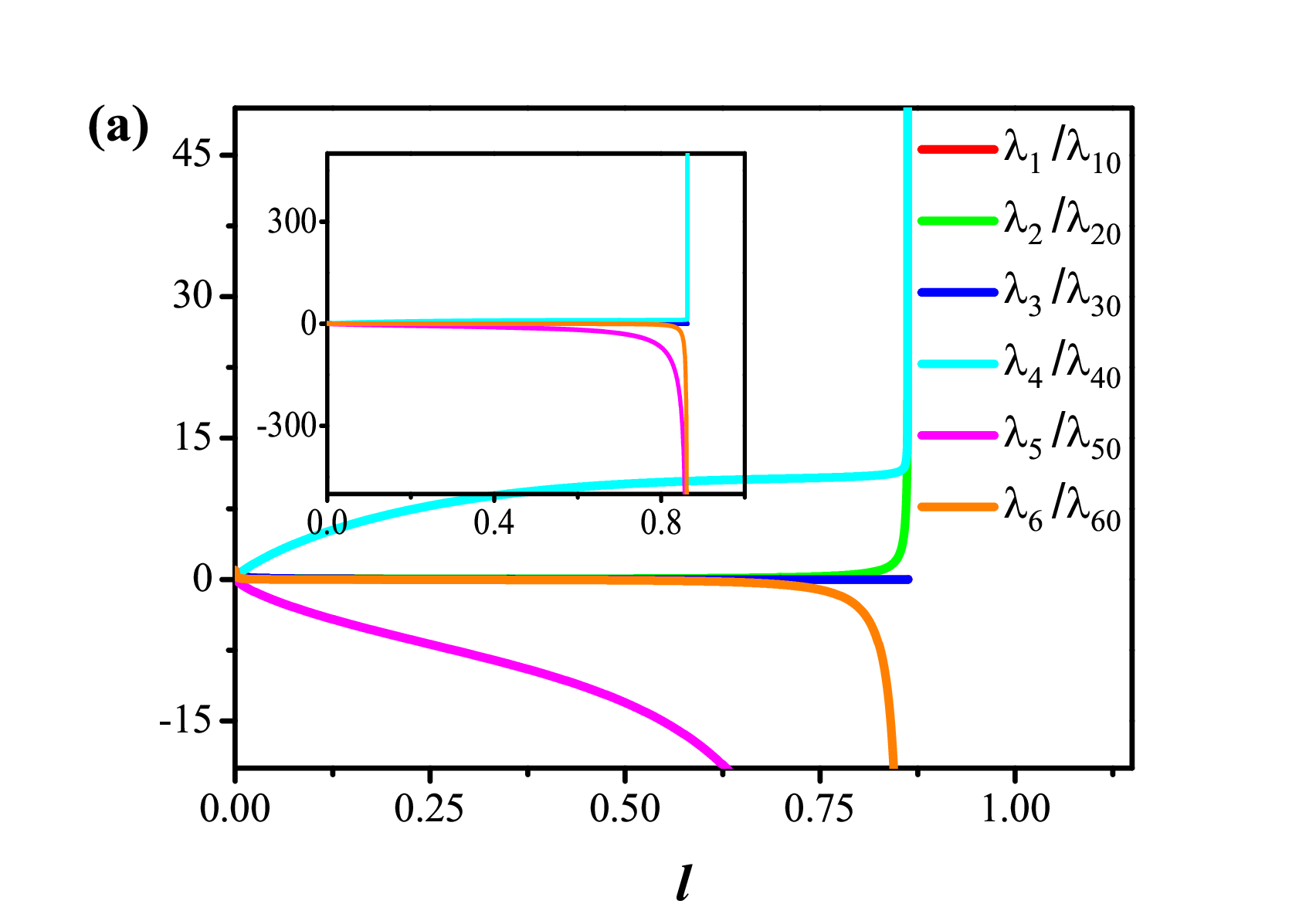}\vspace{-0.3cm}
\includegraphics[width=3in]{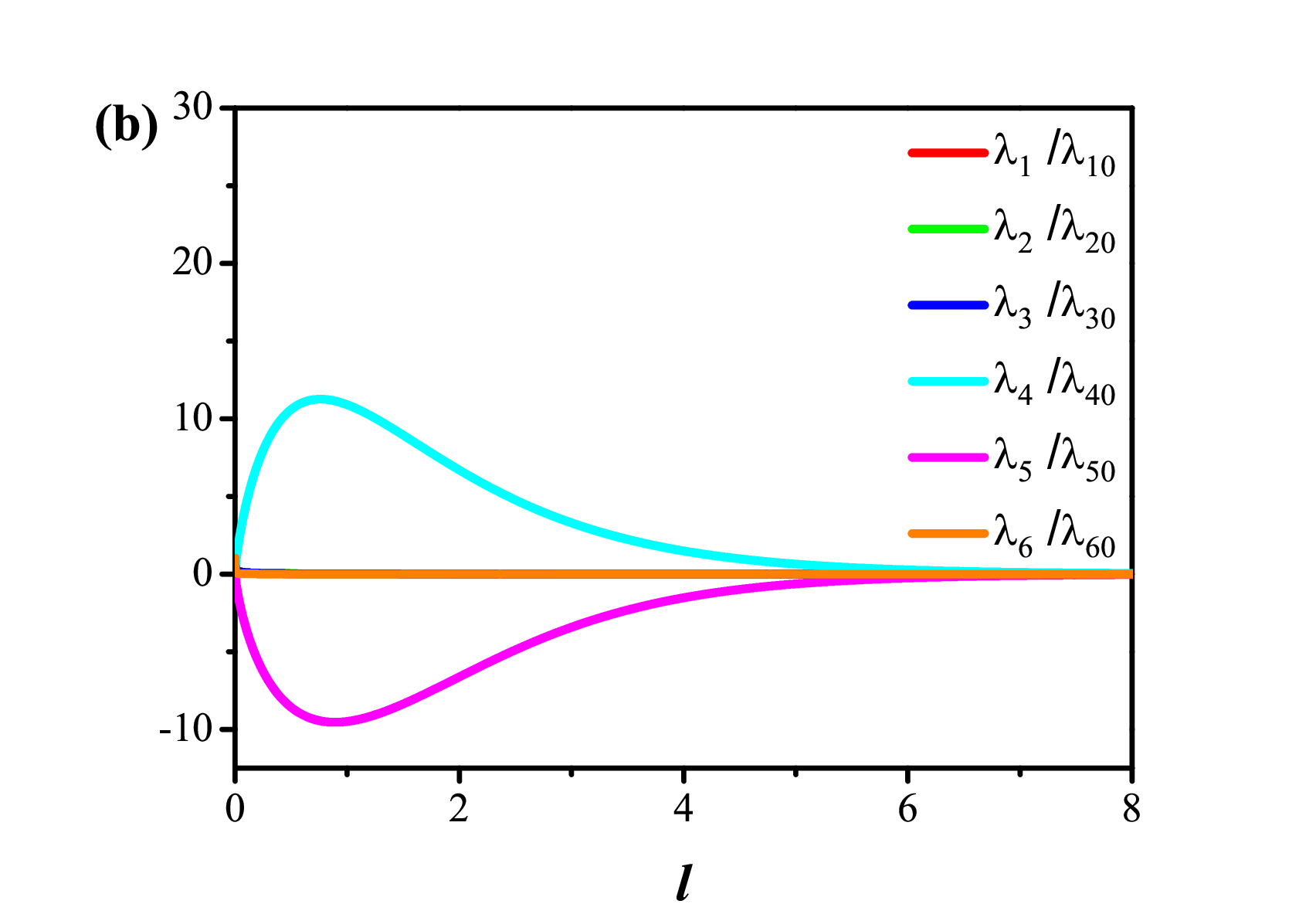}\vspace{-0.3cm}
\includegraphics[width=3in]{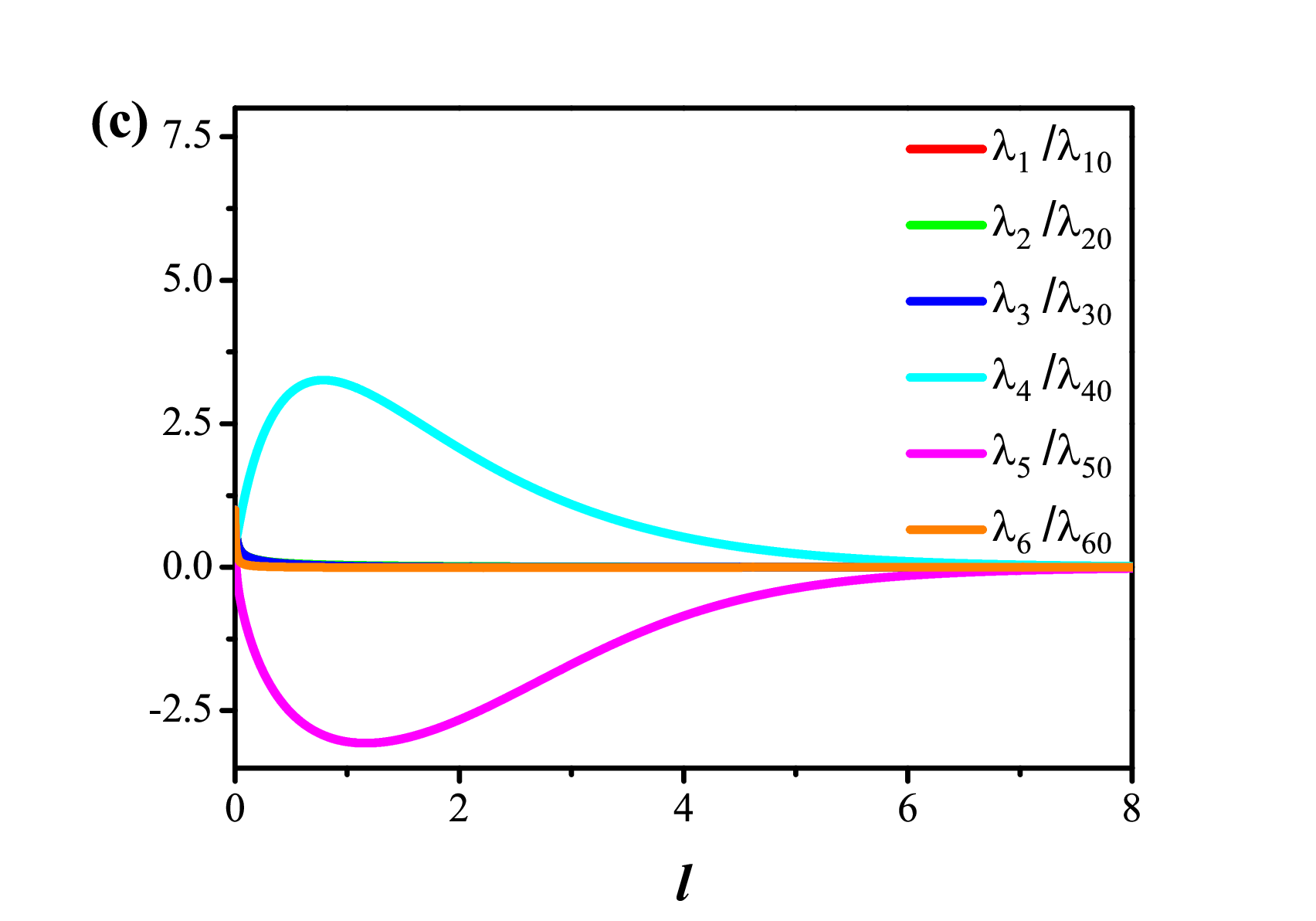}\\
\vspace{-0.1cm}
\caption{(Color online) Energy-dependent evolutions of fermion-fermion
couplings under different initial conditions: (a) $v_{z0}/v_{p0}=0.1$, $\lambda_{i0}=10^{-3}$ and
$\Delta_{20}=10^{-3}$; (b) $v_{z0}/v_{p0}=0.1$, $\lambda_{i0}=10^{-4}$, $\Delta_{20}=10^{-3}$;
and (c) $v_{z0}/v_{p0}=0.5$, $\lambda_{i0}=10^{-3}$ and $\Delta_{20}=10^{-3}$.}\label{dis2-f-f-interaction}
\end{figure}

\begin{figure}
\includegraphics[width=3in]{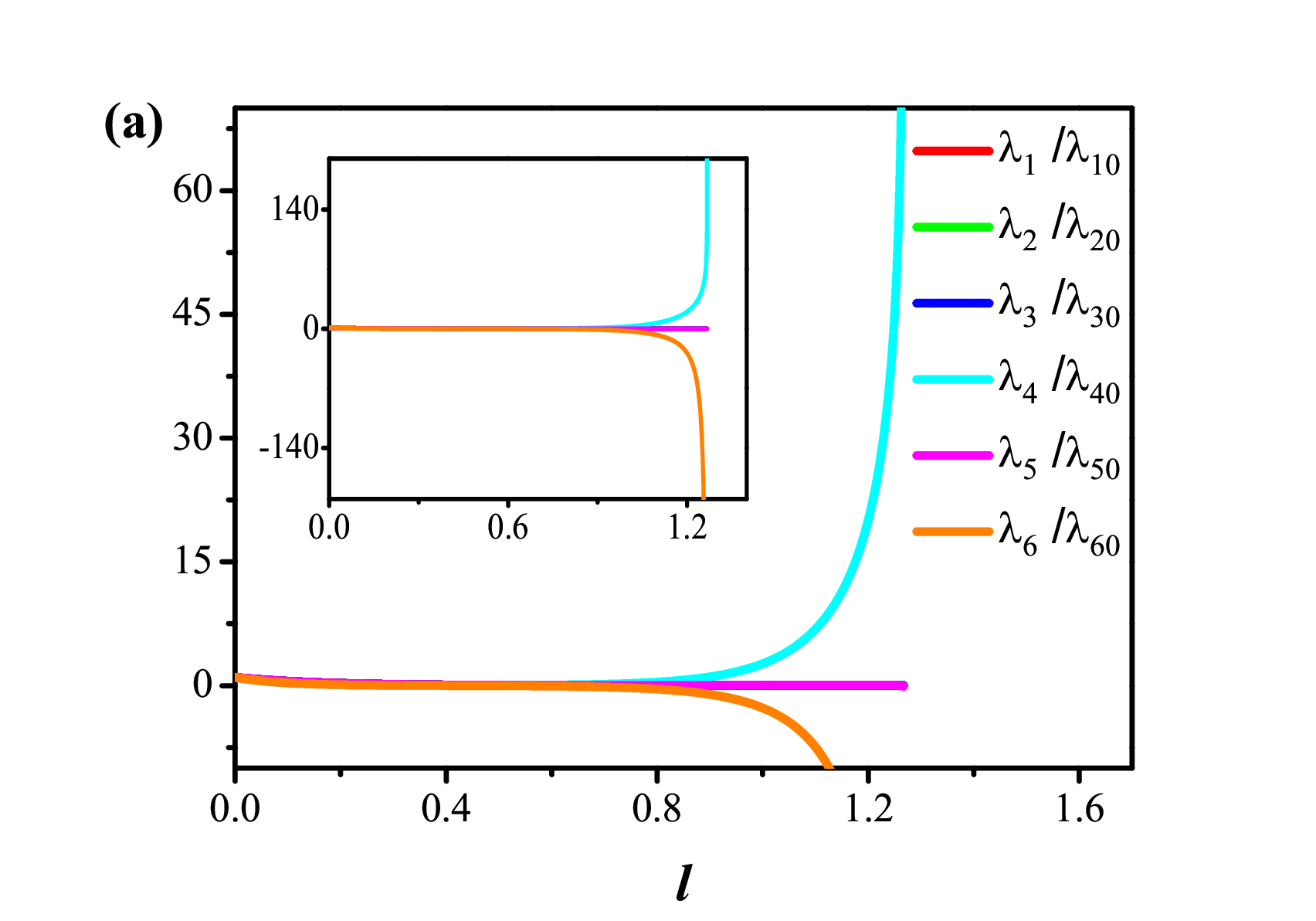}\vspace{-0.3cm}
\includegraphics[width=3in]{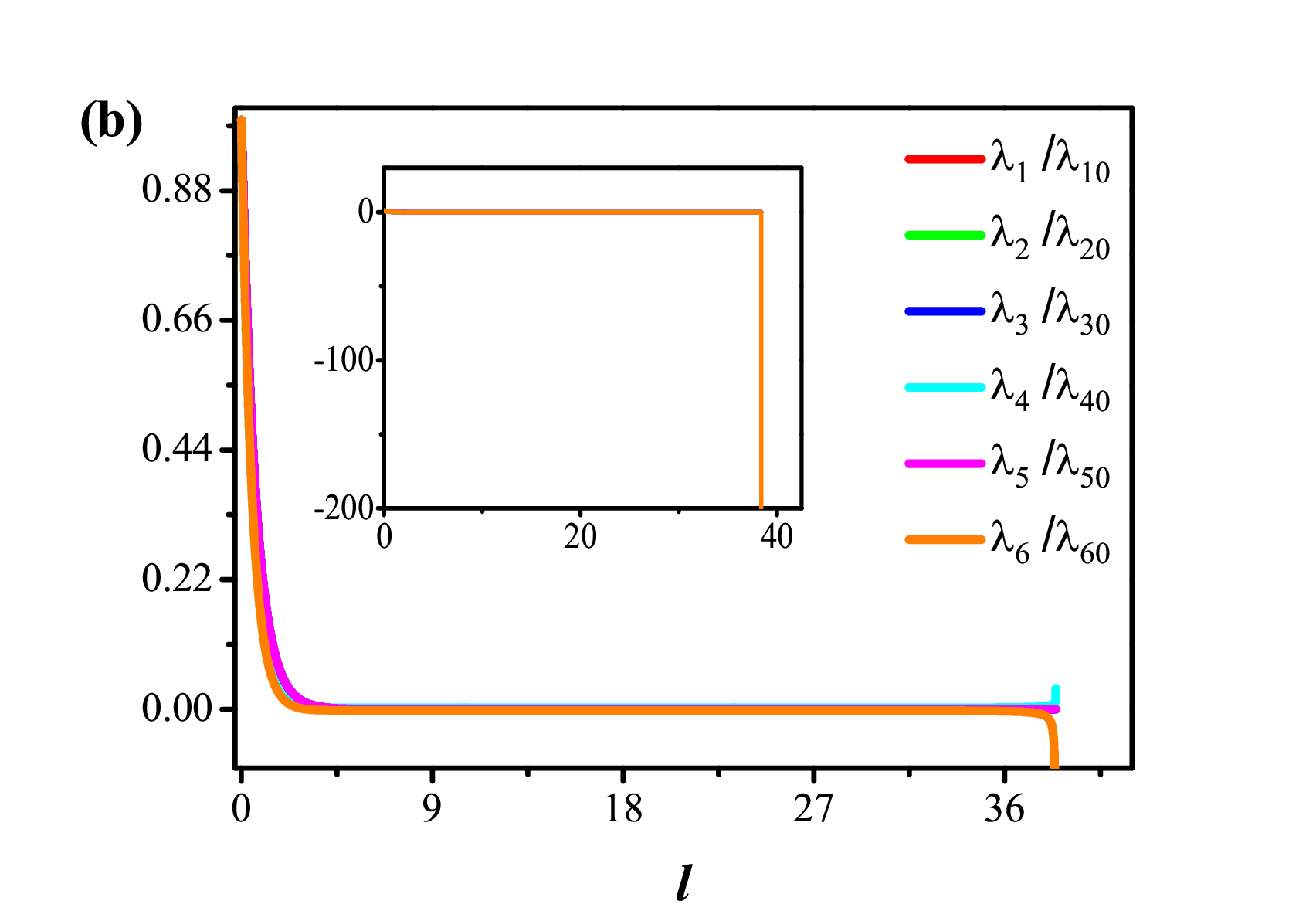}\vspace{-0.3cm}
\includegraphics[width=3in]{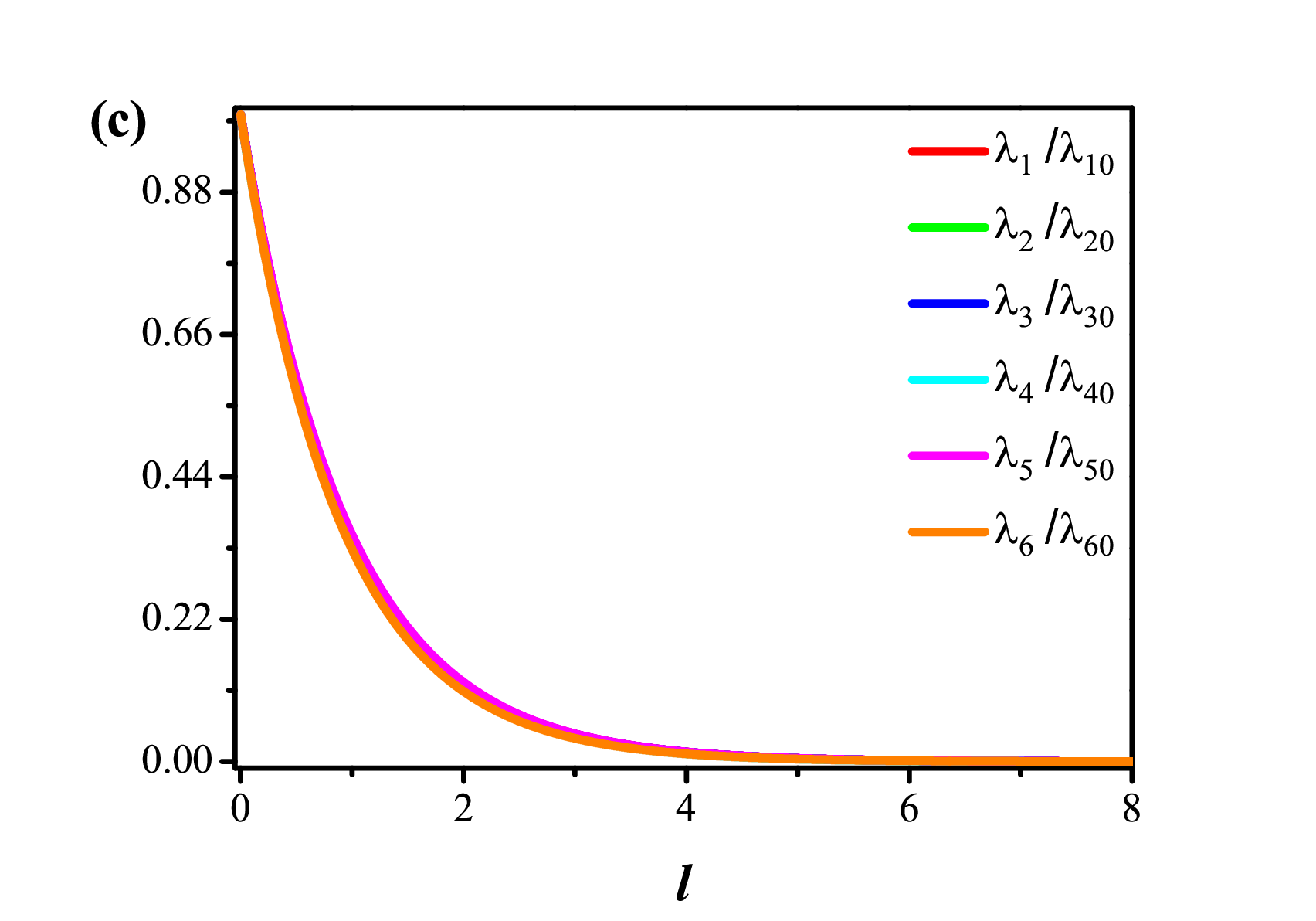}\\
\vspace{-0.1cm}
\caption{(Color online) Energy-dependent evolutions of fermion-fermion
couplings for anisotropy $v_{z0}/v_{p0}=0.1$ and fermion interaction $\lambda_{i0}=10^{-4}$
with different initial strengths of disorder $\Delta_{30}$:
(a) $\Delta_{30}=10^{-5}$, (b) $\Delta_{30}=10^{-6}$, and (c) $\Delta_{30}=10^{-7}$.}\label{dis3-f-f-interaction}
\end{figure}

\subsubsection{Single type of disorder}\label{Subsubsec_ff-dis-single-type}

At first, we take into account the presence of only one type of disorder and defer the general multiple
types of disorders to the subsequent Sec.~\ref{Subsubsec_ff-dis}.

To this end, let us assume only one of disorder equations exists and participates in the coupled
equations~(\ref{Eq_RG_v_z})-(\ref{Eq_RG_Delta}).  After carrying out the numerical calculations
and comparing them with the clean-limit behaviors of fermion-fermion interactions shown in
Sec.~\ref{sub-Sec_clean-A}, we realize that the low-energy tendencies of fermion-fermion interactions
are considerably dependent upon concrete kind of disorder.
Specifically, the tendencies of fermion-fermion interactions cannot be altered by the sole presence of
disorder $\Delta_4$ as depicted in Fig.~\ref{dis4-f-f-interaction} compared to their clean-limit counterparts.
But rather, Figs.~\ref{dis1-f-f-interaction}-\ref{dis3-f-f-interaction} indicate that
the presence of any one of $\Delta_1$, $\Delta_2$, or $\Delta_3$ can be capable of
dramatically changing the fates of fermion-fermion interactions. Colloquially, the
fermion-fermion interactions are driven to be divergent at the critical energy $l_c$ by the
contributions from disorder scattering, which may cause certain instability and make the
system unstable.

In principle, these results are in well consistent with the behaviors of disorders in
Sec.~\ref{Sub-Sec_dis-A} and fermion velocities in Sec.~\ref{Subsec_behaviors_fermion-velocities}, which
in other words, inherit from both the fates of disorders and fermion velocities. As to the disorder $\Delta_1$,
its unique low-energy behaviors presented in Sec.~\ref{Sub-Sec_dis-A}
make a profound effect to the fermion-fermion interactions. Learning from Fig.~\ref{dis1-f-f-interaction}(a),
the sole presence of $\Delta_1$ completely hinders the decrease in fermionic couplings at clean limit
case and makes them climb up quickly, which goes toward divergence at certain critical energy scale denoted
by $l_c$. Additionally, one can notice from Fig.~\ref{dis1-f-f-interaction}(b)
and Fig.~\ref{dis1-f-f-interaction}(c) that this basic result is insusceptible
to the starting conditions of all interaction parameters which are
prone to modifying the critical scale $l_c$ with a bigger $\Delta_{10}$ and a moderate $v_{z0}/v_{p0}$
corresponding to a smaller $l_c$. In comparison, although the sole presence of $\Delta_4$ shares
the similar behaviors with $\Delta_1$ as shown in Sec.~\ref{Sub-Sec_dis-A}, Fig.~\ref{dis4-f-f-interaction} suggests that
it contributes negligibly to the fermion-fermion interactions.
In particular, such result is insensitive to the initial values of other parameters as shown
in Fig.~\ref{dis4-f-f-interaction}(b) and Fig.~\ref{dis4-f-f-interaction}(c).
Several analytical comments are necessary to clarify this point.
Taking the $\lambda_6$ for instance (other fermion-fermion interactions are analogous),
its RG equation~(\ref{Eq_RG_lambda_6}) with the equal initial values at clean limit reads
\begin{eqnarray}
\frac{d\lambda_6}{dl}
&=&2\lambda_6\Bigl[
-\frac{1}{2}+\mathcal{C}_1
(-\lambda_5-3\lambda_6)\nonumber\\
&&-(\mathcal{C}_3\lambda_2+\mathcal{C}_4
\lambda_1)\Bigr]+2\mathcal{C}_1
\lambda_2\lambda_5. \label{Eq_lambda_6-ff}
\end{eqnarray}
Based on the definitions of coefficients $\mathcal{C}_i$ in Eqs.~(\ref{Eq_coeff-C1})-(\ref{Eq_coeff-C7}),
one can figure out that the first (term1) and second (term2) terms of right hand side
(RHS) of Eq.~(\ref{Eq_lambda_6-ff}) satisfy  $\mathrm{term1} < 0$ and $\mathrm{term2} > 0$ with
$|\mathrm{term2}|<|\mathrm{term1}|$, which implies that $\frac{d\lambda_6}{dl}<0$ and hence
$\lambda_6$ decreases and tends toward zero at the lowest-energy limit. Switching on $\Delta_4$,
its RG equation is henceforth reformulated as
\begin{eqnarray}
\frac{d\lambda_6}{dl}
&=&2\lambda_6\Bigl[
-\frac{1}{2}+\mathcal{C}_1
(-\lambda_5-3\lambda_6)-
(\mathcal{C}_3\lambda_2+\mathcal{C}_4
\lambda_1)\nonumber\\
&&-
\mathcal{C}_2(\Delta_{41}+\Delta_{42})
\Bigr]+2\mathcal{C}_1
\lambda_2\lambda_5.
\end{eqnarray}
This suggests that the disorder-contributed term is negative, and thus
the disorder term is helpful to decrease the $\lambda_6$ as also displayed
in Fig.~\ref{dis4-f-f-interaction}(b) that the larger initial value of $\Delta_4$
corresponds to the faster decrease of the fermionic interaction.

\begin{figure}
\includegraphics[width=3.35in]{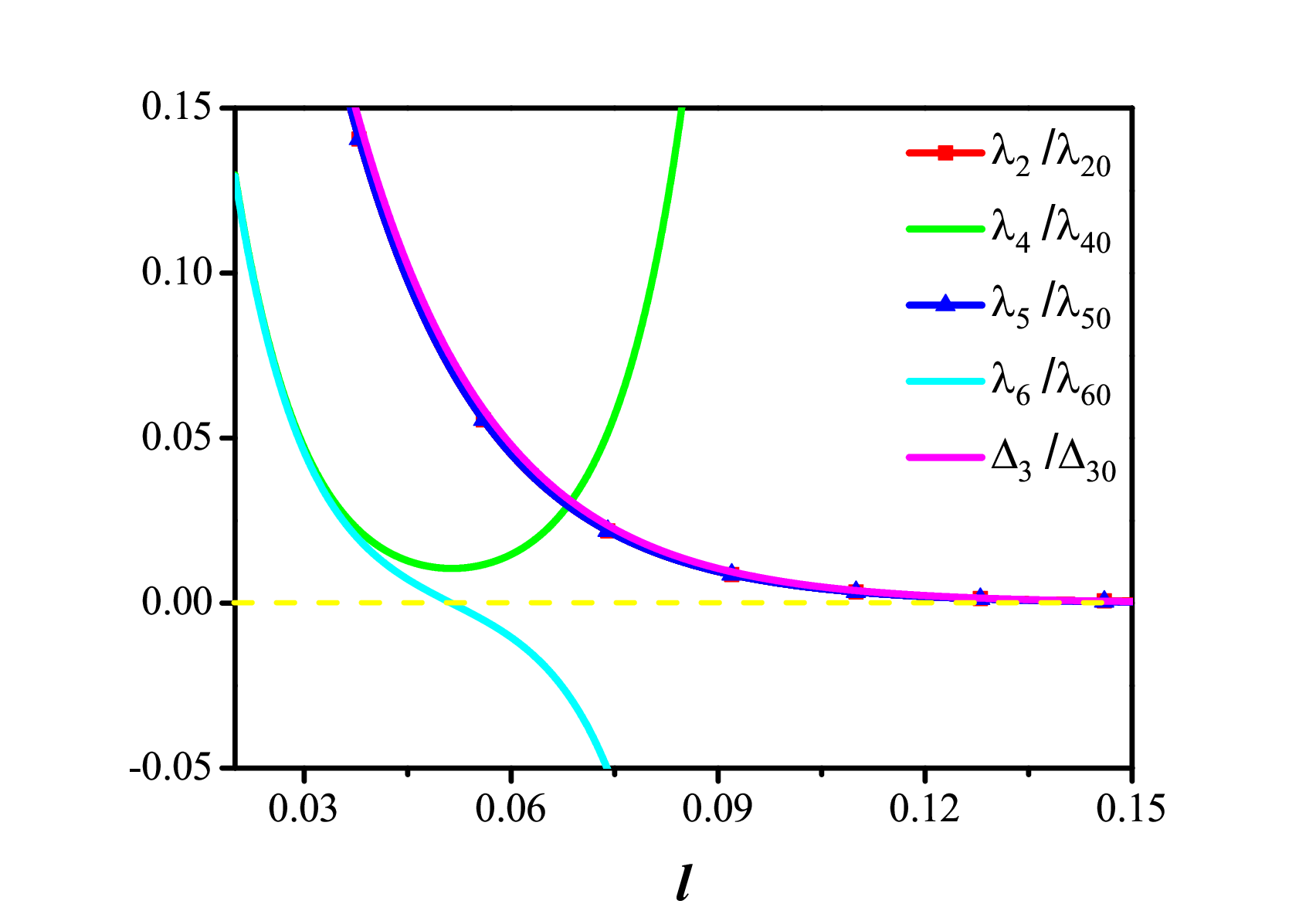}
\vspace{-0.39cm}
\caption{(Color online) Part region for energy-dependent flows for $\lambda_6$
in the sole presence of $\Delta_3$ at $v_{z0}/v_{p0}=0.1$, $\lambda_0=10^{-4}$,
and $\Delta_{30}=10^{-4}$.}\label{compare-Delta-3}
\end{figure}

Then, we move to the disorder $\Delta_2$ or $\Delta_3$.
With respect to these two kinds of disorders that are unimportant in the low-energy
regime for themselves as shown in Sec.~\ref{Sub-Sec_dis-A}, it is of particular importance to point out that
they behave like certain catalyst to arouse and ignite the fermion-fermion interactions
to exhibit the distinct fates compared to clean limit's
as displayed in Figs.~\ref{dis2-f-f-interaction}-\ref{dis3-f-f-interaction}.
For the $\Delta_3$, Fig.~\ref{dis3-f-f-interaction} indicates that
the fate of fermionic couplings can be altered as long as the $\Delta_{30}$
is suitable. In other words, the initial value of $\Delta_3$ takes a leading
responsibility and other parameters play subordinate roles.
Compared to the disorder $\Delta_3$, the sole presence of
$\Delta_2$ cannot trigger the divergence of fermionic couplings itself.
However, it is able to make them divergent under the suitable initial fermion-fermion strengths
in tandem with the proper anisotropy of fermion velocities ($v_{z0}/v_{p0}\approx 0.1$)
that enter into the coefficients of $\mathcal{C}_i$.
In analogy with the analysis for the $\Delta_4$, these can be roughly understood as follows.
For the sake of simplicity, we only consider the $\lambda_6$ in the presence of $\Delta_3$
whose RG equation is reduced to
\begin{eqnarray}
\frac{d\lambda_6}{dl}
&=&
2\lambda_6\left[
-\frac{1}{2}+\mathcal{C}_1
(-\lambda_5-3\lambda_6)-
(\mathcal{C}_3\lambda_2+\mathcal{C}_4
\lambda_1)\right]\nonumber\\
&&+2\mathcal{C}_1
\lambda_2\lambda_5-4\mathcal{C}_5\lambda_4\Delta_3.\label{Eq_lambda_6-ff_2}
\end{eqnarray}
According to the evolution information in Fig.~\ref{compare-Delta-3}, the RHS of $\lambda_6$'s equation including
the $\Delta_3$'s contribution keeps negative at $l>l^*$ where $l^*$ is defined by $\lambda_6(l^*)=0$ once the second-line RHS
of Eq.~(\ref{Eq_lambda_6-ff_2}) $(2\mathcal{C}_1
\lambda_2\lambda_5-4\mathcal{C}_5\lambda_4\Delta_3)$ is less than zero. As a result, $\lambda_6$ continues to decrease and
finally goes toward divergence at the critical energy scale.
Paralleling above analysis for $\Delta_2$ gives rise to the similar results.

\subsubsection{Multiple types of disorders}\label{Subsubsec_ff-dis}

\begin{figure}
\includegraphics[width=3in]{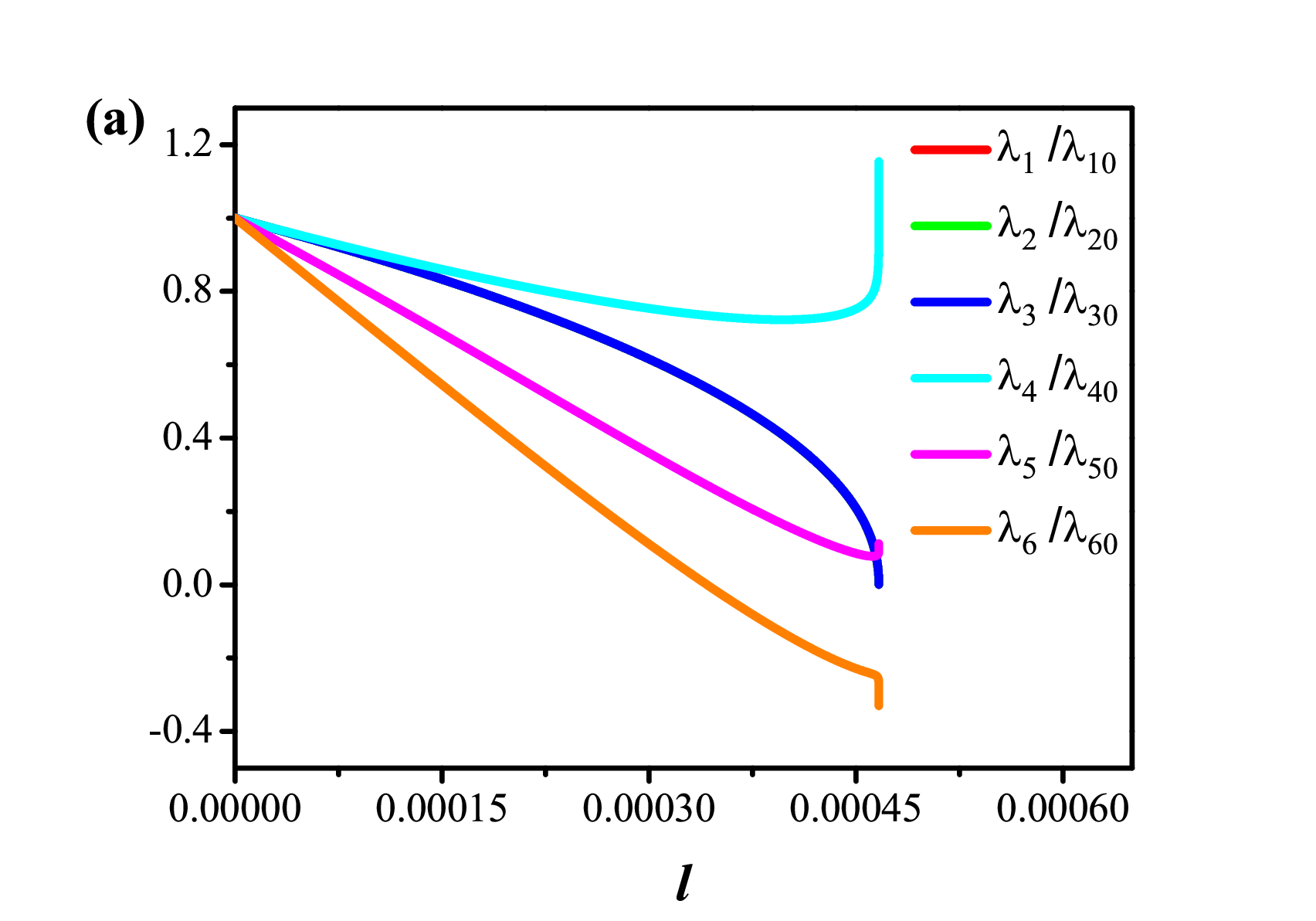}\vspace{-0.3cm}
\includegraphics[width=3in]{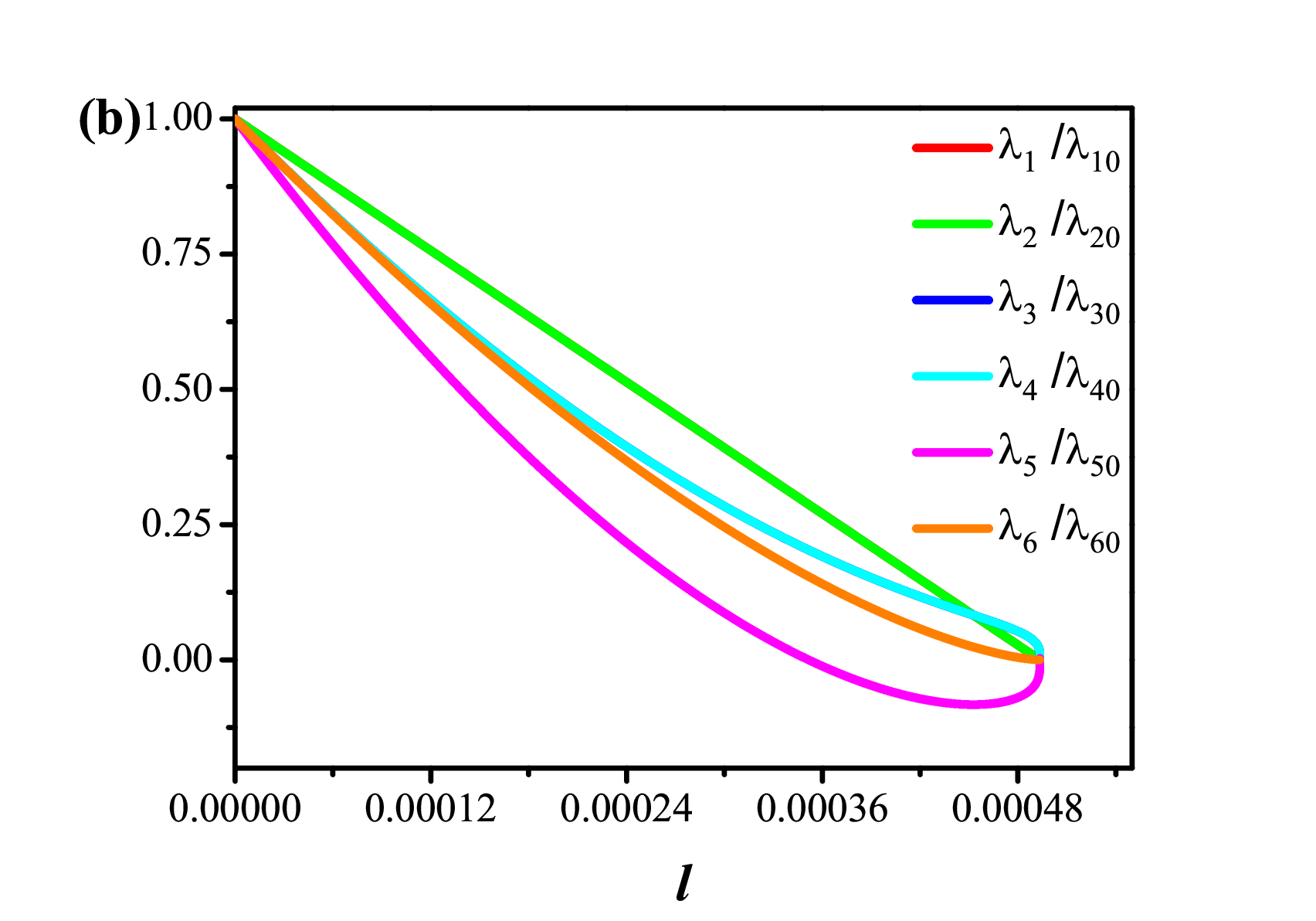}\\
\vspace{-0.1cm}
\caption{(Color online) Energy-dependent evolutions of fermion-fermion
couplings $\lambda_i/\lambda_{i0}$ at $v_{z0}/v_{p0}=0.1$ and $\lambda_{i0}=10^{-3}$ with $i=1-6$ for
(a)$\Delta_{1,3}=10^{-3}$ and (b)$\Delta_{2,4}=10^{-3}$.}\label{many-dis}
\end{figure}

Subsequently, let us briefly address the results for the simultaneous presence of multiple types of disorders.

Following the similar strategy in Sec.~\ref{Subsubsec_ff-dis-single-type} to carry
out the analogous numerical analysis of RG equations~(\ref{Eq_RG_v_z})-(\ref{Eq_RG_Delta}), we find that
different types of disorders would intimately entangle and compete with each other. Such ferocious competition
among distinct sorts of disorders prefers to render only two kinds of scenarios for the low-energy
behaviors of fermion-fermion couplings, which are in sharp contrast with their counterparts in
the sole presence of disorder.

On the one hand, the fermion-fermion couplings are prone to flowing toward certain finite values as delineated in
Fig.~\ref{many-dis}(a) where $\Delta_{1}$ and $\Delta_{3}$ are present initially.
It is worth stressing that the basic conclusion of Fig.~\ref{many-dis}(a) is always
stable as long as the disorder $\Delta_1$ exists no matter it presents at the starting point or is generated by
other types of disorders as investigated in Sec.~\ref{Subsec_multi_disorder} (the corresponding
results are analogous and hence not shown hereby for simplicity).
As a consequence, this also supports that the disorder $\Delta_1$ plays a more important role in the
competition among other types of disorders, which is in well agreement with our analysis in
Sec.~\ref{Sub-Sec_dis-A}. On the other hand, the fermion-fermion strengths $\lambda_i$ with $i=1-6$ gradually
decrease and eventually vanish at the low-energy limit once
the disorder $\Delta_1$ is absent in the competition as displayed in Fig.~\ref{many-dis}(b)
for taking the existence of $\Delta_2$ and $\Delta_{4}$ as an example.

To be brief, distinct types of disorders together with their unique furious competition
yield different fates of fermion-fermion couplings in the low-energy. Concretely,
the disorder $\Delta_1$ takes in charge the dominant contribution and disorders
$\Delta_{2,3}$'s effects are subordinate to $\Delta_1$'s but rather
the disorder $\Delta_4$ provides a negligible impact.

\subsection{Behaviors of fermion velocities}\label{Subsec_behaviors_fermion-velocities}

Furthermore, armed with the energy-dependent properties of disorders and fermion-fermion couplings,
we subsequently move to study how the fermion velocities behave with lowering the energy
scale in the presence of disorders.

\begin{figure}
\centering
\includegraphics[width=3in]{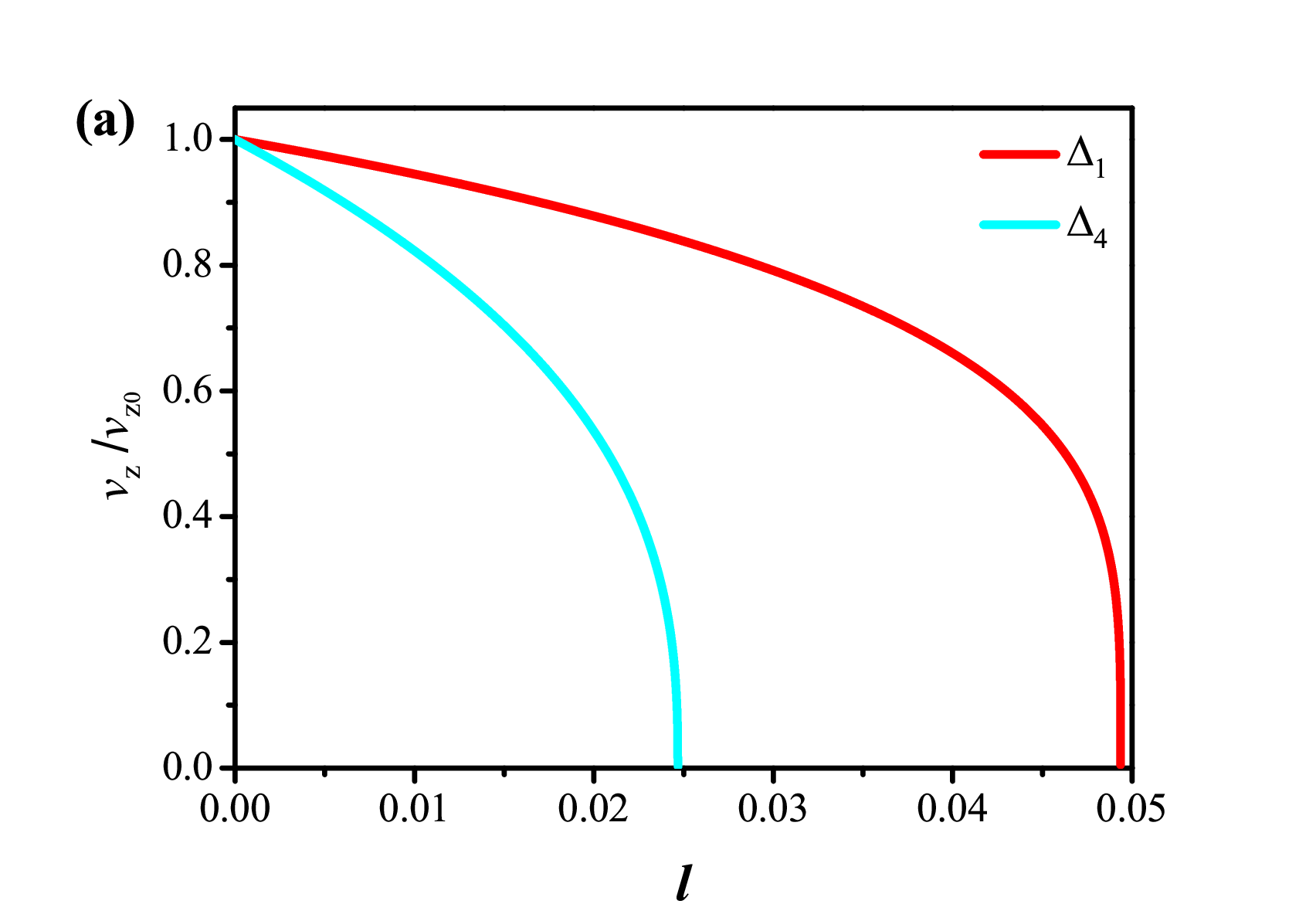}\\ \vspace{-0.3cm}
\includegraphics[width=3in]{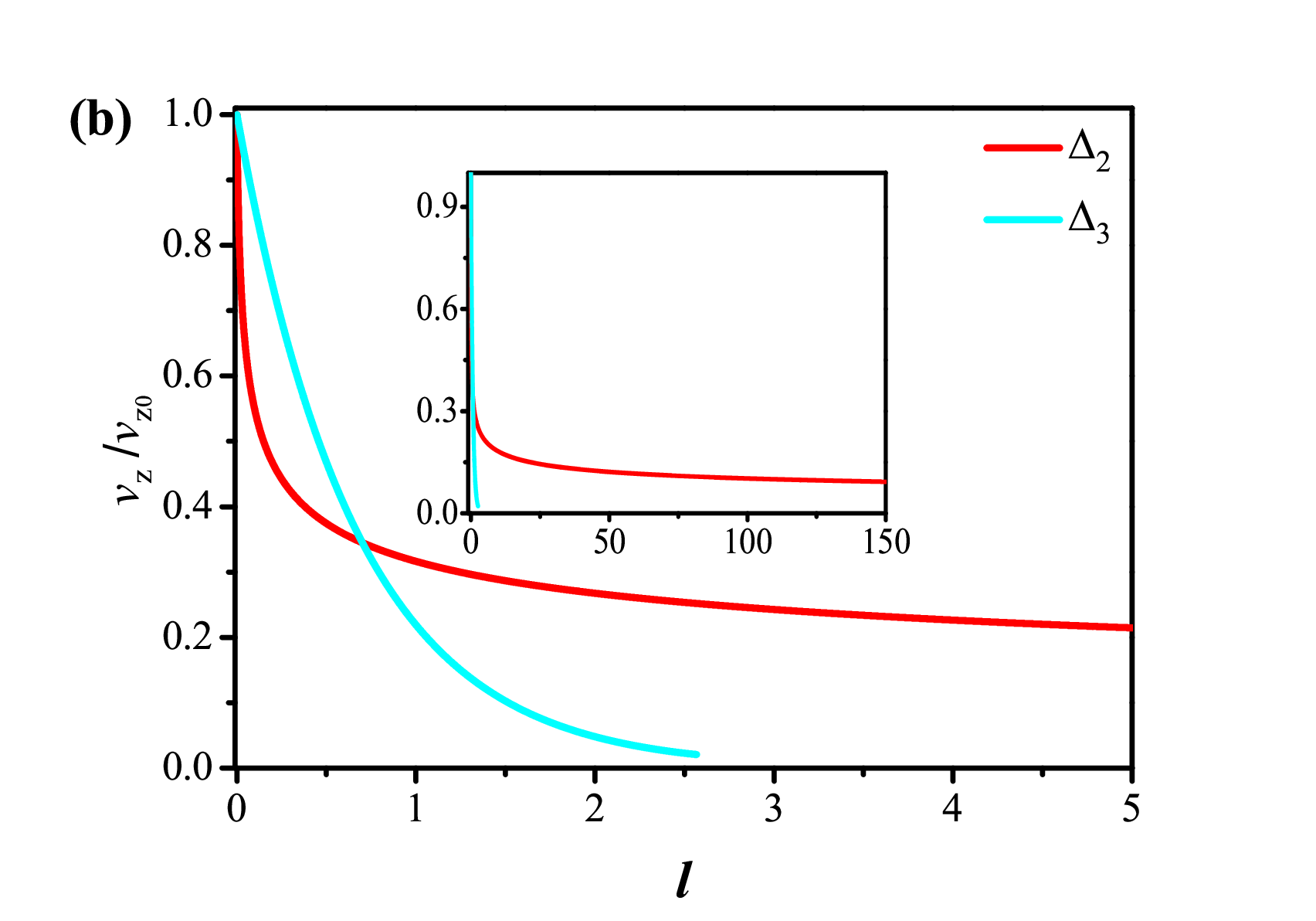}\\
\vspace{-0.1cm}
\caption{(Color online) Energy-dependent evolutions of $v_{z}/v_{z0}$ in the presence of
disorder (a) $\Delta_1$ or $\Delta_4$ with $v_{z0}/v_{p0}=0.5$
and (b) $\Delta_2$ or $\Delta_3$ with $v_{z0}/v_{p0}=0.1$
at $\lambda_0=10^{-4},\Delta_0=10^{-4}$
(the qualitative results are insensitive to the
initial values of interaction parameters).}\label{12}
\end{figure}

\begin{figure}
\centering
\includegraphics[width=3in]{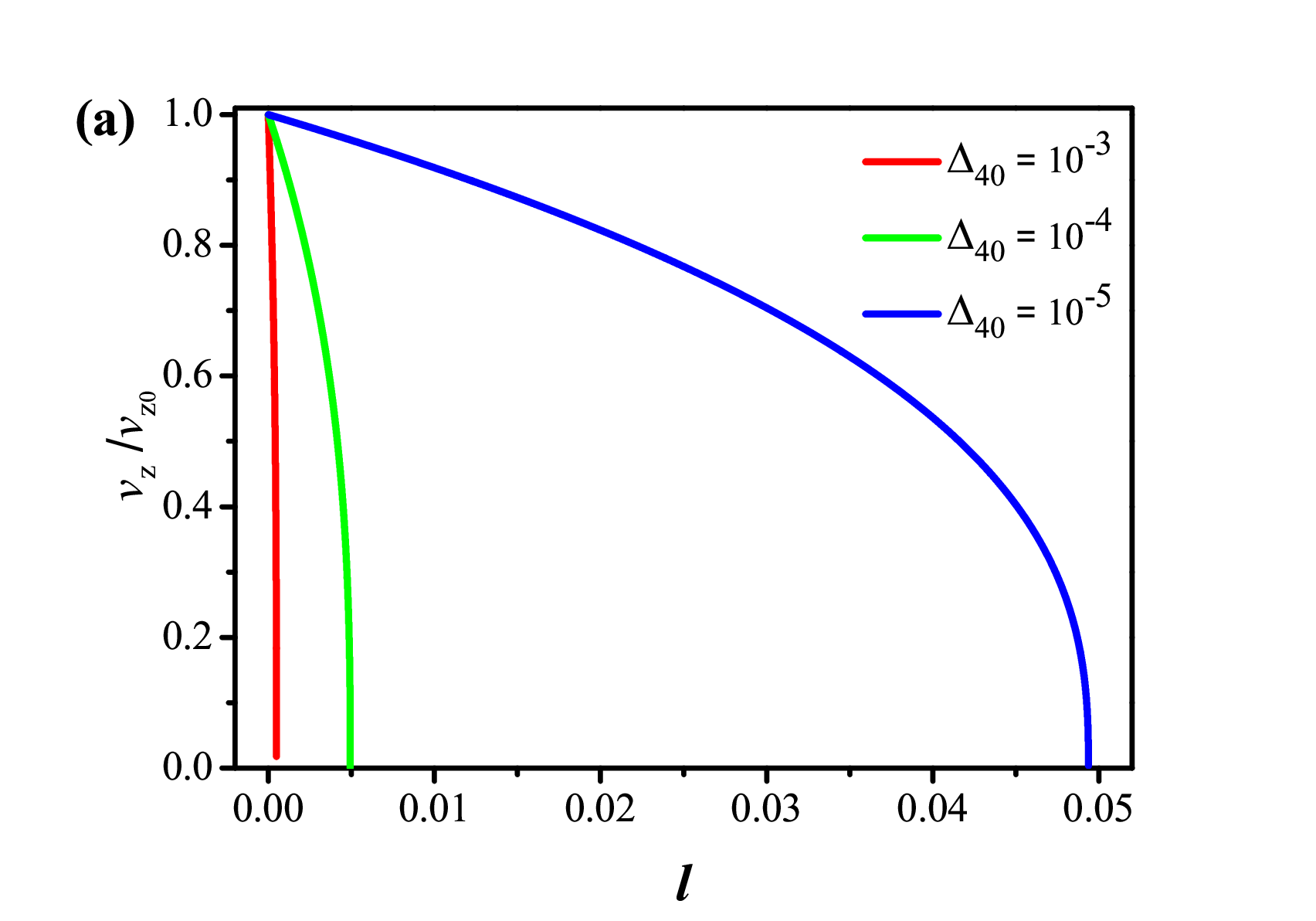}\\ \vspace{-0.3cm}
\includegraphics[width=3in]{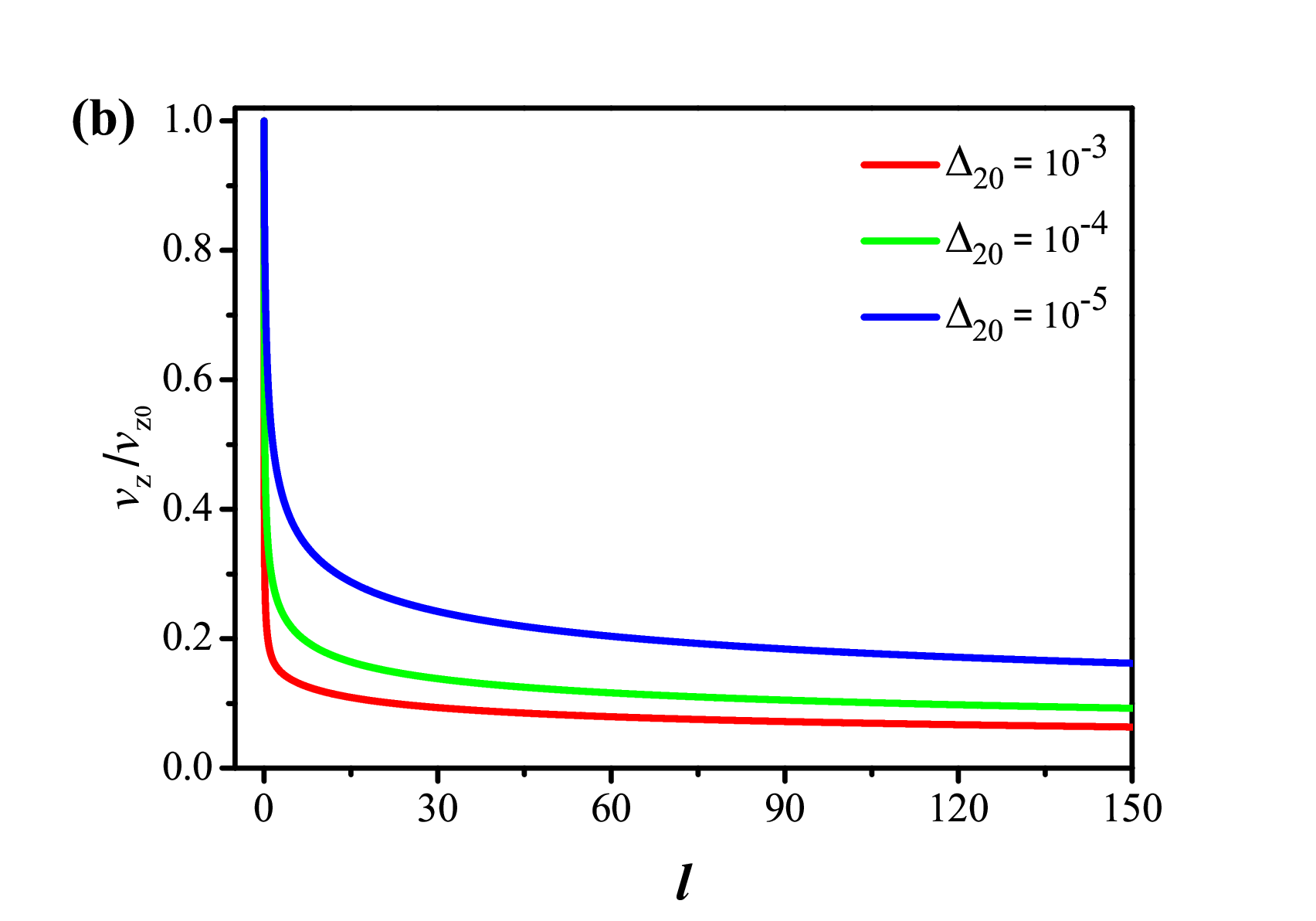}\\
\vspace{-0.1cm}
\caption{(Color online) Energy-dependent evolutions of $v_z/v_{z0}$
with several representative values of initial disorder strengths at
$v_{z0}/v_{p0}=0.1$ and $\lambda_0=10^{-4}$ for the presence of disorder
(a) $\Delta_4$ and (b) $\Delta_2$, respectively.}\label{13}
\end{figure}

\begin{figure}
\centering
\includegraphics[width=3in]{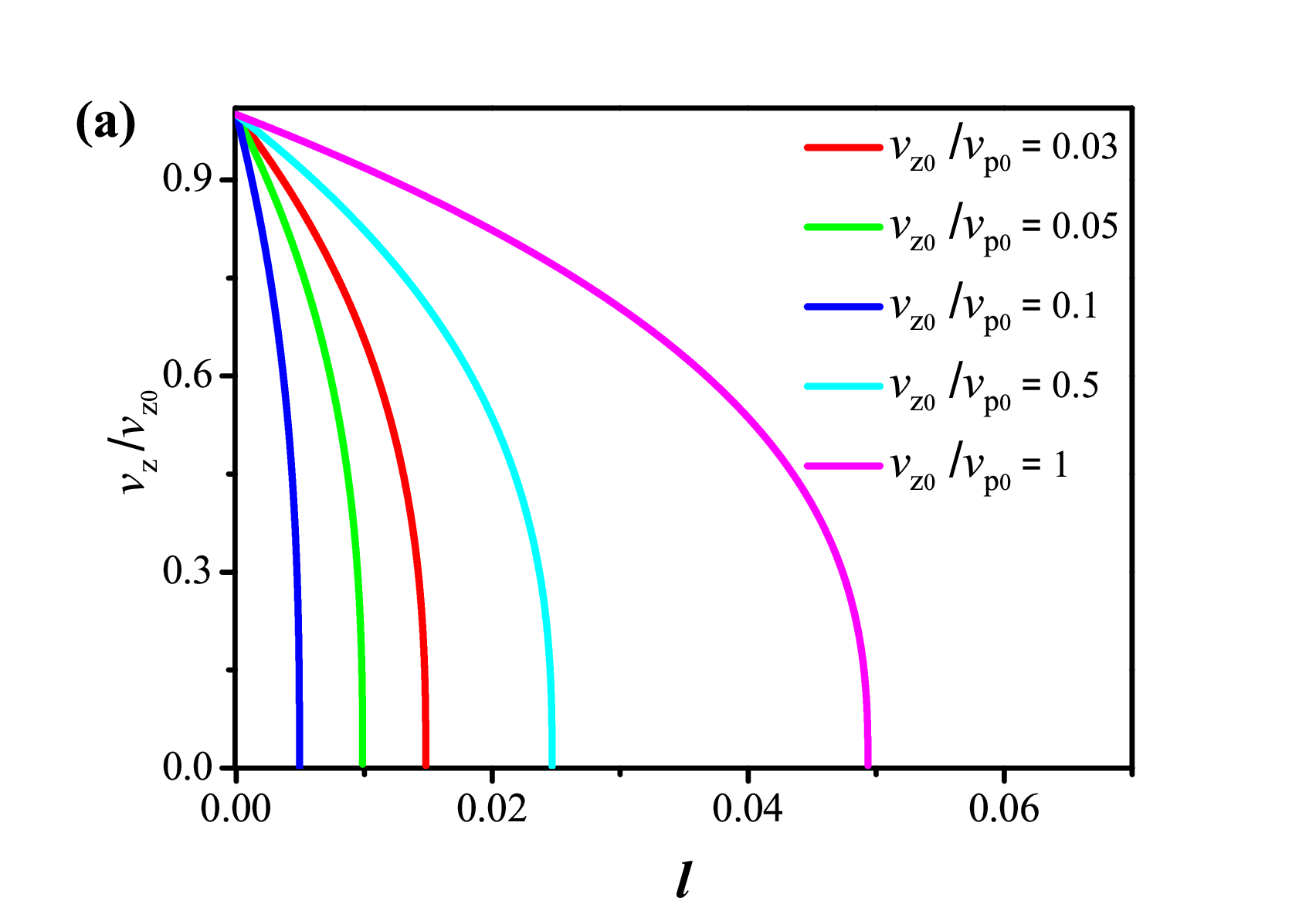}\\ \vspace{-0.3cm}
\includegraphics[width=3in]{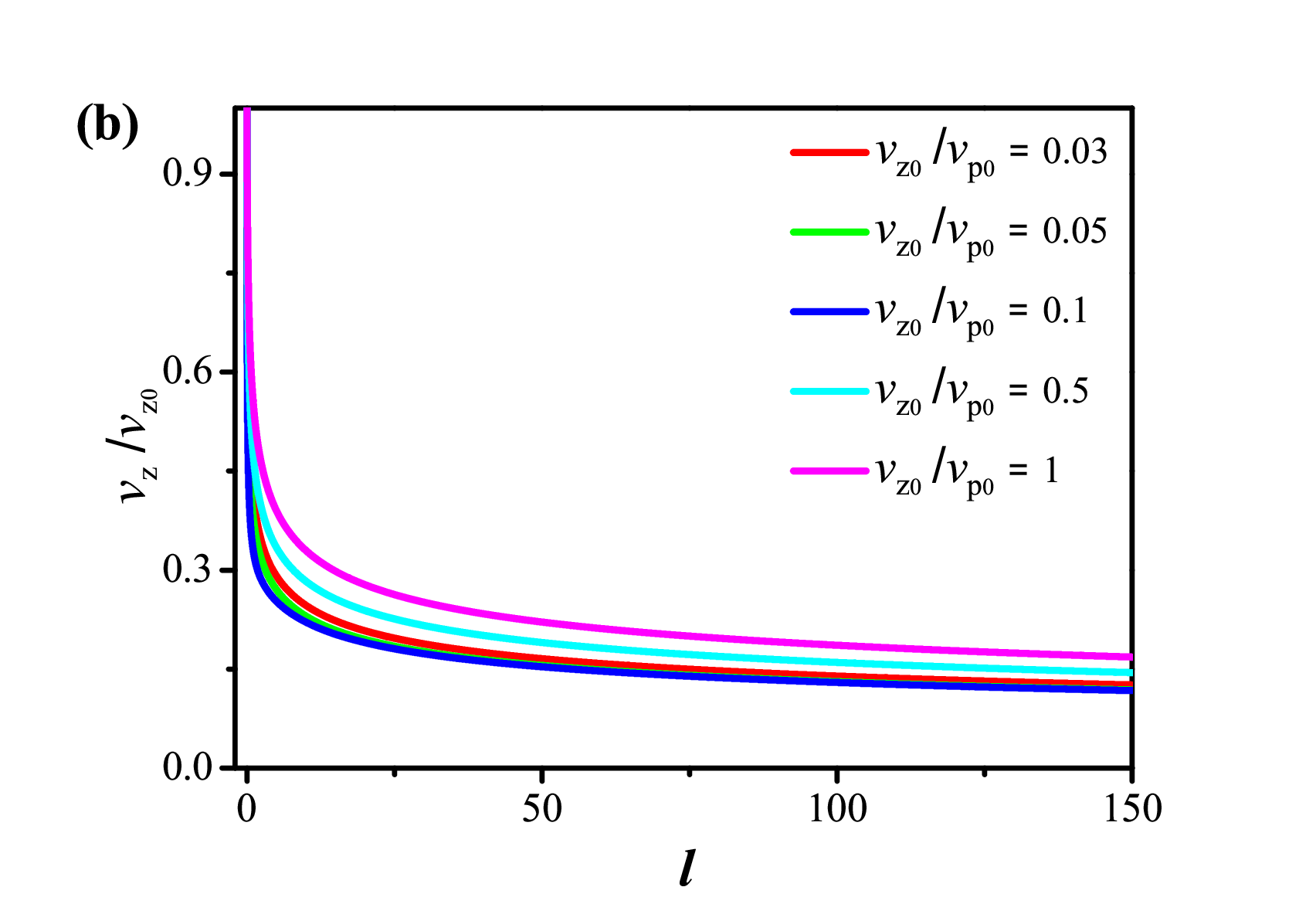}\\
\vspace{-0.1cm}
\caption{(Color online) Energy-dependent evolutions of $v_z/v_{z0}$ under several
representative groups of initial anisotropy $v_{z0}/v_{p0}$ for the presence of
disorder (a) $\Delta_4$ with $\Delta_{40}=10^{-4},\lambda_0=10^{-4}$, and
(b) $\Delta_2$ with $\Delta_{20}=10^{-4}$ and $\lambda_0=10^{-3}$, respectively.}\label{14}
\end{figure}

\begin{figure}
\centering
\includegraphics[width=3in]{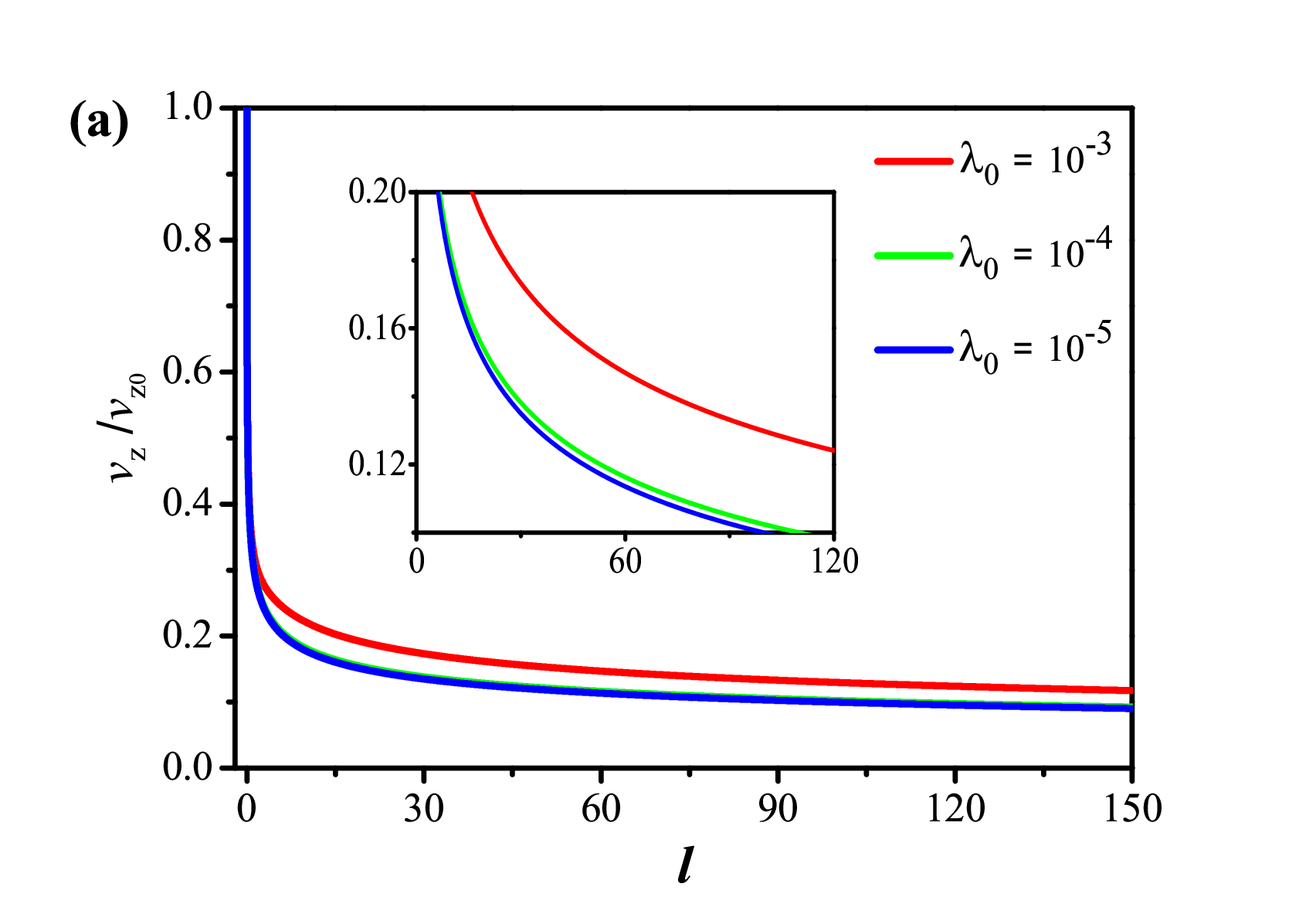}\\ \vspace{-0.3cm}
\includegraphics[width=3in]{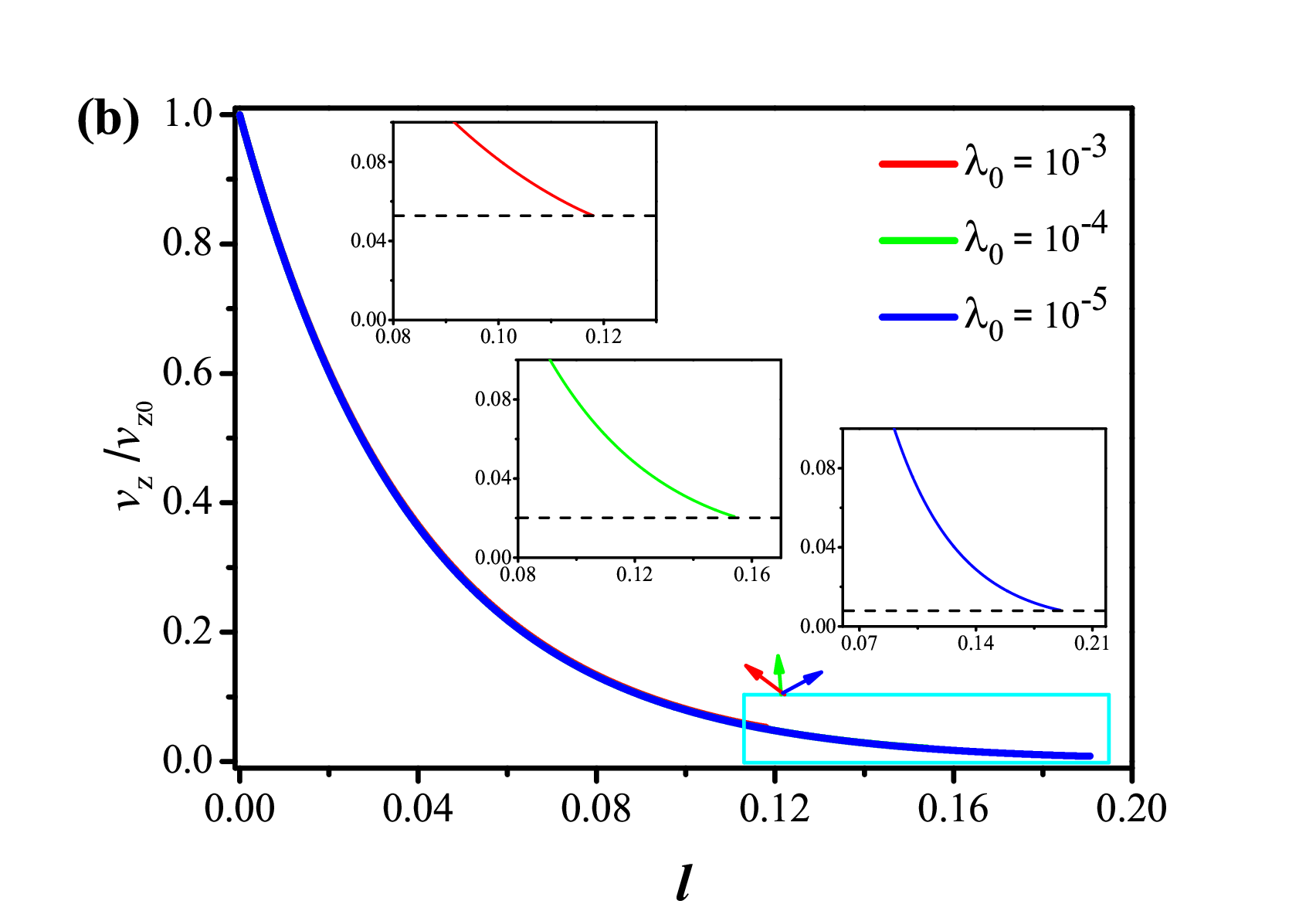}\\
\vspace{-0.1cm}
\caption{(Color online) Energy-dependent evolutions of $v_z/v_{z0}$ under several
representative groups of initial fermion-fermion couplings at $v_{z0}/v_{p0}=0.1$
for the presence of disorder (a) $\Delta_2$ with $\Delta_{20}=10^{-4}$ and
(b) $\Delta_3$ with $\Delta_{30}=10^{-4}$, respectively.}\label{15}
\end{figure}

\begin{figure}
\centering
\includegraphics[width=3in]{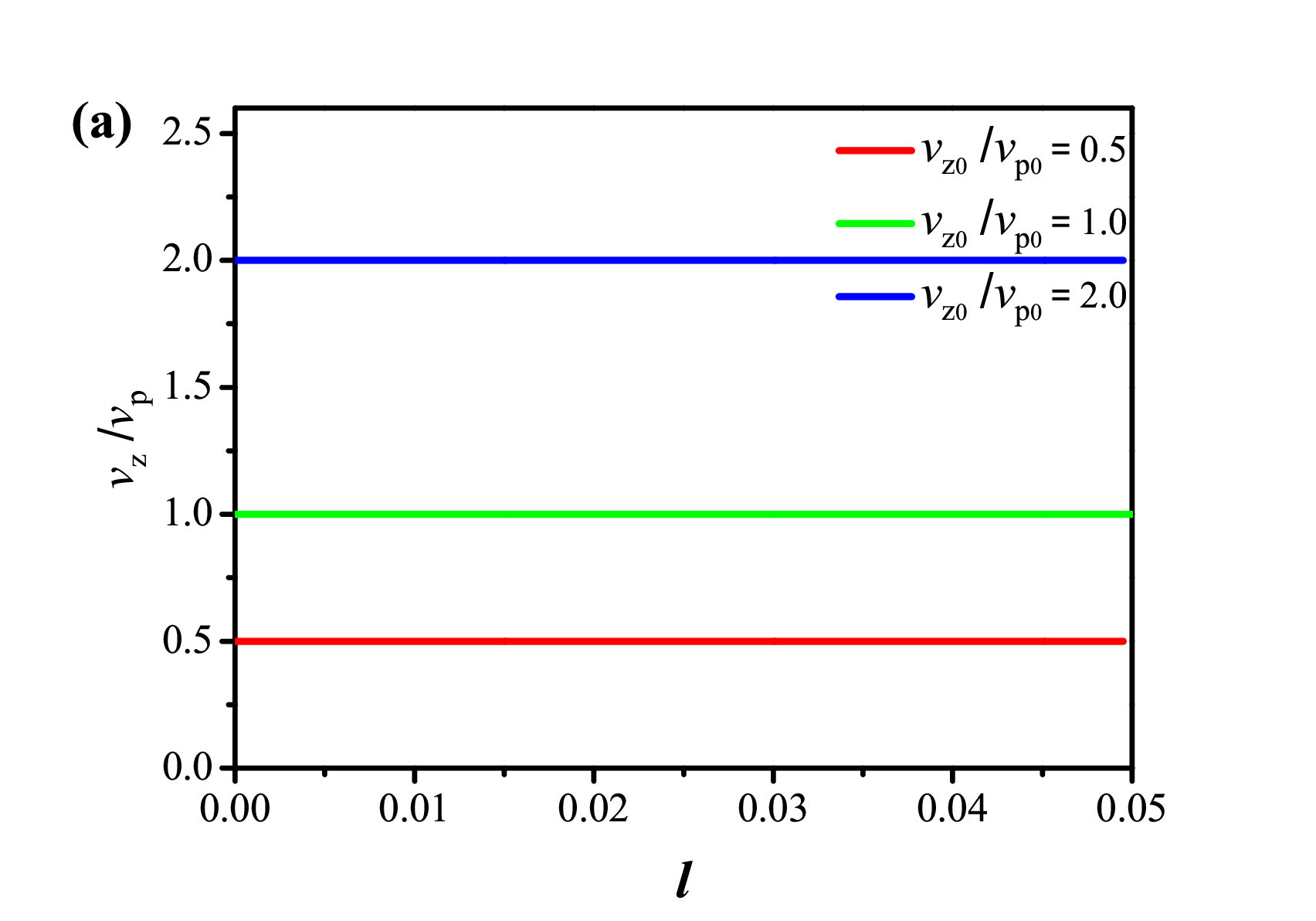}\\ \vspace{-0.3cm}
\includegraphics[width=3in]{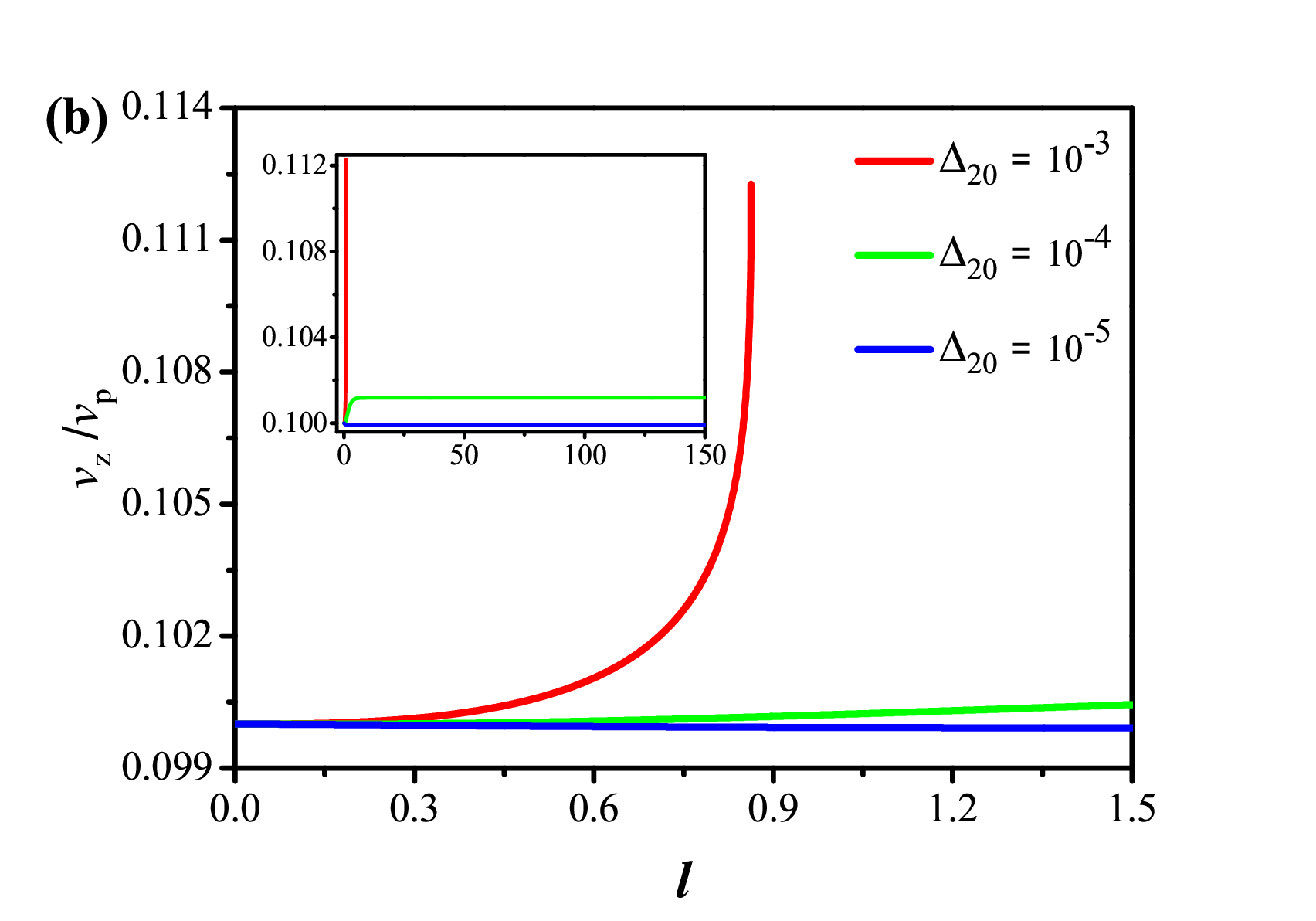}\\
\vspace{-0.1cm}
\caption{(Color online) Energy-dependent evolutions of anisotropy $v_{z}/v_{p}$
under the presence of disorder (a) $\Delta_1$ with several representative values of $v_{z0}/v_{p0}$ at
$\lambda_0=10^{-3}$ and $\Delta_{10}=10^{-4}$ (the basic results are insensitive to
the concrete values of $\Delta_{10}$), and (b) $\Delta_2$ with
several representative values of initial strengths of $\Delta_{20}$ at $v_{z0}/v_{p0}=0.1$
and $\lambda_0=10^{-3}$, respectively.}\label{17}
\end{figure}

As addressed in Sec.~\ref{Subsec-velocities-clean}, the fermion-fermion interactions
are helpful to increase the fermion velocities. In comparison, the disorders are
prone to reducing them after learning  from Eq.~(\ref{Eq_RG_v_z}) and Eq.~(\ref{Eq_RG_v_perp}).
Principally, the energy-dependent coupled RG equations~(\ref{Eq_RG_v_z})-(\ref{Eq_RG_Delta})
codify the intimate competition between disorder scatterings and
fermionic interactions. In order to investigate the finial fates of fermion velocities under
the competition between disorder scatterings and fermionic interactions, we perform the
numerical analysis and present our main results
in Figs.~\ref{12}-\ref{17}, which indicate that disorders win the
competition and hence dominate the tendencies of fermion velocities
in the low-energy regime.

To be concrete, we find that $v_z$ and $v_p$ gradually decrease upon lowering the energy scale
and their final fates heavily hinge upon the disorders. For convenience, we at first inspect
the energy-dependent evolution of $v_z$ in details and then briefly examine the ratio $v_z/v_p$ from
which the behavior of $v_p$ can be extracted. Inheriting from Sec.~\ref{Sub-Sec_dis-A}, different
types of disorders own distinct fates and, as a result, correspond to distinct sorts of evolutions
depicted in Figs.~\ref{12}-\ref{15}.

Considering several representative groups of initial conditions,
one can read from Fig.~\ref{12}(a) that $v_z$ clearly decreases and goes
toward zero in the sole presence of $\Delta_1$ or $\Delta_4$ due to the
divergence of disorder strength. It is interesting to notice
that $\Delta_4$ is more harmful to $v_z$ than $\Delta_1$, which
implies that $v_z$ vanishes at some higher energy scale if there only exists $\Delta_4$.
In marked contrast, Fig.~\ref{12}(b) suggests that $v_z$ quickly decreases and is eventually
driven to certain finite value when only $\Delta_2$ or $\Delta_3$ is present. Compared to $\Delta_3$,
$v_z$ with sole presence of $\Delta_2$ features a bigger critical value which is linked to
a much lower critical energy scale. In order to further elucidate this point, we deliver some
coarse comments on the $v_z$ at the lowest-energy limit. Concerning the presence of disorder
$\Delta_2$ for an instance, the energy-dependent equation
of $v_z$~(\ref{Eq_RG_v_z}) reduces to $dv_z/dl=-\mathcal{C}_2\Delta_2 v_z$.
As presented in Sec.~\ref{Sub-Sec_dis-A}, $\Delta_2$ finally vanishes at the lowest-energy limit and, accordingly,
$v_z$ is saturated at certain nonzero value once one assumes all other parameters to be energy-independent.

Besides these qualitative conclusions, Figs.~\ref{13}-\ref{15} illustrate the quantitative
contributions from the distinct starting conditions.
With variations of the initial values of disorders, Fig.~\ref{13} shows that the bigger initial strengths
lift up the critical energy scale a little and are in favor of the decrease in
fermion velocities. Fig.~\ref{14} exhibits that
there exists a moderate anisotropy with $v_{z0}/v_{p0}\approx0.1$ to facilitate the decrease in
fermion velocities but instead the strong anisotropy of fermion velocities is preferable to
hamper the drop of $v_z$ or $v_p$. Particularly, the initial values of fermion-fermion interactions
play an important role as well. We can learn from Fig.~\ref{15}(a) for the presence
of $\Delta_2$ that the bigger initial strengths of fermionic interactions inhibit
the decrease in fermion velocities. Fig.~\ref{15}(b) shows that the
saturated values of fermion velocities are
slightly modified in the sole presence of $\Delta_3$.
Before moving further, it is of necessity to briefly address that the basic results for the presence of
more than one sorts of disorders bear similarities to the conclusions for the sole type of disorder presented above
and henceforth are not shown for brevity. Colloquially, $v_z$ decreases and is eventually driven to zero
as long as the disorders dominate over the fermionic couplings and become divergent
in the low-energy regime. Otherwise, it goes toward some finite value.

\begin{figure}
\centering
\includegraphics[width=3.3in]{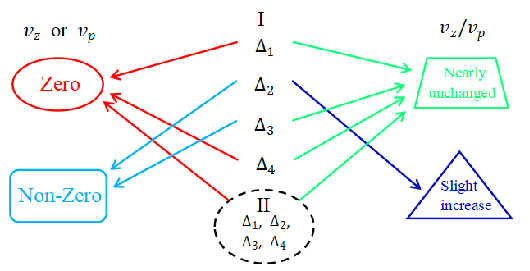}\\
\vspace{0.10cm}
\caption{(Color online) Schematic diagrams present the primary results of fermion velocities and their ratio.
Hereby, the accompanied arrows point to the final fates of related quantities at the lowest-energy
limit. In addition, the areas $\mathrm{I}$ and $\mathrm{II}$
correspond to the presence of only one type of disorder
and multiple types of disorders, respectively.}\label{Fig_velocity-summary}
\end{figure}

In addition to the behavior of $v_z$, let us investigate the evolution of the anisotropy of fermion
velocities $v_z/v_p$, from which the information of $v_p$ can be derived together with $v_z$'s tendencies.
Generally, its low-energy tendency clusters into two distinct scenarios.
On the one side, while the fermion-fermion interactions are subordinate to the disorder contribution and
become less and less important with lowering the energy scale, the ratio $v_z/v_p$ is preferable to hardly fluctuate and
nearly keep invariant, which are concomitant to the clean-limit case contributed by the irrelevant
fermion-fermion interactions in Sec.~\ref{Subsec-velocities-clean}.
Taking the sole presence of $\Delta_1$ for instance, we find from Fig.~\ref{17}(a)
that the ratio of fermion velocities is baldly insusceptible to the variation of energy scales regardless
of both the starting ratio and disorder strength. In other words,
$v_z$ shares the similar energy-dependent evolutions with $v_p$ in this situation.
On the other side, once the disorder strength is irrelevant
and rapidly vanishes via lowering the energy scale, the fermion-fermion interaction is expected
to play a more significant role. In such circumstance, the $v_z$ would obtain much more
supports than $v_p$, evincing that the anisotropy of fermion velocity deviates largely from isotropy.
As illustrated in Fig.\ref{17}(b), we notice that the ratio of fermion velocities roughly gets a 12
percent increase with suitable initial conditions. Again, we stress that the qualitative conclusions
are analogous for the presences of more than one sorts of disorders. To be brief, Fig.~\ref{Fig_velocity-summary}
catalogs our primary conclusions for the low-energy fates of fermion velocities under the competition
between fermionic interactions and disorder scatterings.

\section{Summary}\label{Sec_summary}

In summary, we utilize the powerful momentum-shell RG
approach~\cite{Wilson1975RMP,Polchinski9210046,Shankar1994RMP}
to unbiasedly investigate the low-energy physical consequences generated by
short-ranged fermion-fermion interactions
and disorder scatterings as well as their competitions
in the 3D line-nodal superconductors.
The coupled RG evolutions of both fermion-fermion and
fermion-impurity strengths that contain the low-energy physics
are derived via practicing the standard RG analysis on the
basis of one-loop corrections. Performing the numerical analysis
of these coupled RG equations indicates that disorder
strengths and fermion-fermion interactions as well as fermion velocities
exhibit a number of interesting properties in the low-energy regime.

For the sake of completeness, we commence with the clean-limit circumstance.
In such a situation, the fermion-fermion interactions are forced to decrease and finally vanish
with lowering the energy scale. This causes that fermion velocities $v_{z,p}$ increase
in the beginning and finally flow toward some saturated constants. Meanwhile, the
anisotropy of fermion velocities $v_z/v_p$ can be also revised under the
contributions from fermion-fermion interactions. With the initial conditions
$v_{z0}/v_{p0}<1$ and $v_{z0}/v_{p0}>1$, it would receive a slight decrease and
increase, respectively. Subsequently, we put our focus on the consequences of
competitions between fermion-fermion couplings and disorder scatterings.
At the outset, we assume that there only exists a single type of disorder and find
that the disorder strength monotonously decreases for $\Delta_{2,3}$ and goes toward
divergence for $\Delta_{1,4}$ with lowering the energy scale, respectively.
In particular, the relevant disorder may result in a disorder-dominated diffusive metallic state~\cite{Fradkin1986PRB,Foster2008PRB,Kobayashi2014PRL,Lai-1409.8675,Goswami2011PRL,Moon-1409.0573,Wang2015PLA}.
With respect to the presence of multiple types of disorders,
we figure out that only $\Delta_1$ holds its sole-presence property and remains relevant.
In sharp contrast, the fates of other three types of disorders are sensitive to the
competitions among distinct types of disorders. On the one hand, the $\Delta_4$ can be
driven to be irrelevant but instead $\Delta_2$ and $\Delta_3$
being changed to be divergent. On the other hand, certain dynamical-generated disorder
which is absent initially can be induced due to the disorder competitions once two
sorts of disorders $\Delta_1$, $\Delta_2$, and $\Delta_3$ are present at the starting point.
Then, the effects of disorders on fermion-fermion couplings are addressed.
The sole presence of $\Delta_{1}$ ($\Delta_{2}$ or $\Delta_{3}$) is of particular
importance to increase the fermion-fermion interactions
and even drive them toward divergence at certain critical energies although the disorder
$\Delta_4$ provides a negligible impact.
However, the intimate competition among distinct types of disorders can weaken and
neutralize the disorder contributions, which accordingly causes the fermionic couplings
to evolve toward zero or certain finite nonzero values. Further, we briefly deliver the behaviors of fermion velocities
under the interplay between fermion-fermion interactions and disorders.
For the sole presence of $\Delta_{1}$ (or $\Delta_{4}$) or multiple kinds of disorders,
the fermion velocities progressively decrease and vanish at the lowest-energy limit but instead
they are attracted by some certain saturated value for the single presence of $\Delta_{2}$ (or $\Delta_{3}$).
Additionally, the ratio $v_z/v_p$ would receive a little increase while the fermion-fermion interactions
dominate over the disorders. In comparison, it nearly keeps invariant once the latter wins the former in the
low-energy regime. For convenience, we employ Fig.~\ref{Fig_dis_summary} in conjunction with
Fig.~\ref{Fig_velocity-summary} to schematically present our central conclusions.

In principle, the interaction and disorder-induced signatures are closely associated with the fixed points of systems in
the parameter space and henceforth the potential instabilities as well as phase transitions.
As a corollary, the properties of physical quantities around the fixed points may be altered and modified,
and, accordingly, such unusual signatures may be indirectly probed by detecting the physical quantities including the
density of states, spectral function, specific heat, etc., which are expected to
inherit parts of information from the disorder-induced instabilities~\cite{Mahan1990Book}.
Lastly, we hope that our results would provide helpful clues for further
experiments to examine the low-energy behaviors of physical implications
that are associated with fermion velocities and phase transitions in the
3D nodal-line superconductors.

\section*{ACKNOWLEDGEMENTS}

W.H.B. thanks Dr. Wen Liu and M.S. Xiao-Yue Ren for the helpful discussions.
J.W. was partially supported by the National Natural
Science Foundation of China under Grant No. 11504360.

\appendix

\section{One-loop corrections}\label{Sec_appendix-one-loop-corrections}

\begin{figure}
\centering
\includegraphics[width=3.25in]{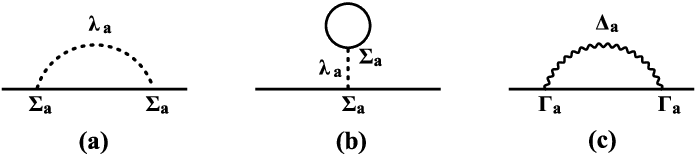}
\vspace{-0.01cm}
\caption{One-loop corrections to the fermion propagator
due to the fermion-fermion interactions (a) and (b)
as well as disorder scatterings (c) (the solid, dashed,
and wavy lines represent the fermion propagator, fermion-fermion interaction,
and disorder scattering, respectively). }\label{Fig_one-loop-self-energy}
\end{figure}

To be convenient, we collect all the one-loop corrections within this appendix.
Starting from our effective theory~(\ref{Eq_S_eff}), the fermionic propagator as well as
fermion-fermion couplings and fermion-disorder strengths would receive the
one-loop corrections due to their intimate interplays as diagrammatically
exhibited in Fig.~\ref{Fig_one-loop-self-energy} and Fig.~\ref{Fig_one-loop-ff-inter},
respectively. After performing tedious but
straightforward calculations~\cite{Vafek2012PRB,Vafek2014PRB,Wang2017QBCP,Wang2018-2019},
we obtain

\begin{widetext}

\begin{figure}
\centering
\includegraphics[width=6.15in]{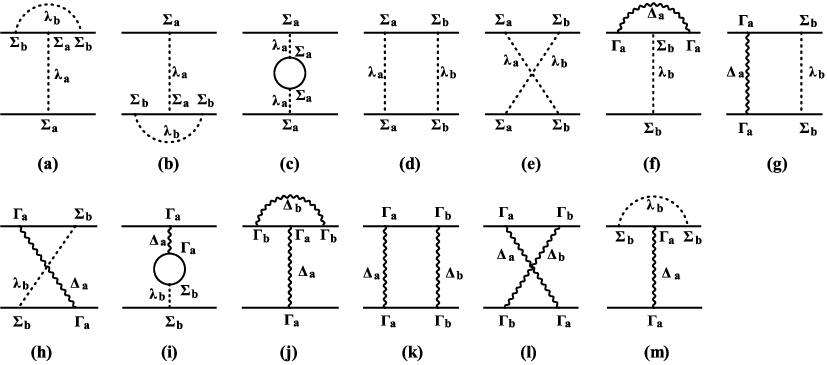}
\vspace{-0.01cm}
\caption{One-loop corrections to the fermion-fermion couplings (a)-(i) and
the fermion-disorder strengths (j)-(m) (the solid, dashed
and wavy lines represent the fermion propagator, fermion-fermion interactions,
and disorder scattering, respectively)~\cite{Wang2020PRB,Roy-Saram2016PRB}.}\label{Fig_one-loop-ff-inter}
\end{figure}


\begin{eqnarray}
\Sigma(i\omega,\mathbf{k})
&=&\int\frac{d^3\mathbf{k}d\omega}
{(2\pi)^4}\psi^{\dag}_{\mathbf{k},\omega}
\{[\mathcal{C}_{2}
(\Delta_{1}+\Delta_{2}+\Delta_{3}
+\Delta_{41}+\Delta_{42}+\Delta_{43})
l](i\omega)\nonumber\\
&&-(4\mathcal{C}_{1}\lambda_{2}l)\upsilon_{z}\delta k_{z}\Sigma_{03}-
(4\mathcal{C}_{1}\lambda_{1}l)\upsilon_{p}\delta k_{\perp}
\Sigma_{01}\}\psi_{\mathbf{k},\omega},\label{Eq_self-energy}
\end{eqnarray}
for the one-loop corrections to the noninteracting
fermionic propagator. In addition, the fermion-fermion couplings appearing in
Eq.~(\ref{Eq_S_int-2}) would receive the one-loop corrections from Fig.~\ref{Fig_one-loop-ff-inter} as follows
\begin{eqnarray}
\delta\lambda_1\!&=&\!\!
\lambda_{1}[\!2\mathcal{C}_{3}
(-3\lambda_{1}-2\lambda_{2}-\lambda_{3}+\lambda_{4}
+\lambda_{5}-\lambda_{6})+
2\mathcal{C}_{7}(\Delta_{1}
-\Delta_{2}
+\Delta_{3}-\Delta_{41}
-\Delta_{42}
-\Delta_{43}
-8\Delta_3)\!]l,\label{Eq_one-loop-corrections-lambda-1}\\
\delta\lambda_2\!&=&\!\!\lambda_{2}[\!2\mathcal{C}_{4}
(-2\lambda_{1}-3\lambda_{2}+\lambda_{3}+\lambda_{4}
-\lambda_{5}-\lambda_{6})+2\mathcal{C}_{7}
(-\Delta_{1}+\Delta_{2}
+\Delta_{3}-\Delta_{41}
-\Delta_{42}
-\Delta_{43}
)\!]l,\\
\delta\lambda_3
\!&=&\!\!\{\!\lambda_{3}[2\mathcal{C}_{4}
\left(
-2\lambda_{1}+\lambda_{2}-3\lambda_{3}-\lambda_{4}
+\lambda_{5}+\lambda_{6}
\right)
+2\mathcal{C}_7
(\Delta_2-\Delta_1
+\Delta_3
+\Delta_{41}
+7\Delta_{42}
+\Delta_{43})]
-4\mathcal{C}_6\lambda_5\Delta_{41}\!\}l,\\
\delta\lambda_4\!&=&\!\!
[\!2\lambda_{4}(
\mathcal{C}_2\Delta_1
+\mathcal{C}_2\Delta_2
+\mathcal{C}_2\Delta_3
-\mathcal{C}_2\Delta_{41}
-\mathcal{C}_2\Delta_{42}
+\mathcal{C}_2\Delta_{43})
-
2(2\mathcal{C}_5\lambda_5\Delta_2
+2\mathcal{C}_5\lambda_6\Delta_3
+\mathcal{C}_{4}\lambda_{2}\lambda_{6}
+\mathcal{C}_{3}\lambda_{1}\lambda_{6})
\!]l,\\
\delta\lambda_5\!&=&\!\!
\{\!\lambda_5
[2\mathcal{C}_3
(\lambda_1-2\lambda_2+\lambda_3
+\lambda_4-3\lambda_5-\lambda_6
)+
2(\mathcal{C}_7\Delta_1
-\mathcal{C}_7\Delta_2
+\mathcal{C}_7\Delta_3
+\mathcal{C}_7\Delta_{41}
+\mathcal{C}_7\Delta_{42}
-\mathcal{C}_7\Delta_{43})]
\nonumber\\
&&+2\mathcal{C}_1\lambda_2\lambda_6
-4(\mathcal{C}_6\lambda_3\Delta_{41}
+\mathcal{C}_5\lambda_4\Delta_2
+\mathcal{C}_5\lambda_6\Delta_1)\!\}l,\\
\delta\lambda_6\!&=&\!\!
\{\!\lambda_6
[2\mathcal{C}_1
(-\lambda_1-\lambda_2+\lambda_3
+\lambda_4-\lambda_5-3\lambda_6)
-2(\mathcal{C}_3\lambda_2+\mathcal{C}_4\lambda_1)
+2(-\mathcal{C}_2\Delta_1
-\mathcal{C}_2\Delta_2
+\mathcal{C}_2\Delta_3\nonumber\\
&&
-\mathcal{C}_2\Delta_{41}
-\mathcal{C}_2\Delta_{42}
+\mathcal{C}_2\Delta_{43})]
+2\mathcal{C}_1\lambda_2\lambda_5
-4\mathcal{C}_5(\lambda_4\Delta_3
+\lambda_5\Delta_1)\!\}l.
\end{eqnarray}
Furthermore, the one-loop corrections from Fig.~\ref{Fig_one-loop-ff-inter} contributing to
the disorder scatterings involved in Eq.~(\ref{Eq_S_dis}) take the form of~\cite{Roy-Slager2018PRX,Roy2018PRX}
\begin{eqnarray}
\delta \Delta_1\!&=&\!\!
\{\!\Delta_1
[4\mathcal{C}_2
(\Delta_1+\Delta_2+\Delta_3
+\Delta_{41}+\Delta_{42}+\Delta_{43}
)]
+
8\mathcal{C}_5\Delta_2\Delta_3
\!\}l,\\
\delta \Delta_2\!&=&\!\!
\{\!\Delta_2
[
4\mathcal{C}_2
(\Delta_3-\Delta_1-\Delta_2
+\Delta_{41}+\Delta_{42}+\Delta_{43}
)+
\mathcal{C}_1
(\lambda_1+\lambda_2+\lambda_3
-\lambda_4+\lambda_5-\lambda_6)
]+8\mathcal{C}_5\Delta_1\Delta_3\!\}l,\\
\delta \Delta_3\!&=&\!\!
\{\!\Delta_3
[4\mathcal{C}_7
(\Delta_1-\Delta_2+\Delta_3
-\Delta_{41}-\Delta_{42}-\Delta_{43})
+
\mathcal{C}_3
(\lambda_2-\lambda_1+\lambda_3
-\lambda_4-\lambda_5+\lambda_6)]+
8\mathcal{C}_5\Delta_1\Delta_2\!\}l,\\
\delta \Delta_{41}\!&=&\!\!
\{\!\Delta_{41}
[4\mathcal{C}_7
(\Delta_2-\Delta_1+\Delta_3
-\Delta_{41}+\Delta_{42}+\Delta_{43})
+
\mathcal{C}_4
(\lambda_1-\lambda_2+\lambda_3
+\lambda_4-\lambda_5-\lambda_6)]+
8\mathcal{C}_5\Delta_{42}\Delta_{43}\!\}l,\\
\delta \Delta_{42}\!&=&\!\!
\{\!\Delta_{42}[4\mathcal{C}_7
(\Delta_2-\Delta_1+\Delta_3
+\Delta_{41}-\Delta_{42}+\Delta_{43})
+
\mathcal{C}_4
(\lambda_1-\lambda_2-\lambda_3
+\lambda_4-\lambda_5-\lambda_6)]+
8\mathcal{C}_5\Delta_{41}\Delta_{43}\!\}l,\\
\delta \Delta_{43}\!&=&\!\!
\{\!\Delta_{43}[4\mathcal{C}_7
(\Delta_2-\Delta_1+\Delta_3
+\Delta_{41}+\Delta_{42}-\Delta_{43})
+
\mathcal{C}_4
(\lambda_1-\lambda_2+\lambda_3
-\lambda_4+\lambda_5+\lambda_6)]+8\mathcal{C}_5\Delta_{41}\Delta_{42}\!\}l,\label{Eq_one-loop-corrections-dis}
\end{eqnarray}
where the coefficients $\mathcal{C}_i$ with $i=1-7$ in above
results~(\ref{Eq_self-energy})-(\ref{Eq_one-loop-corrections-dis})
have already been designated in Eqs.~(\ref{Eq_coeff-C1})-(\ref{Eq_coeff-C7}).

\vspace{0.5cm}

\end{widetext}



\end{document}